\documentclass[12pt,twoside]{report} 
\usepackage{feynmf,epsf} 
\newcommand{\be}{\begin{eqnarray}}
\newcommand{\ee}{\end{eqnarray}}
\newcommand{\bc}{\begin{center}}
\newcommand{\ec}{\end{center}}

\begin{document}
\setlength{\unitlength}{1mm}
\begin{fmffile}{oneloop}

\begin{flushright}
{\normalsize NUC-MINN-01/13-T}
\end{flushright}
\vspace*{1cm}

\pagestyle{empty}
\bc
{\Large Non-Equilibrium Aspects of Chiral Field Theories}\\
\vspace*{2cm}
A DISSERTATION\\
PRESENTED TO THE FACULTY OF THE GRADUATE SCHOOL\\
OF THE UNIVERSITY OF MINNESOTA\\
BY\\
\vspace*{2cm}
{\large \'Agnes M\'ocsy}\\
\vspace*{2cm}
IN PARTIAL FULFILLMENT OF THE REQUIREMENTS\\
FOR THE DEGREE OF\\
DOCTOR OF PHILOSOPHY\\
\vspace*{2cm}
Professor Joseph I. Kapusta, Advisor\\
\vspace*{1cm}
September 2001
\ec
\newpage

\bc
\copyright  2001 by \'Agnes M\'ocsy 
\ec

\newpage
\pagestyle{plain}
\pagenumbering{roman}

\bc
{\bf Acknowledgements}
\ec

During my tenure as a graduate student at the University of Minnesota, I enjoyed assistance and support from a number of people, some of whom I wish to recognize here. 

I sincerely express my gratitude to Prof. Joseph Kapusta, my thesis advisor, for his leadership throughout this investigation. I have been inspired by his scientific expertise and have enjoyed his guidance and friendship. His oral and written comments were always extremely perceptive, helpful, and appropriate. He influenced the way I perceive physics and research.

Prof. Igor Mishustin is especially recognized for his enthusiastic supervision and help throughout our collaboration. Part of this thesis is a result of this collaboration. Our discussions and his advice will not be forgotten. I want to thank Prof. Paul Ellis for having his office open for me anytime I wished to discuss physics or other issues. I also thank him for proof reading my dissertation, providing not only useful remarks but many stylistic suggestions, and also for chairing the defense itself. My thanks go out to Prof. L\'aszl\'o Csernai for discussions, his unceasing help and friendship. I thank all of the above scientists for demonstrating their support also with letters of recommendation during my search for employment. The Nuclear Theory Group is acknowledged for the financial support provided through the Department of Energy Grant No. De-FG02-87ER40328 throughout most of my graduate school years. My thanks is also extended to the Graduate School of the University of Minnesota for awarding me the L.T. Dosdall Fellowship for the academic year 2000/2001.

I also wish to acknowledge Prof. Cynthia Cattell and Prof. Diether Dehnhard for taking time from their busy schedules to serve on my final oral committee. I am likewise grateful to Prof. Leonard Kuhi who has additionally acted as a reviewer. 

I would like to acknowledge valuable discussions, advice and friendship with Dr. Rob Pisarski, Dr. Raju Venugopalan, Dr. Edmond Iancu, and Dr. Greg Carter. I thank Dr. Ove Scavenius and Dr. Dirk Rischke for their collaboration.

Many people on the faculty of the School of Physics and Astronomy assisted and encouraged me in various ways during my course of studies. I would like to express my special thanks and deepest respect to Prof. Larry McLerran. My early years in graduate school would not have been the same without Dr. Sangyong Jeon, who provided selfless help at anytime. My graduate studies were also influenced by the social and academic challenges, excitement and cooperation provided by all my graduate student and postdoctoral colleagues. I wish to thank my special friends, Ramin G. Daghigh for discussions, Azwinndini Muronga for volunteering to critically read the manuscript, and Kenichi Torikoshi for teaching me about computers all through these years. My special thanks to Aaron Wynveen for invigorating physics discussions and friendship that I greatly value. I extend my appreciation also to Heather Brown, Mohamed Belkacem, Steven Wong, Igor Shovkovy, Ted Vidnovic, Luis Hernandez, Jeff Bjoraker, Klaus Halterman, Vince Kuo, Hassib Amini, Greg Wagner and any others that I may have left out. 

Finally, I am most personally grateful and forever indebted to my mother, Ildik\'o, and father, L\'aszl\'o, for their endless patience and encouragement when it was most required. They have always been my strongest inspiration and source of strength.  Advice and support from my brother, Laci, has been priceless. I thank my husband, Brad Dahlgaard, not only for believing in me and trying to bear with me even when I did not make things easy, but also demonstrating a great deal of support by moving with me to a different continent. 

\newpage

\bc
{\it Sz\"uleimnek. For my parents.}
\ec

~\clearpage
\bc
{\bf Abstract}
\ec

Different aspects of the non-equilibrium chiral phase transition as well as in-medium properties of the scalar and pseudoscalar mesons is investigated using quantum field theoretical descriptions at nonzero temperatures and densities. The chiral phase transition at nonzero temperature and baryon chemical potential is studied at mean field level in the sigma model that includes quark degrees of freedom explicitly.  For small bare quark masses the critical point separating the first order phase transition line and the smooth crossover region is determined. The behavior of quark and meson masses with changing temperature and chemical potential is examined. Adiabatic lines are computed showing that the critical point does not serve as focusing point in the adiabatic expansion. The linear sigma model without quarks but with field fluctuations beyond the mean field is used to derive coarse-grained equations of motion for the soft homogeneous and inhomogeneous chiral condensate fields coupled to a heath bath. Multiple effects of the heat bath on the chiral condensate fields is studied based on linear response theory. The self-consistently solved gap equation is used to determine the order of the phase transition and the temperature-dependence of meson masses. Using these masses the decay widths of pions and sigma mesons are computed at nonzero temperature at one- and two-loop order in perturbation theory. Numerical results show that in the phase transition region not only is the damping of sigma mesons significant, but also that of the pions. The in-medium modification of the pion dispersion relation is examined. A deviation at finite temperature of the velocity of massless pions from the speed of light is found. Separate investigation is carried out for the model with exact and explicitly broken chiral symmetry, making sure Goldstone's Theorem is fulfilled when  chiral symmetry is spontaneously broken.
~\clearpage
\tableofcontents
\listoffigures
\newpage
\pagenumbering{arabic}

\chapter{Introduction \label{sect-intro}}

According to the Big Bang Theory, the Universe started from an extremely hot 
and dense singularity. As it expanded, it cooled rapidly, passing through 
several phase transitions. The GUT (Grand Unified Theory) transition at about 
$10^{-34}$ seconds after the Big Bang, and at a temperature of about $10^{29}~$
K was followed by the EW (Electro-Weak) transition at about $10^{-12}$s and 
$10^{15}~$K. When the Universe was about $10^{-6}$ seconds old, and the 
temperature was about $150-200~$MeV $\sim 10^{12}~$K the confinement phase 
transition took place. This has great significance since it is due to this 
transition that the world we live in, the hadronic world, has been formed.

The theory that governs the nature of the confinement transition is QCD 
(Quantum Chromodynamics), the fundamental theory of strong interactions. 
QCD is a relativistic quantum field theory and as such very complicated. 
Its finite temperature properties cannot be determined analytically. Therefore,
 instead of direct calculations, physicists need to rely on other approaches. 
Two major trends are lattice QCD and phenomenological effective field theories.
 Throughout this dissertation the focus is on effective field theories.

Why is it so difficult to analyze QCD directly? The usual approach when 
studying a field theory is to perform a perturbative expansion in the coupling,
 keep the dominant terms of the expansion, and then look at the ground state 
and the spectrum of excitations. Such perturbative treatment is reasonable in 
the asymptotically free region of QCD where the coupling is weak, $\alpha_s\ll 
1$, for large momentum transfers ($q^2\gg (100\mbox{MeV})^2$) or small 
distances \cite{gross,politzer}. In this region quarks and gluons, the basic 
degrees of freedom in QCD, interact weakly. In our world, though, quarks and 
gluons appear only in confined states. Confined by the strong interaction, they
 form hadrons (mesons and baryons). At hadronic energy scales ($q^2\leq 
(100\mbox{MeV})^2$), or large distances, the QCD coupling constant is large,
 $\alpha_s\gg 1$, making a perturbative expansion not feasible.   

As an important feature of QCD, it has been predicted and now widely accepted 
that at temperatures about $150~$MeV a deconfinement of hadrons happens 
according to which quarks and gluons move freely. Here perturbative approaches 
should be quite reasonable. When studying QCD at finite temperature, however, 
one encounters infrared divergences which causes the breakdown of perturbation 
theory \cite{dolan}. A rather successful non-perturbative method is putting QCD 
on a lattice and performing numerical analysis. The studies done in the pure 
gauge sector (only gluon degrees of freedom) are of great importance for 
understanding high temperature QCD. Developments including fermion dynamics 
(quarks) have begun, but improvement of the present algorithms is required. 
The most recent papers by the lattice-community \cite{lattice,latticelong} 
report promising results at finite temperature. An alternative way to study 
the fundamental theory is to construct effective field theories. These 
incorporate some aspects of QCD and ignore others. Therefore, by sacrificing 
some of the richness of the full theory, one gains some simplifications that 
in certain cases allow even for analytic solutions. Examples of such theories 
are the $SU(2)_R\times SU(2)_L$ \cite{su2} and $SU(3)_R\times SU(3)_L$ 
\cite{su3} linear sigma models, the nonlinear sigma model \cite{sm}, and the 
Nambu-Jona-Lasinio model \cite{njl}.    

Another important characteristic of the QCD structure is the non-vanishing 
expectation value of quark operators in vacuum. Present at low temperatures, 
the non-zero chiral condensate is nothing else but the Bose condensate of the 
underlying quark degrees of freedom, much like Cooper-pairs in superconductors,
 and liquid $^3$He and electrons in superfluids \cite{bose}. In condensed 
matter physics the normal state does not include condensates. These can 
develop after crossing a phase transition at a characteristic temperature 
$T_c$. In nuclear physics this is the opposite, since the condensate is present
 in the normal state. It is widely believed that above a critical temperature 
of $T_c\sim 150~$MeV it is possible to generate regions where the condensate 
is dissolved.

What is the underlying physics that governs the formation and disappearance of 
condensates? The answer, common in Nature and well-rooted in most major parts 
of physics, is symmetry and breaking of symmetry. These concepts, first 
introduced for explaining superconductivity and superfluidity \cite{bose}, are 
now widely applied in particle and nuclear physics. It is common to all of 
these situations that a continuous symmetry is spontaneously broken at low 
temperatures (ordered phase) and subsequently restored at high temperatures 
(disordered phase) by going through a phase transition. Lattice results suggest
 \cite{oldlattice} that the transition between symmetric and broken phases 
happens together with the hadronization, the confinement of quarks and gluons 
into hadrons. 

A phase transition that happens in thermal equilibrium can always be 
characterized by an order parameter which is nonzero in one phase and vanishes
in the other. For details the reader is referred to \cite{landau}. QCD with 
two massless quarks possesses chiral symmetry, described by the $SU(2)_R\times 
SU(2)_L$ group. This is isomorphic to the $O(4)$ group which describes the 
rotational symmetry in ferromagnets. In both cases the continuous symmetry is 
spontaneously broken at low temperatures, resulting in the emergence of 
Goldstone bosons \cite{goldi}. These are spin-waves in ferromagnets and 
massless pions in the two flavor QCD. The order parameters are the 
magnetization and the chiral condensate, respectively. If the transition is of 
second order universality arguments become rather powerful, providing 
qualitative features of the two massless flavor QCD phase transition 
\cite{piswil}. Nature, however, supplies a different situation. First of all, 
quark masses are non-zero. The small quark masses explicitly break chiral 
symmetry, just as an external magnetic field breaks the rotational symmetry of 
a ferromagnet. Explicit symmetry breaking then alters the nature of the phase 
transition. The second order transition in a ferromagnet becomes a smooth 
cross-over. The change in the order of the chiral phase transition is 
non-trivial (see Chapter \ref{sect-igor} and related references). Second, at 
temperatures of interest three flavor QCD studies become necessary. The effect 
of strange quarks on the phase transition has been recently discussed in 
\cite{lenris} and references therein. Third, the phase transition need not 
happen in local thermal equilibrium circumstances. In case of an 
out-of-equilibrium transition there are no universality arguments one can rely 
on. However, such transitions are known in condensed matter physics (for 
example in superfluids and liquid crystals) and can guide us in nuclear 
physics. There is a major difference, though: while classical approach is 
sufficient for describing a phase transition in condensed matter physics, 
quantum field theoretical treatment is required in nuclear physics. For recent 
advances in equilibrium and non-equilibrium aspects of quantum field theories 
at finite temperature the reader is referred to \cite{workshop}.   

In this dissertation different aspects of the non-equilibrium chiral/confinement phase transition as well as in-medium properties of the scalar and pseudoscalar mesons are investigated within an effective quantum field theoretical model, the linear sigma model. This thesis is structured as follows: In Chapter \ref{sect-igor} the chiral phase transition at nonzero temperature and baryon chemical potential is studied within the linear sigma model including quark degrees of freedom explicitly. In Chapter \ref{sect-dcc} we derive equations of motion for the long-wavelength chiral condensate fields which are in contact with a heat bath. Chapter \ref{sect-masses} presents the temperature-dependence of the equilibrium value of the condensate and that of the self-consistently calculated meson masses. In Chapter \ref{sect-dispersion} the effect of the thermal medium on the velocity of long-wavelength pion modes is discussed and different dispersion relations are examined. The topic of Chapters \ref{sect-dissip1} and \ref{sect-dissip2} is the dissipation of long-wavelength sigma and pion condensate fields at one- and two-loop order, respectively. 

\section{Chiral Symmetry in Quantum Chromodynamics \label{sect-qcd}}

A symmetry in Nature arises whenever a change in a variable leaves the essential physics unchanged. This translates to field theory as invariance of the Lagrangian under field transformations.  Chiral symmetry is characteristic of theories with massless fermions, and as such it is a fundamental symmetry of QCD with two massless quarks. Of course, QCD describes the physics of all six quark flavours. Nevertheless, the low energy limit of QCD is dominated by $u$ and $d$ quarks. The discussion presented below can be naturally generalized from two flavors to three, provided the $s$ quark is considered to be light as well (although this can easily be seen as unreasonable at very low temperatures). Other flavors do not play a role below the natural scale of QCD, $\Lambda_{QCD}\sim 100~$MeV, at which the theory becomes strongly coupled. Mostly in regimes with temperatures $T\gg\Lambda_{QCD}$ and/or baryon density $n_B\gg\Lambda_{QCD}^3$, where QCD describes quarks and gluons that are weakly interacting, is where the physics of heavier quarks becomes relevant.  

The massless QCD Lagrangian is 
\be
L_{QCD} = \bar{q}i\gamma^\mu\left(\partial_\mu+ig\frac{\lambda^a}{2}A_\mu^a\right)q -\frac{1}{4}F_{\mu\nu}^aF^{a\mu\nu}\,~,
\label{lagqcd}
\ee
where $q$ is an $N_f$-dimensional Dirac spinor describing the quark field, $N_f$ being the number of flavors. The index $a$ refers to the color of the gauge field $A_\mu^a$, representing the gluons. The Yang-Mills field strength tensor is 
\be
F_{\mu\nu}^a = \partial_\mu A_\nu^a - \partial_\nu A_\mu^a - gf^{abc} A_\mu^bA_\nu^c\,~.
\ee
Using the chirality projection operators
\be
q_{R,L}=\frac{1}{2}(1\pm\gamma^5)q
\ee
the quark fields can be decomposed into their right- and left-handed components. Then the fermionic part of Lagrangian (\ref{lagqcd}), rewritten as 
\be
L_{QCD}^F = \bar{q_R}i\gamma^\mu\left(\partial_\mu+ig\frac{\lambda^a}{2}A_\mu^a\right)q_R + \bar{q_L}i\gamma^\mu\left(\partial_\mu+ig\frac{\lambda^a}{2}A_\mu^a\right)q_L\,~,
\ee
is invariant under the global transformations
\be
q_{R,L}\rightarrow e^{-i\alpha_{R,L}\gamma^5}q_{R,L}\,~.
\label{chiraltransf}
\ee
This basically corresponds to rotating independently the right- and left-handed components of the spinors that describe the quark fields and are called chiral transformations. The symmetry group of this transformation is $SU(N_f)_R\times SU(N_f)_L$. With every invariance of the Lagrangian under a global transformation there is a conserved current associated with it. This is Noether's Theorem. The Noether current corresponding to the chiral transformations is 
the axial vector current
\be
j^{\mu 5} &=& \bar{q}\gamma^\mu\gamma^5q\nonumber\\
&=& \bar{q}_R\gamma^\mu q_R - \bar{q}_L\gamma^\mu q_L \,~.
\ee
This current is not only conserved when the quarks are massless,
\be
\partial_\mu j^{\mu 5}=0\,~,
\ee
but also does not connect spinors with different chiralities. Dirac bilinears like $\bar{q}q$ and $\bar{q}\gamma^5 q$ have the same quantum numbers as scalar and pseudoscalar mesons, respectively. The vacuum expectation value of the scalar bilinear, what we refer to as the scalar condensate, is nonzero 
\be
\langle\bar{q}q\rangle = \langle\bar{q}_Lq_R\rangle\neq 0\,~.
\ee
This signals that the vacuum mixes left- and right-handed quarks, allowing quarks to acquire an effective mass as they move through the vacuum. We say that the symmetry is spontaneously broken. This means that the Lagrangian maintains invariance under the transformations but the ground state breaks that invariance. For two flavors we say that the global chiral symmetry $SU(2)_R\times SU(2)_L$ is spontaneously broken to a subgroup $SU(2)_{L+R}$. According to Goldstone's Theorem \cite{goldi} the result of spontaneous symmetry breaking is the existence of massless bosons. For two massless quark flavors there exists three massless pseudoscalars. These are the three pions.

The massless quark situation is an idealization. One has to account for small quark masses. The Lagrangian then contains a mass term of the form 
\be
-\bar{q}mq
\ee
where $m$ is a mass-matrix. The presence of this term results in the mixing of left- and right-handed quarks under the chiral transformation (\ref{chiraltransf}):
\be
\bar{q}_Lmq_R + \bar{q}_Rmq_L
\ee
Thus chirality is not preserved anymore. Because the symmetry is not exact the axial current is not conserved either
\be
\partial_\mu j^{\mu 5} = 2im\bar{q}\gamma^5q\,~.
\ee
This non-vanishing divergence measures the degree to which chiral symmetry is broken. The small pion mass is the result of explicit symmetry breaking by small quark masses. A rather comprehensive discussion of this physics is presented in a recent article \cite{hands}.

\section{The Linear Sigma Model \label{sect-sigmamodel}}

Effective field theories are extremely useful tools in understanding QCD physics. In such theories the QCD Lagrangian (\ref{lagqcd}) given in terms of the basic quark and gluon degrees of freedom, is exchanged for Lagrangians built out of confined states of these (baryons, mesons, glueballs). The soundness of this shift in the degrees of freedom is demonstrated in \cite{greg} using the path integral formalism. For details of this procedure the reader is referred to \cite{sm,peskin}.

The linear sigma model \cite{su2} is one of the most instructive of all effective field theories. It describes the low energy non-perturbative sector of QCD extremely well. When using this model one gains access to analytical derivations and accessible numerical solutions, albeit at the expense of losing some properties, such as confinement.
The Lagrangian of the sigma model is
\be
{\cal L} = \bar{\psi}\left[i\gamma^{\mu}\partial_{\mu}-g(\sigma+i\gamma_{5}\vec{\tau}\cdot\vec{\pi})\right]\psi + \frac{1}{2}\left(\partial_{\mu}\sigma\partial^{\mu}\sigma + \partial_{\mu}\vec{\pi}\cdot\partial^{\mu}\vec{\pi}\right) - U(\sigma,\vec{\pi}) \,\, .
\label{lagsigma}
\ee
Here $\psi$ is a fermion field which, in early works has been considered to describe nucleons, but can also be identified with quarks as we shall discuss this in Chapter \ref{sect-igor}. The scalar field $\sigma$ and the pseudoscalar field $\vec{\pi} =(\pi_{1},\pi_{2},\pi_{3})$ together form a chiral field $\Phi =(\sigma,\vec{\pi})$. The usual choice for the potential is 
\be
U(\sigma,\vec{\pi}) = \frac{\lambda}{4}\left(\sigma^{2}+\vec{\pi}^{2}-{\it v}^{2}\right)^{2} - H\sigma \,\,~.
\ee
The coupling constant between mesons is $\lambda$ and between mesons and fermions is $g$ and are dimensionless. The theory thus is renormalizable. The Lagrangian (\ref{lagsigma}) is invariant under chiral $SU(2)_L \times SU(2)_R$ transformations if the explicit symmetry breaking term $H\sigma $ is zero. The parameters of the Lagrangian are usually chosen such that chiral symmetry is spontaneously broken in the vacuum. The potential resembles the bottom of a wine bottle. The usual choice for the expectation values of the meson fields are $\langle\sigma\rangle =v={\it f}_{\pi}$ and $\langle\vec{\pi}\rangle =0$, where ${\it f}_{\pi}=93$ MeV is the pion decay constant. Different choices of ground states can be interchanged by a chiral rotation in the circle that is the bottom of the potential. Defining fluctuations about the vacuum as 
\be
\sigma = v + \sigma'
\ee
and inserting this into (\ref{lagsigma}) results in a mass term for fermions $M=gf_\pi$ and for the sigma $m_\sigma^2=2\lambda f^2_\pi$. These mass terms break the symmetry of the Lagrangian, illustrating that spontaneous breaking of symmetry corresponds to a particular choice of ground state out of the degenerate set. Associated with this mechanism are massless pion excitations. 

The symmetry is explicitly broken for any $H\neq 0$. The vacuum is not degenerate anymore, the potential is tilted toward a unique value, and pions are not massless. The constant $H$ is fixed by the PCAC relation which gives $H=f_\pi m_\pi^2$, where $m_\pi=138$ MeV is the pion mass in vacuum. Then one finds $v^{2}=f^{2}_{\pi}-m^{2}_{\pi}/\lambda$. The coupling constant $\lambda$ is determined by the sigma mass, $m_\sigma^2=2\lambda f^2_\pi + m^2_\pi$, which is about $600~$MeV in vacuum, yielding $\lambda\approx 20$. The coupling constant $g$ is usually fixed by the requirement that the fermion mass is the mass of the nucleon or that of the constituent quark in vacuum, which is about $1/3$ of the nucleon mass.

The linear sigma model was studied quite intensely at finite temperatures (see the pioneering works \cite{linde,dolan}). The isomorphism of the $SU(2)\times SU(2)$ symmetry group with the $O(4)$ group led to studies based on universality arguments \cite{piswil}. Consequently, the order parameter of this model, the quark condensate $\langle\bar{q}q\rangle$, and its temperature dependence is expected to have similar behavior as magnetization in a ferromagnet. The sigma model without including fermions undergoes a second order phase transition which leads to the restoration of broken symmetry above a critical temperature \cite{bochkarev}.

\section{Ultra-Relativistic Heavy Ion Collisions \label{sect-rhic}}

Ultra-relativistic heavy ion collisions offer a possibility to study the properties and the behavior of high temperature hadronic matter with the hope of detection of the deconfinement phase transition. Experimental programs at the European Center for Nuclear Research (CERN) SPS and Brookhaven National Laboratory (BNL) AGS have focused on this research up until 1999. In 2000 the Relativistic Heavy Ion Collider (RHIC) at BNL started running at never before reached energies. There are plans for even higher energies at the LHC at CERN, which currently is under construction, and will starts its operation in 2007.

In these collisions heavy atoms such as lead (Pb$^{208}$) and gold (Au$^{197}$) are stripped of their electrons, accelerated to relativistic speeds and smashed into each other. In central collisions of two fast ions a considerable compression of matter occurs, leading to a very hot and dense nuclear matter. After such energetic impact ($100~$GeV/nucleon in the center of momentum frame at RHIC) temperatures are expected to reach above the critical temperature ($\sim 150~$MeV), creating a deconfined form of hadronic matter. In this state of matter quarks and gluons are moving freely without the constraint of a confining potential. This soup of the basic constituents of matter is known as quark-gluon plasma (QGP). The possible formation and evolution of QGP is the most stimulating phenomena in relativistic heavy ion physics and has been the subject of intense investigations in the last couple of decades both from theoretical and experimental perspectives. For a review on the advances see \cite{qgpreviews}. 

 Due to the very high internal pressure the system expands and cools rapidly. The initial stages of a QGP are dominated by hard scatterings of quarks and qluons, leading to thermalization \cite{raju}. Once a local thermal equilibrium is achieved a hydrodynamical description is feasible \cite{jd}. During this hydrodynamic evolution the plasma expands and cools adiabatically. When reaching a critical temperature rehadronization sets in. After the newly formed hadrons (mostly pions) freeze-out (cease interacting), these fly to the detectors. 

There is a wast amount of physics originating in the above very schematically described scenario. This is partly because only indirectly, by inferring back from what is recorded in the detectors, can the formation of QGP be detected (see for example \cite{rw,gale}). For any further phenomenological or technical details, or for viewing images of current heavy ion experiments at RHIC see \cite{rhic}.  
\chapter{Chiral Phase Diagram \label{sect-igor}}

In this chapter the order of the chiral symmetry restoring phase transition 
is studied at finite temperatures and baryon densities. The linear sigma 
model coupled to quarks is used in the mean field approximation to verify the 
existence of the tricritical point and to find the line of first order 
transition in the phase diagram. Then the behavior of quark and meson masses 
with changing temperatures and chemical potentials is examined. Finally, the 
adiabatic lines are computed in the $(T,\mu_B)$ plane. 

\section{Review of Previous Results}
 
The idea that the spontaneously broken chiral symmetry of the QCD vacuum is 
restored at high temperatures is well accepted. The order of the transition, 
though, has not been determined unambiguously.   

It is important to determine the order of the chiral transition, as this 
influences the dynamical evolution of the system. It has been shown that a 
first order transition in rapidly expanding matter may manifest itself by 
strong non-statistical fluctuations due to droplet formation \cite{igor}. In 
the case of strong supercooling it may lead to large fluctuations due to 
spinodal decomposition \cite{ms,oa}. In a second order phase transition one 
may expect the appearance of critical fluctuations due to a large correlation 
length \cite{srs}. Experimentally, large-acceptance detectors are now able to 
measure average as well as event-by-event observables, which in principle can 
distinguish between scenarios with a first order, a second order, or merely a 
crossover type of phase transition.

Theoretically, the QCD phase diagram in the $(T,\mu_B)$ plane has recently 
received much attention \cite{srs,halasz,klevansky,arw}. QCD with two flavors 
of massless quarks has a global $SU(2)_L \times SU(2)_R$ symmetry (see 
Chapter \ref{sect-intro}). This symmetry is spontaneously broken in the QCD 
vacuum, such that the order parameter $\langle\bar{q}_L q_R\rangle$ acquires 
a non-vanishing expectation value, where $q$ is the two-flavor quark field. 
At zero baryon chemical potential the effective theory for this order 
parameter is the same as the $O(4)$ model which has a second order phase 
transition. Therefore, by universality arguments \cite{piswil}, the chiral 
transition in two flavor QCD is likely to be of second order at $\mu_B = 0$. 
Nonzero quark masses introduce a term in the QCD Lagrangian which explicitly 
breaks chiral symmetry. Then there is no theoretical argument for the 
existence of a phase transition in the strict sense. In this case the second 
order transition becomes a crossover. 

At nonzero baryon chemical potential it is more difficult to infer the order 
of the chiral transition from universality arguments \cite{hsu}. One commonly 
resorts to phenomenological models to describe the chiral transition in this 
case. Depending on the parameters of these models they may predict a first 
order, a second order, or a crossover transition. However, if there is a s
econd order phase transition for $\mu_B=0$ and nonzero $T$ and a first order 
transition for small $T$ and nonzero $\mu_B$, then there exists a tricritical 
point in the $(T,\mu_B)$ plane where the line of first order phase 
transitions meets the line of second order phase transitions. For nonzero 
quark masses, this tricritical point becomes a critical point. Very recent 
lattice calculations \cite{hu} reported the determination of this critical 
point.  

It has been proposed \cite{srs} that this point could lead to interesting 
signatures in heavy-ion collisions at intermediate energies if the evolution 
went through or close to this critical point. At this point, susceptibilities 
(e.g.\, the heat capacity) diverge, the order parameter field becomes massless
 and consequently exhibits strong fluctuations. This could be detected in 
event-by-event observables. A thorough discussion on how to find the QCD 
critical point is given in \cite{critpoint}.

\section{Thermodynamics}

The thermodynamics of two effective field theoretical models, the linear 
sigma model and the Nambu-Jona-Lasinio model, has been recently discussed and
 compared in \cite{paper}. Here we present the details and results of the 
investigation of the linear sigma model which includes quark degrees of 
freedom. The use of such model is justified since high temperatures are 
dominated by quarks and low temperatures are dominated by mesonic excitations.
 Our model differs from other, more common realizations of the sigma model, 
where quark degrees of freedom are neglected but mesonic excitations are 
included (see for instance reference \cite{randrup}). We believe that our 
approach  is more justified at high $T$ and $\mu$ when constituent quarks 
become light but mesonic excitations are heavy (see below). Of course, due to 
the confining forces, at low $T$ and $\mu$ quarks and antiquarks will 
recombine into mesons, baryons and antibaryons. We can only hope that this 
hadronization process will not change drastically the character of the chiral 
transition which we study here. Let us recall the Lagrangian (\ref{lagsigma}) 
of this model
\be
{\cal L} = \bar{q}\left[ i\gamma ^{\mu}\partial _{\mu}-g(\sigma +i\gamma _{5}\vec{\tau} \cdot \vec{\pi} )\right] q+ \frac{1}{2} \left(\partial _{\mu}\sigma \partial ^{\mu}\sigma + \partial _{\mu}\vec{\pi} \cdot \partial ^{\mu}\vec{\pi}\right)-U(\sigma ,\vec{\pi}) \,\, ,
\label{sigma}
\ee
where the potential is
\be
U(\sigma ,\vec{\pi} )=\frac{\lambda}{4} \left(\sigma^{2}+\vec{\pi}~^{2} -{\it v}^{2}\right)^{2}-H\sigma \,\, .
\ee
Here $q$ is the light quark field $q=(u,d)$, the scalar $\sigma$ field describes the sigma meson and $\vec{\pi} =(\pi_{1},\pi_{2},\pi_{3})$ the three pions. The choice of the parameters of the Lagrangian are described in section \ref{sect-sigmamodel}. When $H=0$ chiral symmetry is spontaneously broken in the vacuum and the expectation values of the meson fields are $\langle\sigma\rangle ={\it f}_{\pi}$ and $\langle\vec{\pi}\rangle =0$. With our model parameters the couplings are $\lambda= 20$ and $g = 3.3$.

Let us consider a spatially uniform system in thermodynamical equilibrium at temperature $T$ and quark chemical potential $\mu\equiv \mu_B/3$, where $\mu_B$ is the baryon chemical potential. In thermodynamics the most important function is the partition function, because from it every other thermodynamical quantity can be determined. In general, the grand partition function reads \cite{kapusta}:
\be
{\cal Z}={\rm Tr} \, \exp \left[-\left(\hat{\cal H}-\mu \hat{\cal N}\right)/T \right] = \int{\cal D}\bar{q}\, {\cal D}q\, {\cal D}\sigma\, {\cal D}\vec{\pi}\; \exp\left[\int_x \left({\cal L}+\mu \, \bar{q} \gamma^0 q\right) \right]\,~,
\ee
where $\int_x \equiv \int_0^{\beta} d\tau \int_V d^3 {\bf x}$ and $V$ is the volume of the system. $\hat{\cal H}$ and $\hat{\cal N}$ are the Hamiltonian and the number operator. In the following we adopt the mean-field approximation. This consists of replacing $\sigma$ and $\vec{\pi}$ in the exponent by their expectation values. In other words, we neglect both quantum and thermal fluctuations of the meson fields and retain only quarks and antiquarks as quantum fields. Then, up to an overall normalization factor:
\be
{\cal Z} & = & \exp\left(-\frac{VU}{T}\right) \;\int{\cal D}\bar{q}\, {\cal D}q\; \exp\left\{ \int_x \bar{q} \left[ i\gamma ^{\mu}\partial _{\mu}-g(\sigma +i\gamma _{5}\vec{\tau} \cdot \vec{\pi})\right]q + \mu \bar{q} \gamma^0 q \right\} \nonumber \\
& = & \exp\left(-\frac{VU}{T}\right){\rm det}_p \left\{ \left[ p_{\mu}\gamma^{\mu} + \mu \gamma^0 -g(\sigma +i\gamma_{5}\vec{\tau} \cdot \vec{\pi}) \right]/T\right\}\,\, .
\label{meanz}
\ee
The grand canonical potential is
\be
\Omega (T,\mu)=-\frac{T\ln {\cal Z}}{V}=U(\sigma,\vec{\pi})+ \Omega_{q\bar{q}}\,~,
\label{ptotsig}
\ee
where the contribution of quarks and antiquarks follows from equation (\ref{meanz}) and is
\be
\Omega_{q\bar{q}}(T,\mu) = -\nu_{q} \int \frac{d^3{\bf p}}{(2\pi)^3}\left\{E+T\, \ln \left[1+e^{(\mu-E)/T}\right] + T\, \ln \left[1+e^{(-\mu-E)/T} \right] \right\} \,~.\nonumber\\
\label{pqq}
\ee
Here $\nu_q = 2 N_c N_f = 12$ is the number of internal degrees of freedom of the quarks with three colors $N_c = 3$, and $E=\sqrt{p^2+M^2}$ is the energy of quarks and antiquarks. The constituent quark/antiquark mass, $M$, is defined to be:
\be
M_q^2=g^2(\sigma^2+\vec{\pi}~^2)\,\, .
\label{qmass}
\ee
The divergent first term in equation (\ref{pqq}) comes from the negative energy states of the Dirac sea. As follows from the standard renormalization procedure \cite{peskin} it can be partly absorbed in the coupling constant $\lambda$ and the constant $v^2$. However, a logarithmic correction from the renormalization scale remains. This term is neglected in the following calculations. After integrating equation (\ref{pqq}) by parts the pressure, $P=-\Omega$, due to quarks and antiquarks can be written as 
\be
P_{q\bar{q}}(T,\mu) = \frac{\nu_q}{6\pi^2}\int^{\infty}_0 dp \frac{p^4}{E}\left [n_q(T,\mu)+n_{\bar{q}}(T,\mu)\right ],
\ee
where $n_q$ and $n_{\bar{q}}$ are the quark and antiquark occupation numbers,
\be 
n_q(T,\mu) &=& \frac{1}{1+e^{(E-\mu)/T}} \nonumber\\
n_{\bar{q}}(T,\mu) &=& n_q(T,-\mu)\,\, .
\label{occunum}
\ee
The baryon-chemical potential is determined by the net baryon density
\be
n_B = -\frac{1}{3}\frac{\partial \Omega}{\partial \mu} =\frac{\nu_q}{6\pi^2}\int p^2dp [n_q(T,\mu)-n_{\bar{q}}(T,\mu)]\,~.
\ee
The net quark density is $n=3n_B$. The values for the $\sigma$ and $\vec{\pi}$ fields and thereby the quark masses in equation (\ref{qmass}) are obtained by minimizing the potential $\Omega$ with respect to $\sigma$ and $\vec{\pi}$. The results are equations of motion for the meson fields:
\be
\frac{\partial \Omega }{\partial \sigma}=\lambda(\sigma^2+\vec{\pi}^2-v^2)\sigma-H+g\rho_{s}=0\,\, ,
\label{smmass}
\ee
and
\be
\frac{\partial \Omega }{\partial \vec{\pi}}=\lambda(\sigma^2+\vec{\pi}^2-v^2)\vec{\pi}+g\vec{\rho}_{ps}=0\,\, .
\label{pmmass}
\ee
The scalar and pseudoscalar densities of quarks and antiquarks can be expressed as \cite{csernai}:
\be\label{scalardensity}
\rho_s=\langle\bar{q}q\rangle=g \sigma \nu_q \int \frac{d^3{\bf p}}{(2\pi)^3}\frac{1}{E}[n_q(T,\mu)+n_{\bar{q}}(T,\mu)] \,\, ,
\label{scaldens}
\ee
\be
\vec{\rho}_{ps}=\langle\bar{q}i\gamma_5\vec{\tau}q\rangle=g \vec{\pi} \nu_q \int \frac{d^3{\bf p}}{(2\pi)^3}\frac{1}{E}[n_q(T,\mu)+n_{\bar{q}}(T,\mu)] \,\, .
\ee
These densities generate the source terms in the equations of motions for the meson fields (\ref{smmass}), (\ref{pmmass}). The minima of $\Omega$ defined by eqs. (\ref{smmass}), (\ref{pmmass}) correspond to stable or metastable states of matter in thermodynamical equilibrium where the pressure is maximal $P_{max}=-\Omega_{min}$.

 This version of the sigma model was used earlier in \cite{csmm} for thermodynamical calculations at nonzero $T$ and $\mu=0$, and at nonzero $\mu$ and $T=0$. Some useful formulae for the case of a small quark mass are given in Appendix \ref{app-igor}.

\section{Phase Diagram}

In our model everything is determined when the gap equation (\ref{smmass}) is 
solved in the $(T,\mu)$ plane. Below we present results of our numerical 
calculations.
\begin{figure}[htb]
\begin {center}
\leavevmode
\hbox{%
\epsfysize=10cm
\epsffile{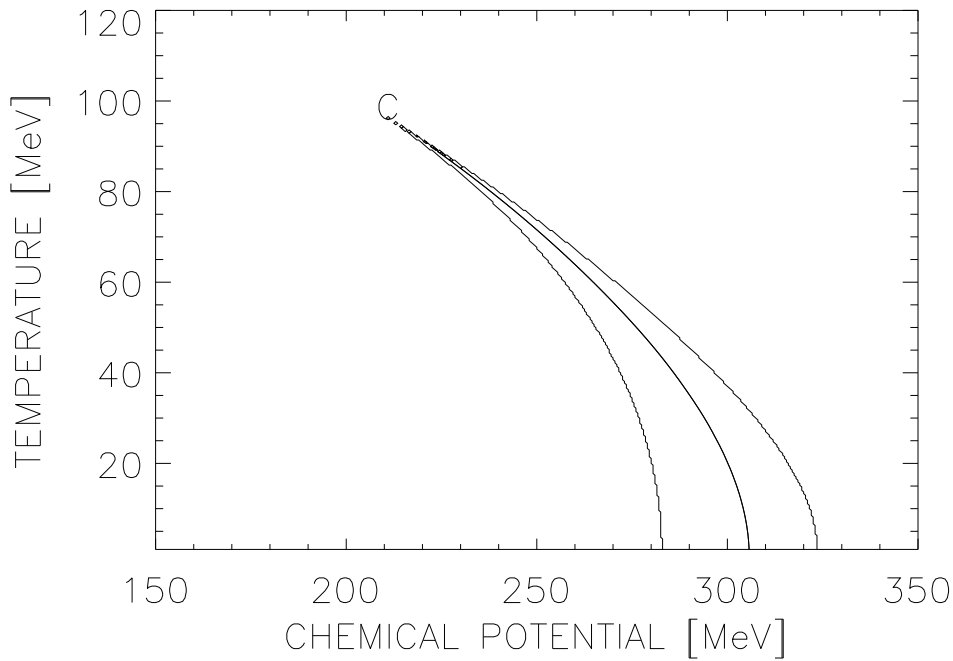} 
}
\caption{\small{The phase diagram in the ($\mu,T$) plane. The middle curve is 
the critical line and the outer lines are the lower and upper spinodal lines. 
C is the critical point.}\label{ptdia}}
\end{center}
\end{figure}
As our calculations show, the model exhibits a first order chiral phase 
transition at $T<T_c$ and nonzero chemical potential. In figure \ref{ptdia} 
we present the resulting phase diagram in the $(T,\mu)$ plane. The middle line
 corresponds to states where two phases co-exist in the first order phase 
transition. 
\begin{figure}[htp]
\begin {center}
\leavevmode
\hbox{%
\epsfysize=10cm
\epsffile{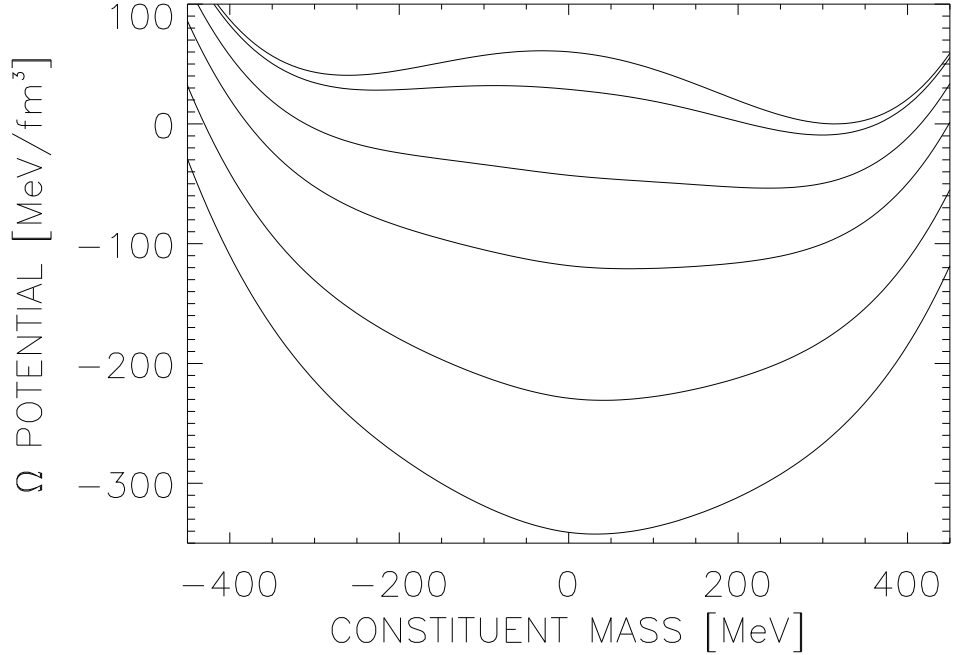} 
}
\caption{\small{The thermodynamical potential $\Omega$ for $\mu=0$. 
The levels correspond to (starting from the top): 
$T=[0,100,135,155,175,190]~$MeV}\label{pott}}
\end{center}
\end{figure}
Along this line the thermodynamical potential $\Omega$ has two minima of equal
  depth separated by a potential barrier whose height grows towards lower 
temperatures. At the critical point C the barrier disappears and the 
transition is of second order. The other lines in figure \ref{ptdia} are 
spinodal lines which constrain the regions of spinodal instability where $
\left(\partial n_B/\partial\mu\right)_T<0$. Information about the timescales 
of this instability is important for dynamical simulations of, for example, 
disoriented chiral condensate formation \cite{oa} or quark-antiquark droplet 
formation \cite{ms}.

\begin{figure}[htp]
\begin {center}
\leavevmode
\hbox{%
\epsfysize=10cm
\epsffile{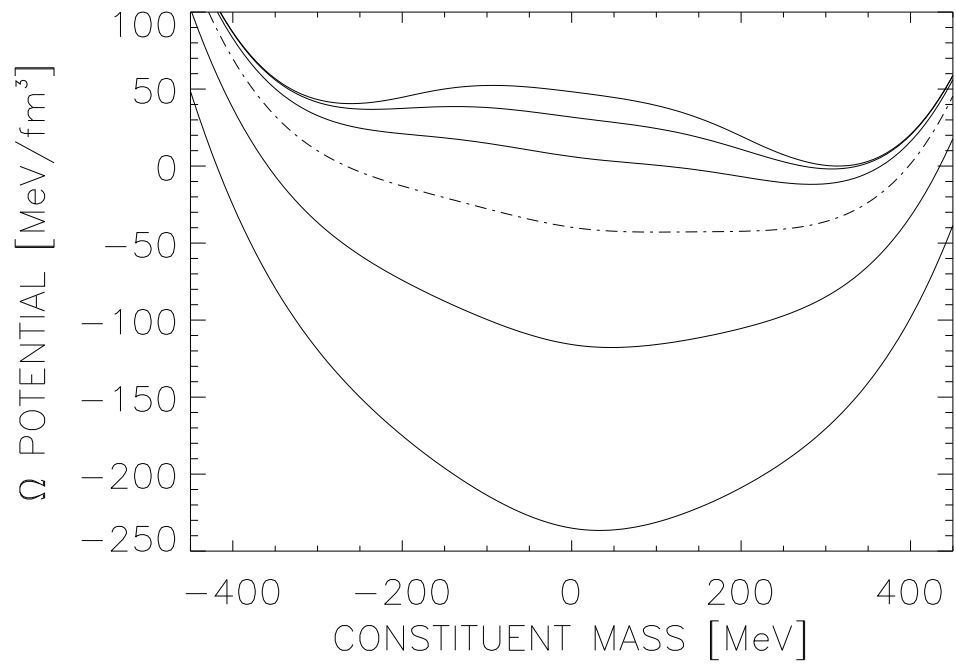} 
}
\caption{\small{The thermodynamical potentials $\Omega$ for $\mu=207~$MeV. 
The levels correspond to (starting from the top): 
$T=[0,50,75,100,125,150]~$MeV.}\label{potcrit}}
\end{center}
\end{figure}

It is instructive to plot the thermodynamic potential as a function of the 
order parameter for various values of $T$ at $\mu=0$, and for various values 
of $\mu$ at $T=0$. The first case is shown in figure \ref{pott}. One clearly 
sees the smooth crossover of the symmetry breaking pattern. 
\begin{figure}[htp]
\begin {center}
\leavevmode
\hbox{%
\epsfysize=10cm
\epsffile{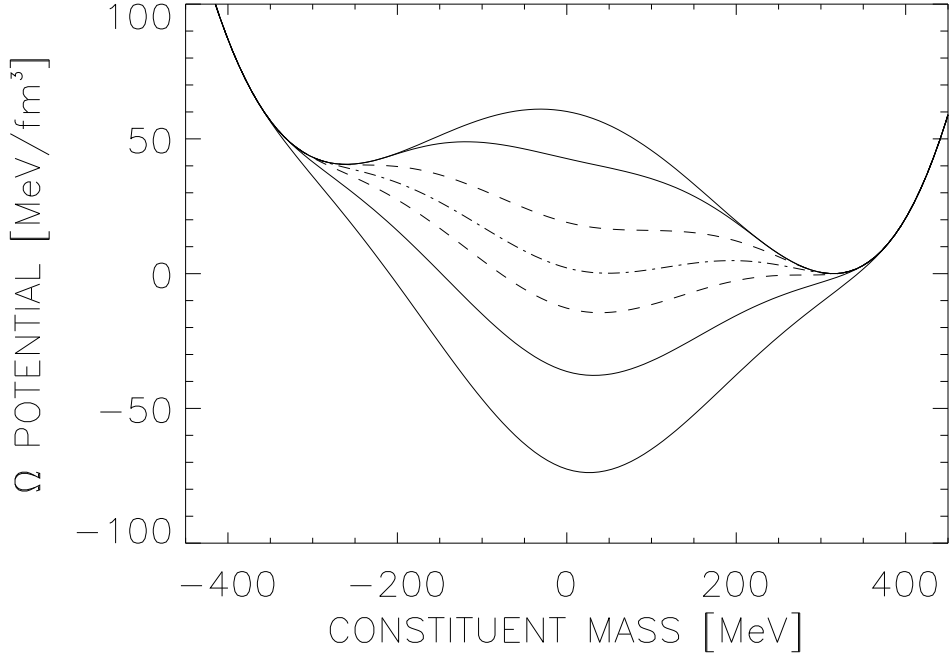} 
}
\caption{\small{The thermodynamical potentials $\Omega$ for $T=0$. The levels 
correspond to (starting from the top): $\mu=[0,225,279,306,322,345,375]~$MeV.}
\label{potmu}}
\end{center}
\end{figure}
The energy difference between the global minimum and the local maximum of the 
potential in vacuum, otherwise known as the bag constant, is about 
$60$ MeV/fm$^3$. The temperature corresponding to the crossover transition is 
about 140 MeV. In figure \ref{potmu} the same plot is shown for $T=0$ and a 
nonzero $\mu$. Here, one clearly observes the pattern characteristic for a 
first order phase transition: two minima corresponding to phases of restored 
and broken symmetry separated by a potential barrier. It now follows that 
somewhere in between these two extremes, for some $\mu_c$ and $T_c$, there 
exists a critical point for a second order phase transition. Indeed, this 
point is found  and shown in figure \ref{ptdia}. The corresponding values 
are $(T_c,\mu_c)\simeq (99,207)$ MeV. The behavior of the thermodynamic 
potential at $\mu=\mu_c$ and various $T$ is shown in figure \ref{potcrit}. 
One can see that the potential has only one minimum which is flattest at the 
critical point.
%

\section{Effective Masses}

The masses of the sigma meson and pion are determined by the curvature of the 
potential at the global minimum:
\be
M^2_{\sigma} = \frac{\partial^2\Omega}{\partial \sigma^2}~~~~\mbox{and}~~~~~
M^2_{\pi_i} = \frac{\partial^2\Omega}{\partial \pi_i^2}\,~,~~~i=1,2,3\,~.
\label{mmass}
\ee
Explicitly they are given by the expressions
\be
M^2_{\sigma}& =& m^2_\pi+\lambda\left (3\frac{M^2}{g^2}-f^2_\pi\right)
\nonumber \\
&+& g^2\, \frac{\nu_q}{2\pi^2}\int dp\,p^2\;\left[\frac{p^2}{E^3}
\left(\frac{1}{1+e^{(E+\mu)/T}}+\frac{1}{1+e^{(E-\mu)/T}}\right)\right.
\nonumber \\
&-&\left.  \frac{M^2}{T\,E^2} \left(\frac{1}{2(1+\cosh[(E+\mu)/T])}
+\frac{1}{2(1+\cosh[(E-\mu)/T])} \right) \right]
\ee
and
\be
M^2_{\pi} &=& m^2_{\pi} + \lambda\left(\frac{M^2}{g^2}-f^2_\pi\right)
\nonumber \\
&+& g^2\, \frac{\nu_q}{2\pi^2}\int dp\, p^2 \; \frac{1}{E}\left[\frac{1}
{1+e^{(E+\mu)/T}}+\frac{1}{1+e^{(E-\mu)/T}}\right]\,~.
\ee
Since we have set the expectation value of the pion field to zero, 
$\langle\vec{\pi}\rangle = 0$, the quark mass is given by
\be
M_q^2=g^2\sigma^2\,~.
\ee
\begin{figure}[htp]
\begin {center}
\leavevmode
\hbox{%
\epsfysize=11cm
\epsffile{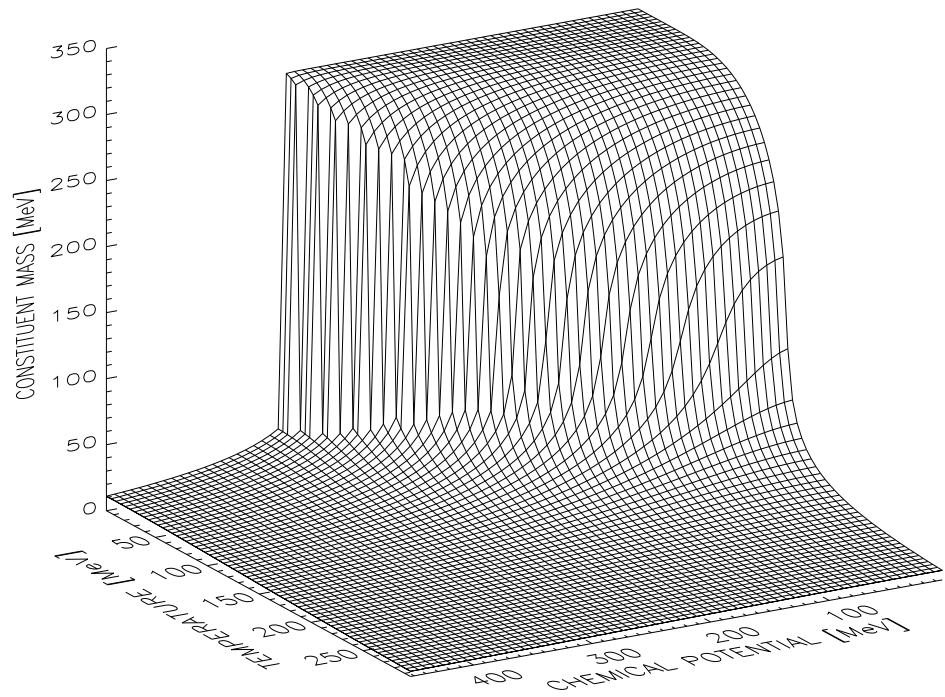} 
}
\caption{The constituent quark/antiquark mass as a function of $\mu$ and $T$.
\label{qmassfig}}
\end{center}
\end{figure}

Figure \ref{qmassfig} shows the constituent quark mass as a function of $T$ 
and $\mu$. This plot, of course, shows the same phase structure as discussed 
in the previous section. At $\mu=0$ the quark mass falls smoothly from its 
vacuum value and approaches zero as $T$ goes to infinity. One could define a 
crossover temperature as corresponding to a steepest descent region in the 
variation of $M$. This again gives a temperature of about $140-150~$MeV. At 
$T=0$ and nonzero $\mu$ the constituent quark mass shows a discontinuous 
behavior reflecting a first order chiral transition.

\begin{figure}[htp]
\begin {center}
\leavevmode
\hbox{%
\epsfysize=10cm
\epsffile{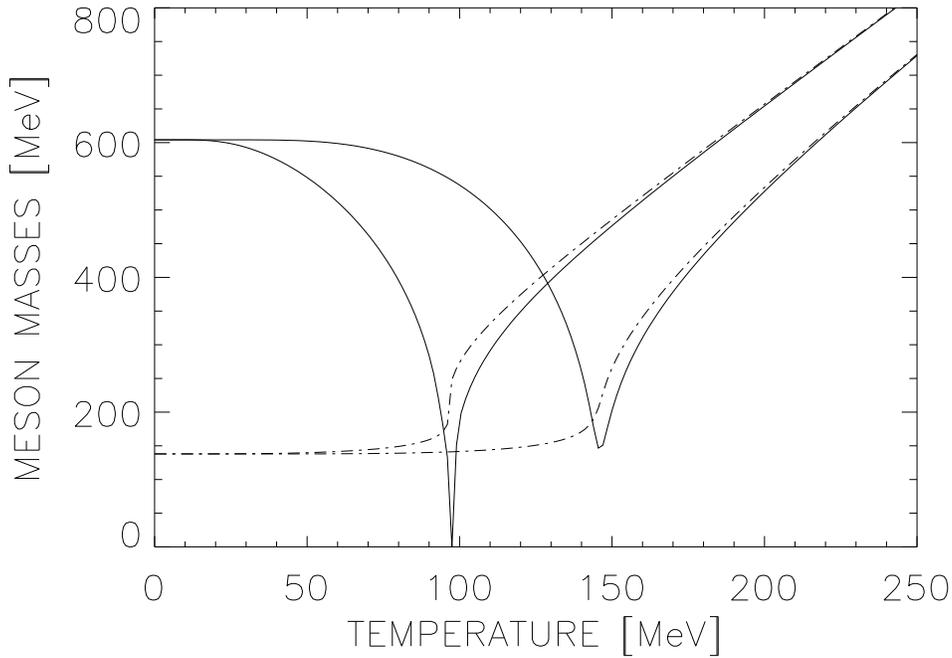} 
}
\caption{\small{The sigma mass (solid line) and pion mass (dashed line) as 
functions of temperature for $\mu=0$ (right pair) and for $\mu=\mu_{c}$ (
left pair).}\label{mesmasst}}
\end{center}
\end{figure}

\begin{figure}[t]
\begin {center}
\leavevmode
\hbox{%
\epsfysize=10cm
\epsffile{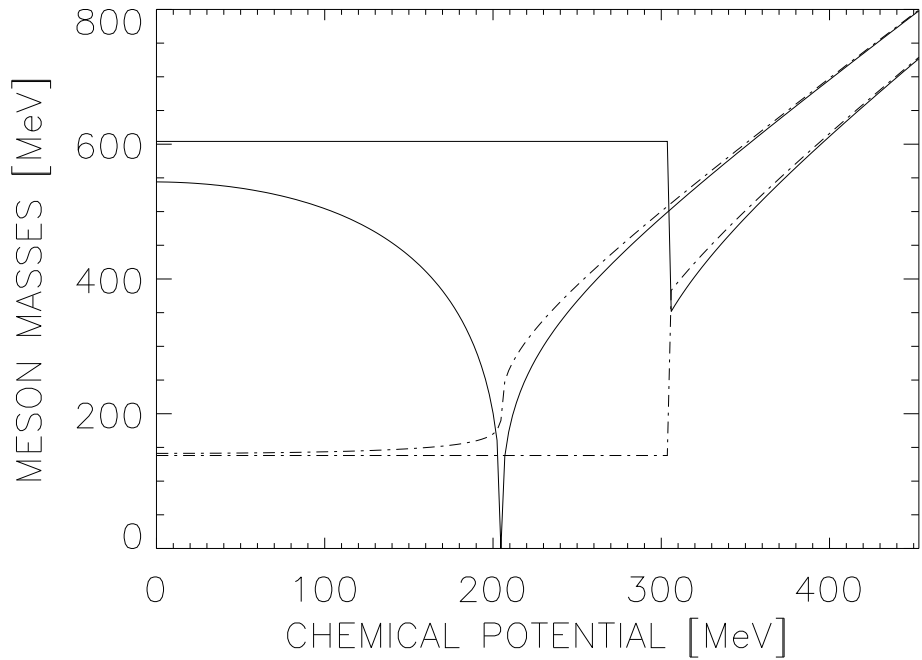} 
}
\caption{\small{The sigma mass (solid line) and pion mass (dashed line) in 
the as functions of chemical potential for $T=0$ (right pair) and for 
$T=T_{c}$ (left pair).}\label{mesmassmu}}
\end{center}
\end{figure}

The sigma and pion masses for various $T$ and $\mu$ are shown in figures 
\ref{mesmasst} and \ref{mesmassmu}. One can see that the sigma mass first 
decreases smoothly, then rebounds and grows again at high $T$. The pion mass 
does not change much at temperatures below $T_c$ but then 
increases rapidly, approaching the sigma mass and signaling the restoration 
of chiral symmetry.
At large $T$ the masses grow linearly with $T$ (see 
Appendix \ref{app-igor}). The $\mu=\mu_c$ case is especially interesting. 
Since the sigma field is the order parameter of the chiral phase transition, 
its mass must vanish at the critical point for a second order phase 
transition, $M_q\sim\sigma\sim\langle \bar{q}q\rangle$. This means that 
$\Omega$ has zero curvature at this point. Figure \ref{mesmasst} indeed shows 

In Fig. \ref{mesmassmu} the masses are plotted as function of $\mu$ for 
$T=0$ and $T=T_c$. For $T=0$ one clearly sees discontinuities in the behavior 
of the masses, characteristic for the first order phase transition.
\begin{figure}[htbp]
\begin {center}
\leavevmode
\hbox{%
\epsfysize=8.5cm
\epsffile{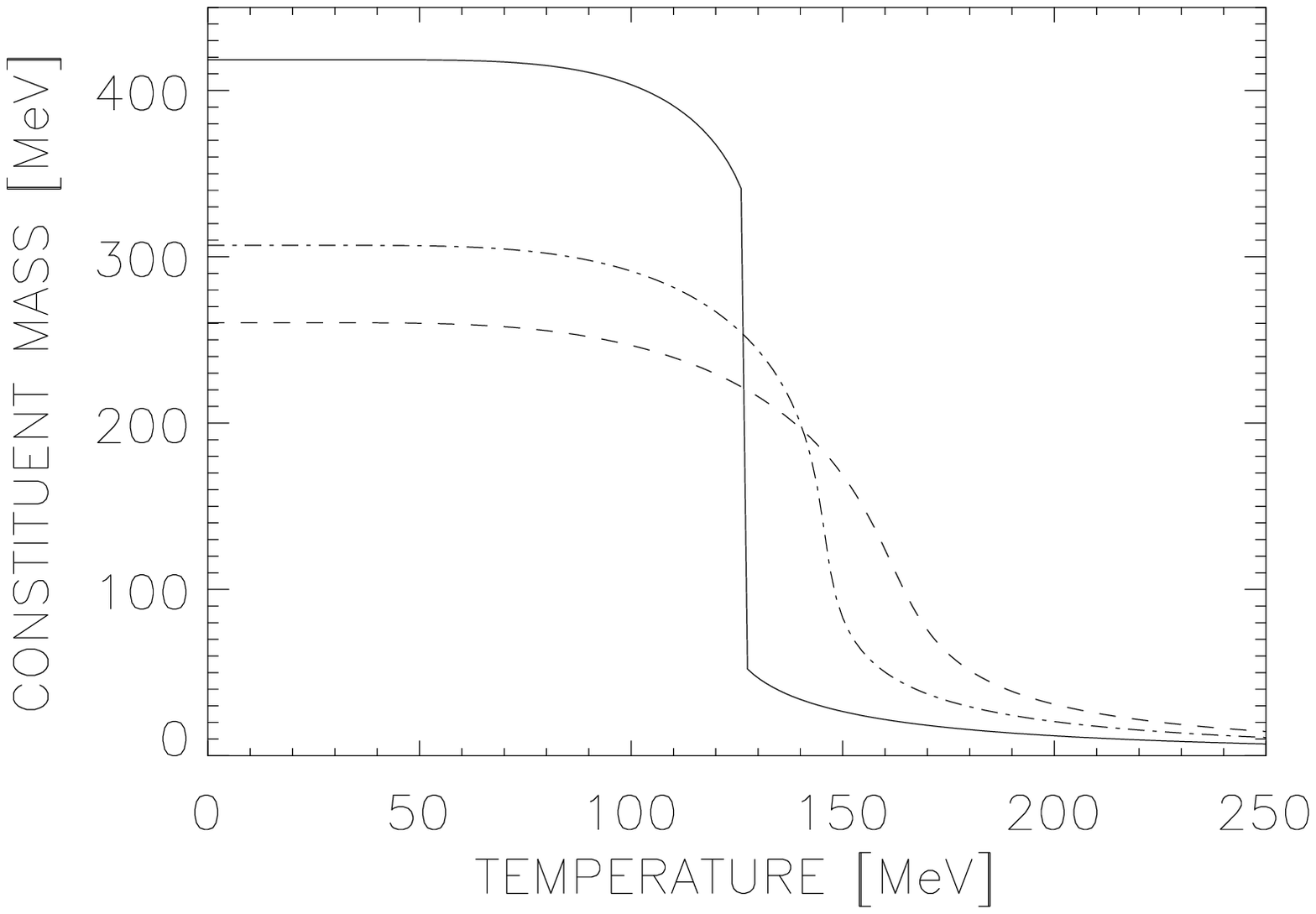} 
\hspace{-1.0cm}}
\hbox{%
\epsfysize=8.5cm
\epsffile{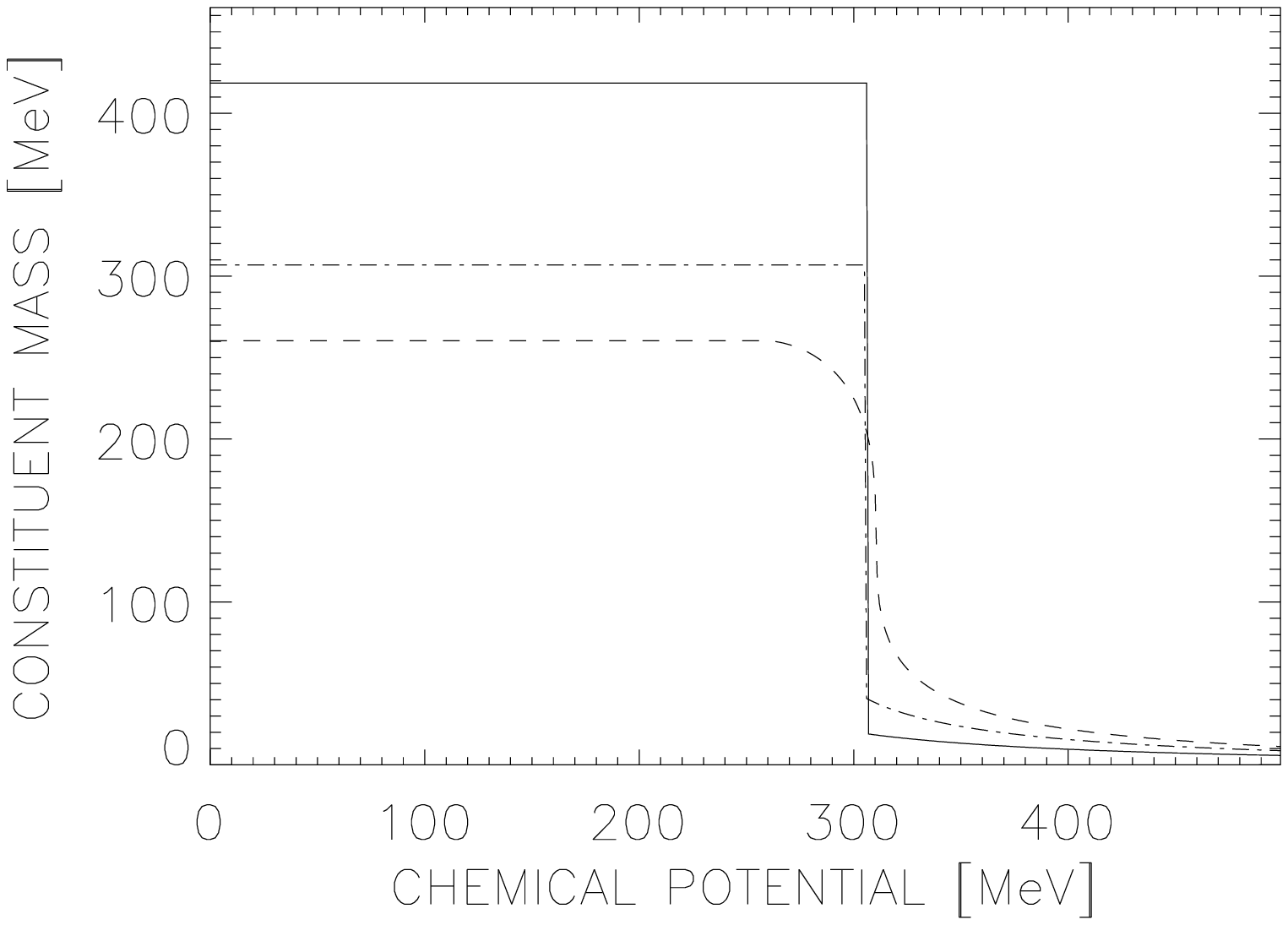}}
\caption{\small{The constituent quark/antiquark mass as a function of $T$ for 
$\mu=0$ (upper) and as a function of $\mu$ (lower) for $T=0$. The solid line 
represents the mass for $g=4.5$, the dashed-dotted line for $g=3.3$ and the 
dashed line for $g=2.8$.}\label{coup}}
\end{center}
\end{figure}

An interesting point is that
the quark mass assumes its vacuum value all the way up to the chiral 
transition, and then drops to a small value in the phase where chiral 
symmetry is restored (see figure \ref{qmassfig}). This behavior is related to 
the appearance of a bound state at zero pressure. Within the linear sigma 
model this bound state was found by Lee and Wick a long time ago 
\cite{leewick}. Recently, it was shown in \cite{mishusat} that a similar 
bound state appears also in the NJL model. This behavior, however, depends on 
the value of the coupling constant $g$. In general, if the coupling constant 
is sufficiently large, the attractive force between the constituent quarks 
becomes large enough to counterbalance the Fermi pressure, thus giving rise 
to a bound state. To demonstrate this we have varied the coupling constant 
$g$ for the sigma model within reasonable limits. The results are shown in 
figure \ref{coup}. It is seen in figure \ref{coup} that, indeed, one can 
change the smooth crossover for $\mu=0$ into a first order transition by 
increasing the coupling constant (left panel) and change the first order 
transition in the case\ of $T=0$ into a smooth crossover (right panel). In 
this way a heavy quark phase comes into existence as the coupling constant is 
decreased and the bound state disappears.

\section{Adiabats}

In hydrodynamical simulations the entropy per baryon, $S/A$, is a fundamental
 quantity. One can calculate it using standard thermodynamic relations,
\be
\frac{S}{A}=3\frac{e+p-\mu n}{Tn}\,~,
\ee
where $e$, $p$, $n$ are the energy density, pressure and net density of 
quarks and antiquarks ($n=3n_B$), respectively. By studying this quantity, 
one can check if there is a tendency towards convergence of the adiabats 
towards the critical point as was claimed in \cite{srs}. If this was the case 
it would be easy to actually hit or go close to this point in a hydrodynamical
 evolution.

\begin{figure}[t]
\begin {center}
\leavevmode
\hbox{%
\epsfysize=10cm
\epsffile{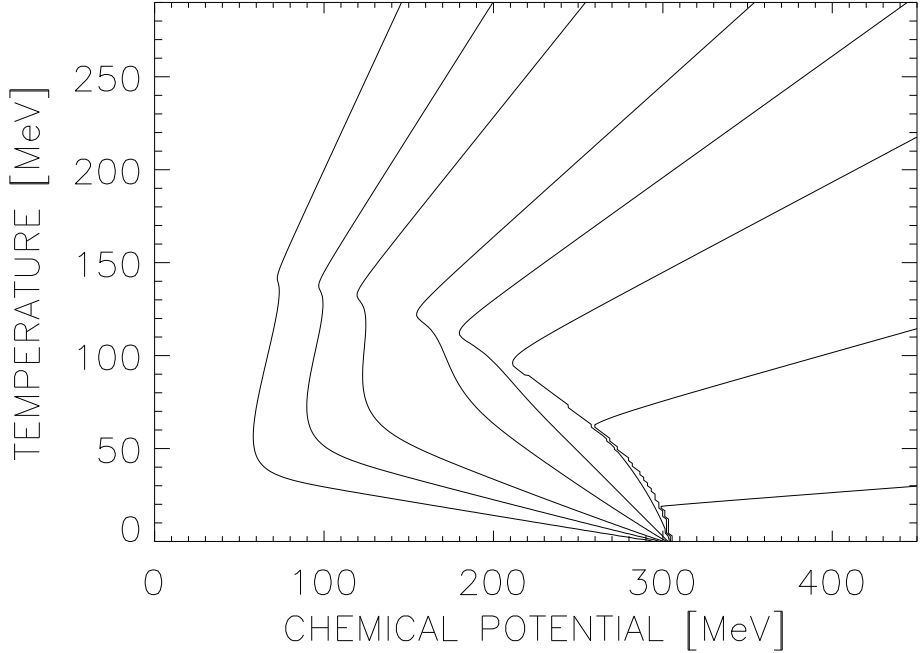} 
}
\caption{\small{The entropy per baryon number $S/A$. The curves correspond to (from left) $S/A$=[28,21,17,13,11,9,6,2].}\label{adiabats}}
\end{center}
\end{figure}

Figure \ref{adiabats} shows the contours of $S/A$ in the $(T,\mu)$ plane. We 
actually observe a trend which is quite opposite to this expectation. It 
turns out that the adiabats turn away from the critical point when they hit 
the first order transition line and bend towards the critical point only when 
they come from the smooth crossover region. 
This is understood as follows. First, note that all adiabats terminate at 
zero temperature and $\mu=M_{vac}$, i.e. the $(T,\mu)$ combination 
corresponding to the vacuum. The reason is that as $T\rightarrow 0$, by the 
third law of thermodynamics also $S\rightarrow 0$, therefore, for fixed $S/A$ 
we have to require that $n\rightarrow 0$ is fulfilled when $\mu=M_{vac}$. For 
our choice of parameters the point $(T,\mu)=(0,M_{vac})$ is also the endpoint 
of the phase transition curve at $T=0$, since the phase transition connects 
the vacuum directly with the phase of restored chiral symmetry. This is 
confirmed by figures \ref{qmassfig} and \ref{coup}. 

This behavior is quite opposite to the case underlying the claim in 
\cite{srs}, where the hadronization of a large number of quark and gluon 
degrees of freedom into relatively few pion degrees of freedom leads to the 
release of latent heat and consequently to a reheating (increase of $T$) 
through the phase transition. Remember, however, that in our case there is 
actually no change in the number of degrees of freedom in the two phases. The 
change is only in the quark masses. Consequently, there is no ``focusing'' 
effect in the linear sigma model with quarks. On the other hand, the behavior 
of the adiabats in figure \ref{qmassfig} is quite typical for an ordinary 
liquid-gas phase transition. Here liquid and gas are represented, 
respectively, by chirally symmetric and broken phases. This analogy was 
further elaborated in \cite{mishusat}.

\section{Summary}

In this chapter the results of our investigations of the thermodynamics of the chiral phase transition within the linear sigma model coupled to quarks were presented. For small bare quark masses we have found a smooth crossover for nonzero temperature and zero chemical potential and a first order transition for zero temperature and nonzero chemical potential. The first order phase transition line in the ($T,\mu$) plane ends in the expected critical point. It has been found that the $\sigma$ mass is zero at the critical point. The behavior of the adiabats shows a pattern opposite to the expectations for the chiral/confinement transition of \cite{srs}. In fact, the phase transition found in this model turned out to be of the liquid-gas type. However, the strength of this transition depends sensitively on the coupling constants of the models.

Certainly it will be interesting to use this model in hydrodynamical simulations in order to confirm or disconfirm possible observable signatures of the phase transition discussed in the introduction. In particular, since the model contains dynamical sigma and pion fields, it would be suitable to study the long wavelength enhancement of the fields at the critical point. 
\chapter{Evolution of Disoriented Chiral Condensates \label{sect-dcc}}

In this chapter, I review the concept of disoriented chiral condensates and the research done so far on this subject. Then I derive coarse-grained equations of motion for out-of-equilibrium condensate fields. Based on linear response theory the fluctuations of the thermal bath, as a response to the presence of the non-thermal condensate, are analyzed. Multiple aspects of the response functions are emphasized, setting the ground work for the chapters that follow.

\section{Review of Previous Results \label{sect-dcc-prev}}

Formation of condensates as a result of a phase transition between a high temperature disordered (symmetric) and a low temperature ordered (symmetry broken) phase is common in different areas of physics. As the disordered phase is cooled, correlation occurs on a microscopic level, reaching a macroscopic size at the critical temperature. In the region of the phase transition small scale physics becomes irrelevant and phase ordering is determined by long wavelength fluctuations. The time-scales involved in the transition play a major role in determining the nature of the phase ordering. When the cooling is much slower than the time taken for microscopic rearrangement (relaxation time), the system is said to be in local thermal equilibrium, and the transition is an equilibrium phase transition. The QCD phase transition in the Early Universe qualifies as such, since the cooling rate around the critical temperature is about $10^{-5}$ seconds, whereas the relaxation time estimated from collisional processes of quarks and gluons has a very different order of magnitude, on the order of $10^{-23}$ seconds. There is also the possibility for the expansion and cooling to be very rapid. The long wavelengths, not adjusting to the fast temperature change, fall out of equilibrium. Because the time-scales involved in a heavy ion collision ($10^{-22}-10^{-23}$ seconds at RHIC, for example) are comparable to the relaxation time, it is now widely believed that the QGP confinement/chiral phase transition happens out of equilibrium. 

In a recent work \cite{boya00} Boyanovsky et al.~discuss in a thorough and comprehensive manner the relaxation of the order parameter and its phenomenological consequences in ultrarelativistic heavy ion collisions and in Early Universe cosmology. The authors focus on the possibility that long wavelength fluctuations are critically slowed down. With decreasing temperature the relaxation time of the long wavelength modes of the order parameter become extremely long. Then these can freeze-out even before reaching the critical point. Such a transition is said to be quenched. There is speculation that such phenomena can be the source of non-equilibrium effects in cosmological phase transitions, which may for example, explain the size and distribution of primordial black holes.

\begin{figure}[htp]
\begin {center}
\leavevmode
\hbox{%
\epsfysize=14cm
\epsffile{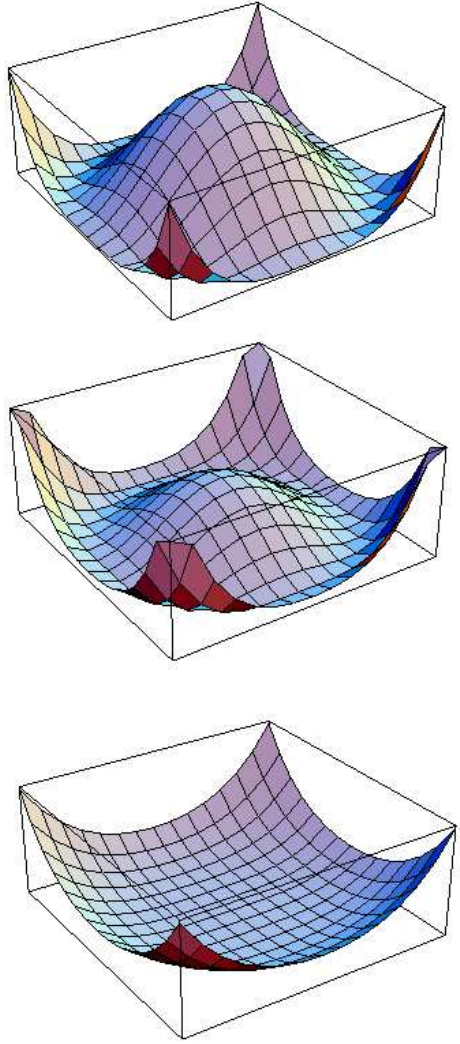} 
}
\caption{{\small TOP: The $T=0$ vacuum potential has the shape of the bottom of a wine bottle and the ground state is degenerate; MIDDLE: For $T>0$ the hill becomes more and more shallow; BOTTOM: At $T\geq T_c$ symmetry is restored and there is a unique ground state. It is assumed that in a heavy ion collision such a state has been achieved. Due to rapid expansion $T<T_c$ potential is TOP again. The symmetric state becomes unstable and will roll down to a stable ground state. This roll down in radial direction can develop configurations in azimuthal direction. These are DCCs.  
}}
\end{center}
\label{rolldown.fig}
\end{figure}

The dynamics of phase ordering is described in terms of domains that form after the temperature has rapidly dropped below its critical value. In heavy ion collisions, after the plasma expands and cools through the non-equilibrium chiral phase transition, such domain formation is expected. These are referred to as disoriented chiral condensates, named shortly as DCC. A DCC is a good candidate for the chiral phase transition to be directly probed. Inside a DCC domain the order parameter is misaligned, meaning that the chiral condensate points in a different direction than in vacuum. The possible physical picture for DCC formation is presented in figure \ref{rolldown.fig} and described in the caption. It has been suggested \cite{rw} that regions in which the chiral condensate is disoriented, corresponding to long wavelength fields, may spontaneously grow and subsequently decay via pion emission. This can be detected through the different pion decay channels: $\pi^0\rightarrow 2\gamma$ and $\pi^{\pm}\rightarrow\mu^\pm\nu_\mu$. DCC decay can lead to experimentally observable anomalies. One is the ratio of neutral to charged soft pions, $r = \frac{n_0}{n_0+n_{ch}}$. Each misaligned region gives pions with a given $r$, but $r$ is different for different regions. The probability distribution is $P(r)=\frac{1}{2\sqrt{r}}$, which is different than what is naively expected ($\frac{1}{3}$). Peculiar events in which such anomalies are present have been observed in cosmic ray experiments \cite{cent1}, and are known as Centauros \cite{cent2}. Ultrarelativistic heavy ion collisions provide a fertile ground for studying fluctuations in $r$. The STAR detector at RHIC \cite{rhic} searches for such nonstatistical or dynamical fluctuations on an event-by-event basis. 

The formation and decay of DCCs have been studied extensively in the last decade \cite{rajagopal} since the pioneering work of Anselm \cite{anselm} in 1989. However, it was first Bjorken \cite{jdb}, and then Blaizot and Krzywicki \cite{blaizot}, who in 1992 made the idea widely popular. In 1993, Bjorken, Kowalski and Taylor proposed a ``baked-Alaska'' model \cite{alaska} which could explain Centauro events. It was Rajagopal and Wilczek \cite{rw}, though, who first used the idea in quark-gluon-plasma physics. In relativistic heavy ion collisions the multiplicity of produced particles is very high. Due to their small mass, the majority of these are pions. Rajagopal and Wilczek proposed that some of these pions are the result of the coherent decay of classical pion fields present in these collisions. These pions would have small momenta and would show large fluctuations in the ratio of produced neutral pions compared to the charged ones. The authors assumed a perfectly quenched scenario for the transition: after the critical temperature is reached, the potential is assumed to take instantaneously its vacuum shape, and the long-wavelength fields then evolve according to zero temperature equations of motion. This original idea stirred much interest and led to the investigation of various aspects of DCC dynamics \cite{gavingosch}-\cite{randrup-dcc}.

In \cite{asakawa} Asakawa, Huang and Wang pointed out that quenching cannot be naturally achieved through expansion. Krzywicki and Serreau in a more recent paper \cite{french} predicted a small probability for an observable DCC signal. However, another recent publication by Scavenius and Dumitru \cite{oa} reports the realization of quenched initial conditions when the chiral phase transition is of first order. 

A quench in the strict sense defined by Rajagopal and Wilczek, is a very drastic assumption which leads to a complete decoupling of soft and hard modes. In a relativistic heavy ion collision, though, pions are produced in a very wide momentum range. Soft pions, with small momentum, are numerous, but carry little energy. Hard modes, having momenta of a few hundreds of MeV, carry most of the energy. Their effect should certainly not be neglected. Bir\'o and Greiner \cite{biro} studied first the effect of hard modes on the non-equilibrium evolution of soft modes. They proposed hard pions as a thermalised background and considered the chiral condensate in this background. The presence of a heat bath leads to dissipation that shows up in the equations for the condensate fields. The result of \cite{biro} is that dissipation suppresses the DCC formation. The problem with their model, though, is that they calculated the dissipation coefficient in $\phi^4$ theory based on \cite{greinermuller}, but this is good only in the symmetric phase. In the broken phase the masses are not degenerate anymore and also the $\phi^3$ interaction is allowed. In \cite{rischke}, a work by Rischke, these problems have been eliminated. He accounted for cubic interaction and so for the possibility of decay, but neglected collisional interactions, arguing that these are higher order in the coupling. Csernai, Ellis, Jeon and Kapusta \cite{cejk} showed that scattering processes can be significant for dissipation.

All of the above mentioned studies looked at homogeneous condensates using different approaches: in \cite{biro} Langevin-type equations of motion; in \cite{boya95} closed time path method; in \cite{rischke} influence functional formalism; in \cite{cejk} linear response theory. The relaxational dynamics of inhomogeneous condensates have been investigated at one-loop level by Boyanovsky, D'Attanasio, de Vega and Holman \cite{boya96}. The authors concentrate on the evolution of the sigma component only, but they assume the vacuum value for the order parameter. We know that the equilibrium value of the condensate is changing with temperature, and their assumption is correct only at low temperatures. Moreover, in their model the mass of the pion is given by hand, so this is not a consequence of symmetry breaking. The effect of the heat bath on the low-frequency modes is discussed at one-loop level by Jakov\'ac, Patk\'os, Petreczky and Sz\'ep \cite{magyar} too.

The work presented in this dissertation was substantially influenced by \cite{cejk}, where the dynamical evolution of the sigma condensate has been followed within the linear sigma model. My analysis is an extension of \cite{cejk} allowing for DCC formation. I present a microscopic, field theoretical description of the non-equilibrium dynamics of soft homogeneous and inhomogeneous sigma and pion field configurations coupled to a thermal bath. To the best of my knowledge, a consistent incorporation of relaxation processes at one- and two-loop level has not been done before in this context. In this model dissipation follows from first principles. This means that the dissipative terms in the equation of motion arise directly from the Lagrangian. Moreover, the temperature-dependence of the order parameter naturally follows, too. A convenient framework for my DCC studies is the linear sigma model. It is convenient because:
\begin{itemize}
\item it is formulated in terms of the chiral field $\phi=(\sigma,\vec\pi)$; 
\item it exhibits symmetry breaking such that in the ground state the scalar condensate is nonzero, $\sigma\sim\langle{\bar{q}}q\rangle\neq 0$, and the pseudoscalar condensate is zero, $\vec{\pi}\sim\langle{\bar{q}}{\vec{\tau}}\gamma_5q\rangle = 0$;
\item it implements a phase transition at finite $T_c$ such that for $T>T_c$ chiral symmetric is restored and $\langle{\bar{q}}q\rangle=0$;
\item it is a renormalizable theory.
\end{itemize}
%

\section{Equations of Motion \label{sect-eom}}

The dynamics of the condensate are completely determined by the evolution equations in space and time for the long-wavelength fields. In the following, the coarse-grained equations of motions are derived. First, the theory with exact chiral symmetry is analyzed. This is an idealized case, which significantly simplifies some of our formulas, and allows for the study of massless bosonic waves, the Goldstone modes. Such modes are present not only in particle physics, but also in condensed matter physics (e.g. spin waves). The low temperature behavior of the pion is well-known from chiral perturbation theory \cite{sm} and, accordingly, the pion is massive and chiral symmetry is explicitly broken. Therefore, for a completely consistent discussion one should work with massive pions. This is the topic of the second section.

\subsection{Chiral Limit \label{sect-eom-chiral}}

The Lagrangian of the O(N) sigma model with no explicit symmetry breaking term is given by
\be
L(\sigma,\vec{\pi}) = \frac{1}{2}(\partial \sigma)^2 +  \frac{1}{2}(\partial \vec{\pi})^2 - U(\sigma,\vec\pi)\,~,
\label{lag}
\ee
where $\sigma$ and $\vec{\pi}$ are the sigma and the $N-1$ component pion fields, and the potential is given by 
\be
U(\sigma,\vec\pi) = \frac{\lambda}{4}(\sigma^2 + \vec{\pi}^2-f_\pi^2)^2\,~.
\ee
Note that the fermionic part of the original sigma model Lagrangian (\ref{lagsigma}) has been neglected here. This is allowed since our focus is on the condensate formed in the symmetry broken phase where quark degrees of freedom are confined already. In the hadronic phase a natural choice would be nucleons. However, we justify their exclusion by their mass, which is much larger than the other scales involved. Therefore, processes that involve nucleons would be strongly suppressed with respect to pure bosonic interactions. 
The evolution of the fields under thermal equilibrium circumstances has been well studied by several authors (see for example \cite{bochkarev}). The usual procedure is to express the fields as 
\be
\sigma(t,\vec x) = v + \sigma_f(t,\vec x)\,~~~ \mbox{and}~~~~~\pi_i(t,\vec x) =\pi_{if}(t,\vec x), ~~i=1,...,N-1\,~.
\label{eq}
\ee
Here $v$ is the chiral order parameter, the thermal average at temperature $T$, chosen to lie along the sigma direction. $\sigma_f$ is the fluctuation about $v$, along the radial direction and $\pi_{if}$ is the fluctuation of the $i$th component of the pion field about zero, in the azimuthal direction.

As explained in section \ref{sect-dcc-prev}, there is good reason to believe that a more realistic picture involves an out-of-equilibrium evolution of the fields. The approach we take is a semiclassical description. The idea is that instead of (\ref{eq}), we decouple the fields into their low and high frequency modes. This can be done, for example, by introducing a momentum scale $\Lambda_c$, as in \cite{bodecker}. Then one integrates out the high frequency modes, and obtains an effective classical theory for the long wavelengths. The fields can be separated as follows: 
\be 
\sigma(x) &=& \tilde{\sigma}(x) + \sigma_f(x)\nonumber\\
\pi_i(x) &=& \tilde\pi_i(x) + \pi_{if}(x)\,~, ~~i=1,...,N-1
\label{slow-fast}
\ee
\begin{itemize}
\item  $\tilde{\sigma}$ and $\tilde{\pi}$ are slowly varying  condensate fields, representing low frequency modes with momentum $\mid\vec{k}\mid<\Lambda_c$, occupied by a large number of particles. These soft modes lie entirely in the non-perturbative regime, and may be treated as classical fields.
\item $\sigma_f$ and $\pi_{if}$ are high frequency, fast modes, with $\mid\vec{k}\mid>\Lambda_c$. These hard modes, representing quantum and thermal fluctuations, constitute a heat bath.
\end{itemize}
The problem to be solved, then, is to describe the evolution of low frequency classical fields that are embedded into a thermal bath. In our approach, the soft modes follow classical equations of motion, whereas the effect of the hard thermal modes is taken into account in a perturbative manner. We are well aware that any perturbative treatment can be a target of criticism, since the sigma model is a strongly coupled effective theory. However, one can still determine qualitative aspects using this model.

Euler-Lagrange field equations are derived after inserting (\ref{slow-fast}) in the Lagrangian (\ref{lag}). For the sigma field one obtains
\be
&&\!\!\!\!\!\!\!\!\!\!\!\! \partial^2\tilde\sigma +\lambda\tilde\sigma^3 + 3\lambda\sigma_f^2\tilde\sigma + \lambda\tilde\sigma\sum_{i=1}^{N-1}\tilde\pi_i^2 + 3\lambda\sigma_f\tilde\sigma^2 + \lambda\tilde\sigma\sum_{i=1}^{N-1}\pi_{if}^2 + 2\lambda\tilde\sigma\sum_{i=1}^{N-1}\tilde\pi_i^2\pi_{if} - \lambda f_\pi^2\tilde\sigma  \nonumber \\
&+& \lambda\sigma_f^3 + \lambda\sigma_f\sum_{i=1}^{N-1}\tilde\pi_i^2 + \lambda\sigma_f\sum_{i=1}^{N-1}\pi_{if}^2 + 2\lambda\sigma_f\sum_{i=1}^{N-1}\tilde\pi_i^2\pi_{if} - \lambda f_\pi^2\sigma_f = 0\,~,
\label{01}
\ee
and for the $i$th component of the pion field
\be
&&\!\!\!\!\!\!\!\!\!\!\!\! \partial^2\tilde\pi_i + \lambda\tilde\pi_i^3 + 3\lambda\pi_{if}^2\tilde\pi_i + 3\lambda\tilde\pi_i^2\pi_{if} + \lambda\tilde\pi_i\sum_{j\neq i=1}^{N-1}\tilde\pi_j^2 + \lambda\tilde\pi_i\sum_{j\neq i=1}^{N-1}\pi_{jf}^2 + 2\lambda\tilde\pi_i\sum_{j\neq i=1}^{N-1}\tilde\pi_j\pi_{jf}  \nonumber \\
&+& \lambda\pi_{if}^3 + \lambda\pi_{if}\sum_{j\neq i=1}^{N-1}\tilde\pi_j^2 + \lambda\pi_{if}\sum_{j\neq i=1}^{N-1}\pi_{jf}^2 + 2\lambda\pi_{if}\sum_{j\neq i=1}^{N-1}\tilde\pi_j\pi_{jf} +\lambda\tilde\pi_i\tilde\sigma^2 +\lambda\tilde\sigma^2\pi_{if} \nonumber \\
&+& \lambda\sigma_f^2\tilde\pi_i + \lambda\sigma_f^2\pi_{if} + 2\lambda\tilde\sigma\sigma_f\tilde\pi_i + 2\lambda\tilde\sigma\sigma_f\pi_{if} - \lambda f_{\pi}^2\tilde\pi_i - \lambda f_{\pi}^2\pi_{if} = 0 \,\, .
\label{02}
\ee
Now we average the equations of motion over time and length scales that are short compared to the scales characterizing the change in the slow fields, but long relative to the scales of the quantum fluctuations. This is known as coarse-graining. The thermal average of high frequency fluctuations is $\langle\sigma_f(x)\rangle = 0$ and $\langle\vec\pi_f(x)\rangle = 0$, while $\langle\tilde\sigma(x)\rangle = \tilde\sigma(x)$ and $\langle\tilde\pi_i(x)\rangle = \tilde\pi_i(x)$. Throughout this work the notation $\langle{\cal O}\rangle$ refers to the non-equilibrium ensemble average of an operator $\cal O$. The equilibrium ensemble average is denoted by $\langle{\cal O}\rangle_{eq}$. It should be noted that we allow for a nonzero ensemble average not only along the sigma direction, but also in the pion direction. In other words, we allow the formation of disoriented chiral condensates. The coarse-grained equations of motion are
\be
&&\!\! \partial^2\tilde\sigma + \lambda\tilde\sigma^3 + 3\lambda\langle\sigma_f^2\rangle\tilde\sigma + \lambda\tilde\sigma\sum_{i=1}^{N-1}\tilde\pi_i^2 + \lambda\tilde\sigma\sum_{i=1}^{N-1}\langle\pi_{if}^2\rangle - \lambda f_\pi^2\tilde\sigma \nonumber\\
&& + 2\lambda\sum_{i=1}^{N-1}\langle\sigma_f\pi_{if}\rangle\tilde\pi_i - \lambda f_\pi^2\tilde\sigma = 0\,~,
\label{1}
\ee
and 
\be
&&\!\! \partial^2\tilde\pi_i + \lambda\tilde\pi_i^3 + 3\lambda\langle\pi_{if}^2\rangle\tilde\pi_i + \lambda\tilde\pi_i\sum_{j\neq i=1}^{N-1}\tilde\pi_j^2 + \lambda\tilde\pi_i\sum_{j\neq i=1}^{N-1}\langle\pi_{if}^2\rangle + 2\lambda\sum_{j\neq i=1}^{N-1}\tilde\pi_j\langle\pi_{if}\pi_{jf}\rangle \nonumber\\
&& + \lambda\tilde\pi_i\tilde\sigma^2 + \lambda\tilde\pi_i\langle\pi_{if}^2\rangle + 2\lambda\tilde\sigma\langle\pi_{if}\sigma_f\rangle - \lambda f_\pi^2\tilde\pi_i = 0\,~.
\label{2}
\ee
The cross correlations between the fluctuations of different fields are nonzero, $\langle\varphi_i\varphi_j\rangle\neq 0$; however, when averaging, all cubic fluctuations of the form $\langle\varphi_i\varphi_j\varphi_k\rangle$ were considered zero, meaning that only {\bf one-loop} order is included. Later in this section the extension to two-loops is presented. 

In the broken symmetry phase, $T<T_c$, we can separate the non-equilibrium condensate fields into their equilibrium value and a slow fluctuation about this. Choose the equilibrium condensate along the sigma direction:
\be
\tilde\sigma(x) &=& v(x) + \sigma_s(x)\nonumber\\ 
\tilde\pi_i(x) &=& \pi_{is}(x),~~~i=1,...,N-1\,~.
\label{philow}
\ee
For now, let us assume that this equilibrium condensate is static and homogeneous, $v(x)=v$. Clearly, this need not be the case, and we will address this issue later. The thermal equilibrium ensemble of the hard fluctuations is affected by the presence of the condensate. Full ensemble averages, which are basically the two-point functions of the thermalized fields, can be written as 
\be
&&\langle\sigma_f^2\rangle = \langle\sigma_f^2\rangle_{eq} + \delta\langle\sigma_f^2\rangle \nonumber \\
&&\langle\pi_{if}^2\rangle = \langle\pi_{if}^2\rangle_{eq} + \delta\langle\pi_{if}^2\rangle \nonumber \\
&&\langle\sigma_f\pi_{if}\rangle = \delta\langle\sigma_f\pi_{if}\rangle \nonumber \\
&&\langle\pi_{if}\pi_{jf}\rangle = \delta\langle\pi_{if}\pi_{jf}\rangle \, .
\label{fluct}
\ee
The deviations in the fluctuations, $\delta\langle\sigma_f^2\rangle$, $\delta\langle\pi_{if}^2\rangle$, $\delta\langle\sigma_f\pi_{if}\rangle$, and $\delta\langle\pi_{if}\pi_{jf}\rangle$, are the responses of the fast modes to the presence of slow $\sigma_s$ and $\vec\pi_s$ background fields. These responses are proportional to the slow modes raised to some positive power. A detailed discussion, including an explicit evaluation using linear response theory, is presented in section \ref{sect-response}. Notice the absence of $\langle\sigma_f\pi_{if}\rangle_{eq}$ and $\langle\pi_{if}\pi_{jf}\rangle_{eq}$. The reason is that the correlation functions in equilibrium, in the absence of a background, are diagonal.

Equations (\ref{1}) and (\ref{2}) must also hold in equilibrium, at fixed temperature. Setting $\sigma_s=0$ and $\vec\pi_s=0$ results in an equation satisfied by the equilibrium condensate: 
\be
\lambda v^3 + 3\lambda v\langle\sigma_f^2\rangle_{eq} + \lambda v\sum_{i=1}^{N-1}\langle\pi_{if}^2\rangle_{eq} - \lambda f_\pi^2v = 0 \,~.
\ee
This has two solutions:
\be
v =0\,~, ~~~~~ T>T_c\,~,
\ee
and
\be
v^2 = f_\pi^2 - 3\langle\sigma_f^2\rangle_{eq} - (N-1)\langle\pi_{if}^2\rangle_{eq}\,~, ~~~T<T_c \,~.
\label{equilib}
\ee
We are interested in the second solution (\ref{equilib}), since this represents the low temperature, symmetry broken phase. Since all components of the pion field are equivalent, we could write the sum as $\sum_{i=1}^{N-1}\langle\pi_{if}^2\rangle_{eq} = \langle\vec\pi_f^2\rangle_{eq} = (N-1)\langle\pi_{if}^2\rangle_{eq}$, where now $\pi_{if}$ is one of the components. This is a convention that we adopt throughout the rest of the paper. The equations describing the evolution of the condensate fields are readily obtained after inserting (\ref{fluct}) and (\ref{philow}) into (\ref{1}) and (\ref{2}):
\be
&&\!\!\!\!\!\!\!\!\!\!\!\!\!\!\!\! \partial^2\sigma_s + 2\lambda v^2\sigma_s + \lambda v\left[3\delta\langle\sigma_f^2\rangle + (N-1)\delta\langle\pi_f^2\rangle\right] = - \lambda\left(\sigma_s^3 + 3v\sigma_s^2 + (N-1)v\pi_s^2 \right. \nonumber\\
&&\!\!\!\! \left. + (N-1)v\sigma_s\pi_s^2 + 2(N-1)\delta\langle\sigma_f\pi_f\rangle\pi_s + 3\sigma_s\delta\langle\sigma_f^2\rangle + (N-1)\sigma_s\delta\langle\pi_f^2\rangle\right)
\ee
and
\be
&&\!\!\!\!\!\!\!\! \partial^2\pi_{is} + 2\lambda\pi_{is}\left[\langle\pi_{if}^2\rangle_{eq} - \langle\sigma_f^2\rangle_{eq}\right] + 2\lambda v\delta\langle\pi_{if}\sigma_f\rangle = - \lambda\left(\pi_{is}^3 + 2v\sigma_s\pi_{is} + \sigma_s^2\pi_{is} \right.\nonumber\\
&&\left. + (N-2)\pi_{is}\pi_s^2 + 2\sigma_s\delta\langle\pi_{if}\sigma_f\rangle + 3\pi_{is}\delta\langle\pi_{if}^2\rangle + \pi_{is}\delta\langle\sigma_f^2\rangle \right.\nonumber\\
&& \left. + (N-2)\pi_{is}\delta\langle\pi_f^2\rangle + 2(N-2)\pi_s\delta\langle\pi_{if}\pi_f\rangle\right)\,~.
\ee
The above set of equations is a coupled non-linear integro-differential equations (see section \ref{sect-response}), and it is hardly possible to solve, even with access to supercomputers. Let us assume only small deviation from equilibrium, $|\sigma_s|, |\pi_s|\ll v$, neglecting higher order terms. The linearized field equations read 
\be
 \partial^2\sigma_s + 2\lambda v^2\sigma_s + \lambda v\left[3\delta\langle\sigma_f^2\rangle + (N-1)\delta\langle\pi_f^2\rangle\right] = 0 \,~,
\label{sigmaeom}
\ee
and
\be
\partial^2\pi_s + 2\lambda\pi_s\left[\langle\pi_f^2\rangle_{eq} - \langle\sigma_f^2\rangle_{eq}\right] + 2\lambda v\delta\langle\pi_f\sigma_f\rangle = 0 \,~.
\label{pioneom}
\ee
Remember, $\pi_s$ and $\pi_f$ refer to one of the $N-1$ components of the pion field. The three physical pions, $\pi^0$ and $\pi^\pm$, are the degrees of freedom in the $N=4$ model. Also, the above results were obtained with the assumption of fixed temperature, and so the equilibrium condensate was constant. In reality, the temperature drops as time elapses, and $v$ changes with it (see previous discussion in section \ref{sect-dcc-prev}). 

We close this section by stating the results when accuracy is extended up to {\bf two-loop} order. Technically this means that, when averaging equations (\ref{01}) and (\ref{02}), we need to keep the three-point functions too. The ensemble average of the product of three fields is given by:
\be
&&\langle\sigma_f^3\rangle = \langle\sigma_f^3\rangle_{eq} + \delta\langle\sigma_f^3\rangle \nonumber\\
&&\langle\sigma_f\pi_f^2\rangle = \langle\sigma_f\pi_f^2\rangle_{eq} + \delta\langle\sigma_f\pi_f^2\rangle \nonumber\\
&&\langle\pi_f^3\rangle = \delta\langle\pi_f^3\rangle \nonumber\\
&&\langle\sigma_f^2\pi_f\rangle = \delta\langle\sigma_f^2\pi_f\rangle \,~.
\ee
On account of cubic couplings one finds nonzero $\langle\sigma_f^3\rangle_{eq}$ and $\langle\sigma_f\pi_f^2\rangle_{eq}$. These are evaluated in the usual way (see section \ref{sect-response}) by performing an expansion in the coupling constant of the exponential of the action \cite{peskin}. When including these terms, the equilibrium value of the condensate is modified compared to its previous expression (\ref{equilib}):
\be
v^2 = f_\pi^2 - 3\langle\sigma_f^2\rangle_{eq} - (N-1)\langle\pi_f^2\rangle_{eq} - \frac{\langle\sigma_f^3\rangle_{eq}}{v} - (N-1)\frac{\langle\sigma_f\pi_f^2\rangle_{eq}}{v}\,~.
\ee
The linearized equations of motion acquire some extra terms compared to equations (\ref{sigmaeom}) and (\ref{pioneom}):
\be
\partial^2\sigma_s + \bar m_\sigma^2\sigma_s + \lambda v\left[3\delta\langle\sigma_f^2\rangle + (N-1)\delta\langle\pi_f^2\rangle\right] + \lambda\delta\langle\sigma_f^3\rangle + \lambda(N-1)\delta\langle\sigma_f\pi_f^2\rangle = 0 \,~,\nonumber\\
\label{sigmaeom2}
\ee
where
\be
\bar m_\sigma^2 = \lambda\left(2v^2 - \frac{\langle\sigma_f^3\rangle_{eq}}{v} - (N-1)\frac{\langle\sigma_f\pi_f^2\rangle_{eq}}{v}\right)\,~,
\ee
and
\be
\partial^2\pi_s + 2\lambda v\delta\langle\sigma_f\pi_f\rangle + \tilde m_\pi^2\pi_s + \lambda\delta\langle\pi_f^3\rangle + \lambda\delta\langle\sigma_f^2\pi_f\rangle= 0\,~,
\label{pioneom2}
\ee
where
\be
\tilde m_\pi^2 = \lambda\left(2\langle\pi_f^2\rangle_{eq} - 2\langle\sigma_f^2\rangle_{eq} - \frac{\langle\sigma_f^3\rangle_{eq}}{v} - (N-1)\frac{\langle\sigma_f\pi_f^2\rangle_{eq}}{v}\right) \,~.
\ee
%

\subsection{Explicit Chiral Symmetry Breaking \label{sect-eom-nochiral}}

The explicit breaking of symmetry is implemented in the Lagrangian through the presence of a term linear in the sigma field (see discussion in \ref{sect-sigmamodel}). The potential term is 
\be
U(\sigma, \vec\pi) = \frac{\lambda}{4}\left(\sigma^2+\vec\pi^2-v_0^2\right) + H\sigma\, ~.
\ee
Since the procedure to obtain the effective equations of motions is identical to the above derived massless pion case, we are going to sketch the major steps of the derivation, giving details only where the physics is different: i) separate the low and high frequency modes of the fields as in (\ref{slow-fast}); ii) write down Euler-Lagrange equations; iii) average these with respect to the statistics of the hard modes; iv) separate the condensate fields into their equilibrium value and the slow fluctuations about these as in (\ref{philow}); v) break the hard fluctuations into a sum of equilibrium fluctuations and deviations from these as in (\ref{fluct}); vi) drop all terms other than linear in the fields.

One of the significant differences between the two cases is in the behavior of the equilibrium condensate, which now at one-loop order satisfies the following equation:
\be
\lambda v^3 + \lambda\left(3\langle\sigma_f^2\rangle_{eq} + (N-1)\langle\pi_f^2\rangle_{eq}\right)v - \lambda v_0^2v - H = 0 \,~.
\label{equilibm}
\ee
Similarly, notice, that unlike before, $v=0$ is not a solution of (\ref{equilibm}). This means that the condensate is never fully dissolved by fluctuations, not even at high temperatures, and the symmetry is never restored. This should come as no surprise, though, since including a finite pion mass means that the $O(N)$ symmetry is only approximate. The linearized equations of motion are
\be
\partial^2\sigma_s + \bar m_\sigma^2\sigma_\sigma + \lambda v\left[3\delta\langle\sigma_f^2\rangle + (N-1)\delta\langle\pi_f^2\rangle\right] = 0 
\label{sigmaeomm}
\ee
\be
\bar m_\sigma^2 = 2\lambda v^2 + \frac{H}{v}\,~,
\label{sigmamassh}
\ee
and
\be
\partial^2\pi_s + \bar m_\pi^2\pi_s + 2\lambda v\delta\langle\sigma_f\pi_f\rangle + 2\lambda\pi_s\left[\langle\pi_f^2\rangle_{eq} - \langle\sigma_f^2\rangle_{eq}\right] = 0 
\label{pioneomm}
\ee
\be
\bar m_\pi^2 = \frac{H}{v} = \frac{f_\pi}{v}m_\pi^2   \,~.
\label{pionmassh}
\ee
When the strength of the symmetry breaking diminishes, $H\rightarrow 0$, the results of the previous section are recovered.

\subsection{Discussion \label{sect-eom-discuss}}

Equations (\ref{sigmaeom}) and (\ref{pioneom}), derived above for the theory with exact chiral symmetry, and (\ref{sigmaeomm}) and (\ref{pioneomm}) when the symmetry is explicitly broken, constitute the basis of the study presented in the rest of this dissertation. Let us enumerate some of the focal points: 
\begin{itemize}
\item The mass of the sigma meson, $\bar m_\sigma$, is given in terms of the equilibrium condensate, $v$, and as such, it is temperature dependent, since $v$ includes thermal fluctuations of the fields. However, the fluctuations themselves are dependent on the sigma mass. Therefore, in order to determine the temperature dependence of the mass, one has to perform a self-consistent evaluation. This analysis is presented in Chapter \ref{sect-masses}.

\item In the theory with exact chiral symmetry the mass of the pion is zero. Goldstone's Theorem requires that it remains zero at every order in perturbation theory. However, equation (\ref{pioneom}) for the pion condensate shows a mass, $\tilde m_\pi^2 = 2\lambda\left[\langle\pi_f^2\rangle_{eq} - \langle\sigma_f^2\rangle_{eq}\right]\neq 0$, that is non-vanishing at finite temperature. This is an apparent violation of Goldstone's Theorem. It is apparent only, because in Chapter \ref{sect-masses} it is shown that Goldstone's Theorem is not violated. 

\item The effect of hard modes on the condensate enters through the deviation of the field fluctuations from their equilibrium values. This is nothing but the response of the fast modes to the presence of the slow ones. In this work these deviations are identified as time-delayed response functions, and are evaluated using linear response theory. The response functions {\sl renormalize} the equations of motion and give rise to {\sl dissipation}. The consequences of such renormalization is the modification of the properties of particles in the condensate, such as mass, and changes in their speed of travel (see Chapter \ref{sect-masses}). Due to the possible interaction with the hard modes decay channels open up and particles can scatter. These processes are responsible for the dissipation of the condensate. For detailed discussion see Chapters \ref{sect-dissip1} and \ref{sect-dissip2}. 

\end{itemize}

\section{Response Functions \label{sect-response}}

Studying the dynamics of the condensates means solving equations (\ref{sigmaeom}) and (\ref{pioneom}) in the chiral limit, and equations (\ref{sigmaeom2}) and (\ref{pioneom2}) for the theory with massive pions, together with some initial conditions. An important step towards this goal is the evaluation of the response functions. 

Linear response theory is a very convenient tool when one is interested in monitoring the effect of an an external field applied to a system initially in thermal equilibrium and this effect is small (for a comprehensive introduction see \cite{walecka}). The response function expresses the difference between the expectation value of an operator before and after an external perturbation has been turned on. Consider a perturbation described by ${\hat H}_{perturbed}$, and turned on at time $t_0$. 
\be
\delta\langle\hat{\cal O}(t)\rangle & = & \langle\hat{\cal O}(t)\rangle_{perturbed} - \langle\hat{\cal O}(t)\rangle \nonumber \\
&=& i\int_{t_0}^t dt'\langle\left[{\hat H}_{pert.}(t'),\hat{\cal O}(t)\right]\rangle_{unperturbed} \,~.
\label{linrespgen}
\ee
The expectation value is evaluated in the unperturbed ensemble. Extension of this expression to spatial coordinates is straightforward.

Our program is to use the result of linear response theory to evaluate the effects of the soft condensate fields on the thermal medium. According to (\ref{linrespgen}), these are just the commutators of the field operators at different powers evaluated at two separate space-time points in the fully interacting, but unperturbed ensemble. The initial conditions are set by the assumption that in heavy-ion collisions the system reaches a state of approximate local thermal equilibrium, then it cools while expanding and reaches the critical temperature. We define the initial time $t=0$ by when the critical temperature is reached. This implies $\sigma_s(0,\vec{x})=0$ and $\vec\pi_s(0,\vec{x})=0$. The general expressions to be evaluated are:
\be
\delta\langle\sigma_f^n(x)\pi_f^m(x)\rangle = i\int_0^t dt'\int d^3x'\langle\left[H_{perturbed}(x'),\sigma_f^n(x)\pi_f^m(x)\right]\rangle_{eq}
\label{linresp}
\ee
The respective powers $n$ and $m$ are identified from (\ref{sigmaeom}), (\ref{pioneom}), (\ref{sigmaeomm}), and (\ref{pioneomm}).
Let us emphasize that the responses should be evaluated in the unperturbed, equilibrium ensemble, which does include all the interactions between the different modes. 

\subsection{Couplings Between Different Modes \label{sect-coupling}}

The possible couplings between low and high frequency modes are determined by evaluating the potential $U(\sigma,\vec\pi)$, with the fields separated into their slow and fast components:
\be
\sigma &=& v + \sigma_s + \sigma_f \,~,\\
\vec\pi &=& \vec\pi_s + \vec\pi_f \nonumber\,~.
\ee
The resulting Hamiltonian contains positive powers of the $\sigma_s$ and 
$\vec\pi_s$ fields. For small departures from equilibrium it is enough to 
keep the dominant, at most linear terms only. With this assumption, the 
relevant couplings are the following\footnote{These couplings refer to the 
massless pion case. The couplings involving massive pions are the same with 
the replacement $f_\pi^2\rightarrow f_\pi^2 - m_\pi^2/\lambda$.}:
\be
&& H_{\sigma_s\sigma_f} = \lambda(3v^2-f_\pi^2)\sigma_s\sigma_f + 3\lambda v\sigma_s\sigma_f^2 + \lambda\sigma_s\sigma_f^3\nonumber\\
&& H_{\pi_s\pi_f} = \lambda(v^2-f_\pi^2)\pi_s\pi_f + \lambda\pi_s\pi_f^3\nonumber\\
&& H_{\sigma_s\pi_f} = \lambda v\sigma_s\pi_f^2\nonumber\\
&& H_{\pi_s\sigma_f\pi_f} =  \lambda\pi_s\sigma_f^2\pi_f  + 2 \lambda v\pi_s\sigma_f\pi_f\nonumber\\
&& H_{\sigma_s\sigma_f\pi_f} = \lambda\sigma_s\sigma_f\pi_f^2
\label{ham}
\ee
As before, $\pi_{s,f}$ refer to one of the $N-1$ components of the pion field. 
Next, we focus on the detailed evaluation of the two-point functions of the thermalized fields. According to the our treatment, these have only linear dependence on the non-thermal fields, and are obtained by inserting (\ref{ham}) in (\ref{linresp}). The technical details of this derivation are uninteresting and so we only state the results:
\be 
\delta\langle\sigma_f^2(x)\rangle = 3i\lambda v\int_0^t dt'\int d^3x'\sigma_s(x')\langle\left[\sigma_f^2(x'),\sigma_f^2(x)\right]\rangle_{eq.} \,~,
\label{phi1}
\ee
\be 
\delta\langle\pi_f^2(x)\rangle = i \lambda v\int_0^t dt'\int d^3x' \sigma_s(x')\langle\left[\pi_f^2(x'),\pi_f^2(x)\right]\rangle_{eq.}\,~,
\label{phi2}
\ee
\be 
\delta\langle\sigma_f(x)\pi_f(x)\rangle = 2i\lambda v\int_0^t dt'\int d^3x' \pi_s(x')\langle\left[\sigma_f(x')\pi_f(x'),\sigma_f(x)\pi_f(x)\right]\rangle_{eq.} \,~.
\label{phi1phi2}
\ee
The three-point functions can readily be read off when inserting (\ref{ham}) into equation (\ref{linresp}). In what follows, the evaluation and physical interpretation of the expectation values is pursued.

\subsection{The Expectation Values}

Obtaining explicit expressions for (\ref{phi1}), (\ref{phi2}), and (\ref{phi1phi2}), requires the evaluation of the expectations values of the commutators. Based on the commutator relation  
\be
[AB,CD] = A[B,C]D + [A,C]BD + CA[B,D] + C[A,D]B \,~,
\ee
the results are
\be
\langle\left[\sigma_f^2(x'),\sigma_f^2(x)\right]\rangle_{eq.} &=& 2 \left(D_\sigma^<(x,x')^2 - D_\sigma^>(x,x')^2\right) \,~,\nonumber\\
\langle\left[\pi_f^2(x'),\pi_f^2(x)\right]\rangle_{eq.} &=& 2 \left(D_\pi^<(x,x')^2 - D_\pi^>(x,x')^2 \right) \,~,\nonumber\\
\langle\left[\sigma_f(x')\pi_f(x'),\sigma_f(x)\pi_f(x)\right]\rangle_{eq.} &=& D_\sigma^<(x,x')D_\pi^<(x,x') - D_\sigma^>(x,x')D_\pi^>(x,x') \,~.\nonumber
\label{expect}
\ee
Keeping in mind that $t'<t$, where $t$ is the time elapsed after switching on the perturbation, and $t'$ is the time-variable that has its values in the $[0,t]$ interval (see (\ref{linrespgen})), the following notation has been introduced:
\be
&& D_\sigma^>(x,x')\equiv \langle\sigma_f(x)\sigma_f(x')\rangle_{eq.}\,~, ~~~ D_\pi^>(x,x')\equiv \langle\pi_f(x)\pi_f(x')\rangle_{eq.}\,~,\nonumber\\
\mbox{and}\nonumber\\
&& D_\sigma^<(x,x')\equiv \langle\sigma_f(x')\sigma_f(x)\rangle_{eq.}\,~, ~~~ D_\pi^<(x,x')\equiv \langle\pi_f(x')\pi_f(x)\rangle_{eq.}\,~.
\label{propag}
\ee
The functions $D^>$ and $D^<$ are known to define the spectral function, which is the inverse Fourier-transform of the thermal average of the commutator: 
\be
\rho(k) &=& D^>(k)-D^<(k) \nonumber\\
&=& \int d^4xe^{ik(x-x')}\left(D^>(x,x') - D^<(x,x')\right) \nonumber\\
&=& \int d^4ke^{ik(x-x')}\langle\left[\varphi(x),\varphi(x')\right]\rangle_{eq.}\,~,
\ee
where $\varphi=\sigma_f,\pi_f$, and $x=(t,\vec x)$ and $k=(k^0,\vec{k})$ are four-vectors in coordinate and momentum space. This spectral density is a crucial ingredient of finite temperature field theories, since it determines the real time propagator, otherwise known as the thermal Green's function: 
\be
D(x,x') &=& \int\frac{d^4k}{(2\pi)^4}e^{-ik(x-x')}D(k)\nonumber\\
&=& \int\frac{d^4k}{(2\pi)^4}e^{-ik(x-x')}\left(\Theta(t-t')+f(k^0)\right)\rho(k)\,~.
\label{relprop}
\ee
Other useful relations are
\be
&& D_i^>(k) = (1 + f(k^0))\rho_i(k)\,~,\nonumber\\
&& D_i^<(k) = f(k^0)\rho_i(k)\,~,~~~i=\sigma,\pi\,~,
\label{><}
\ee
where $f(k^0) = (e^{k^0\beta}-1)^{-1}$ is the Bose-Einstein distribution and $\beta = T^{-1}$. Significant physics is carried in (\ref{relprop}): the poles of this real-time propagator correspond to quasi-particle (collective mode) excitations. The energies and lifetimes of the mesons of our interest are determined in Chapters \ref{sect-masses} and \ref{sect-dissip1} respectively. More on the properties of spectral densities is found in \cite{lebellac}. Another approach to finite temperature field theory, the imaginary time approach, is presented in \cite{kapusta}. 

The final results of this section are the two-point functions of thermalized fields in terms of spectral functions, obtained by the insertion  of (\ref{><}) and (\ref{expect}) in the expressions (\ref{phi1}), (\ref{phi2}), and (\ref{phi1phi2}):
\be
\delta\langle\sigma_f^2(x)\rangle &=& i3\lambda v\int_0^t dt'\int d^3x'\sigma_s(x')2\left(D_\sigma^<(x,x')^2 - D_\sigma^>(x,x')^2\right)\nonumber\\
&=& - i6\lambda v\int d^4x'\sigma_s(x')\int\frac{d^4p}{(2\pi)^4}\int\frac{d^4q}{(2\pi)^4}e^{-i(p+q)(x-x')}\nonumber\\
&& \qquad \qquad \qquad\qquad\times \rho_\sigma(p)\rho_\sigma(q)(1+f(p^0)+f(q^0))
\label{1^2}
\ee
\be
\delta\langle\pi_f^2(x)\rangle &=& - i2\lambda v\int d^4x'\sigma_s(x')\int\frac{d^4p}{(2\pi)^4}\int\frac{d^4q}{(2\pi)^4} e^{-i(p+q)(x-x')}\nonumber\\
&& \qquad\qquad\qquad\qquad\times \rho_\pi(p)\rho_\pi(q)(1+f(p^0)+f(q^0))
\label{2^2}
\ee
\be 
\delta\langle\sigma_f(x)\pi_f(x)\rangle &=& 
 - i2\lambda v\int d^4x'\pi_s(x')\int\frac{d^4p}{(2\pi)^4}\int\frac{d^4q}{(2\pi)^4}e^{-i(p+q)(x-x')} \nonumber\\
&& \qquad \qquad\qquad \times \rho_\sigma(p)\rho_\pi(q)(1+f(p^0)+f(q^0))\,~.
\label{12}
\ee
\chapter{Meson Masses \label{sect-masses}}

As particles propagate through matter, they become dressed by the interactions with the constituents of the medium. Dressed means that they acquire an effective mass, which is the physical mass, and which is different from the one measured in vacuum. In this chapter a self-consistent evaluation of the meson masses at finite temperature is presented in the model with and without explicit symmetry breaking, as introduced in chapter \ref{sect-eom}. Special attention is given to the fulfillment of Goldstone's Theorem in the theory with exact chiral symmetry. 

\section{Sigma Mass \label{sect-sigmamass}}

Recall the equation of motion for the classical sigma field, derived in section \ref{sect-eom-chiral}:
\be
\partial^2\sigma_s + \bar m_\sigma^2\sigma_s + \lambda v\left[3\delta\langle\sigma_f^2\rangle + (N-1)\delta\langle\pi_f^2\rangle\right] = 0 \,~,
\ee
where 
\be
\bar m_\sigma^2 &=& 2\lambda v^2  \,~,
\ee
is the mass of the sigma meson, given in terms of the equilibrium condensate,
\be
v^2 = f_\pi^2 - 3\langle\sigma_f^2\rangle_{eq} - (N-1)\langle\pi_f^2\rangle_{eq} \,~.
\label{selfconsist}
\ee
The sigma mass is temperature dependent since it includes the thermal fluctuations of the fields. These equilibrium field fluctuations, $\langle\sigma_f^2\rangle_{eq}$ and $\langle\pi_f^2\rangle_{eq}$, are determined from the general form for a two-point function \cite{lebellac},
\be
\langle\varphi(x)\varphi(y)\rangle_{eq} = \langle\varphi(x)\varphi(y)e^{S_I}\rangle_0 = \int\frac{d^4p}{(2\pi)^4}e^{-ip(x-y)}D^>(p)\,~,
\label{2point}
\ee
where the time-components satisfy the condition $x^0\geq y^0$. Here $\langle\rangle_{eq}$ means fully interacting equilibrium ensemble average and $\langle\rangle_0$ refers to the non-interacting, free ensemble average. The usual way for finding an explicit expression for (\ref{2point}) is to expand the interaction-action, $S_I$, in powers of the coupling \cite{kapusta}. As a first approximation, one can keep only the leading order term of the perturbative expansion in $\lambda$, meaning that the propagator uses the bare mass. The scalar meson field fluctuation is then
\be
\langle\sigma_f^2\rangle = \int\frac{d^3p}{(2\pi)^3}\frac{1}{2E}(1+2f_\sigma(E))\,~.
\label{sigmafluct}
\ee
Here $f_\sigma$ is the Bose-Einstein distribution function of sigma mesons with energy $E=\sqrt{p^2+m_\sigma^2}$, and $m_\sigma$ is the zero temperature bare mass. The first term in (\ref{sigmafluct}) is the vacuum contribution, while the second term is due to finite temperature effects. The zero temperature part is divergent in the ultraviolet limit. This divergence can and should be removed using vacuum renormalization techniques \cite{peskin}. The finite temperature does not introduce any extra divergence since it is regularized by the distribution function. Therefore, $T=0$ renormalization is enough to obtain finite results (see also \cite[page 43]{kapusta}). Since our focus is on the physics at finite temperatures, in what follows we are going to neglect zero temperature contributions, assuming that these have been renormalized away. 

The pion field fluctuations have a similar form:
\be
\langle\pi_f^2\rangle = \int\frac{d^3p}{(2\pi)^3}\frac{1}{2E}(1+2f_\pi(E)) \,~.
\label{pionfluct}
\ee
It is important to mention that the momentum integration in (\ref{sigmafluct}) and (\ref{pionfluct}) has a lower limit, $\Lambda_c$, due to the restriction to hard momenta, $|\vec{p}|>\Lambda_c$. Therefore, the equilibrium distribution functions have the general form of 
\be
f(E) = \frac{1}{e^{\frac{E}{T}}-1}~\Theta(p-\Lambda_c)\,~, ~~~ E=\sqrt{p^2+m^2} \,~.
\ee
So far $m$ is the zero temperature renormalized mass. A complete treatment, however, involves the full propagator that uses the dressed mass. Accordingly, the fluctuations of the thermalized fields are themselves dependent on the sigma mass. The temperature dependent mass, $\bar m_\sigma$, results as the self-consistent solution of (\ref{selfconsist}) with (\ref{sigmafluct}) and (\ref{pionfluct}), known in the literature as the gap equation. Self-consistent solutions are approximate only, because in general they resum only certain classes of diagrams in the perturbation series. For example, self-consistent resummation of tadpole diagrams is known as the Hartree approximation. At this point, it is worth mentioning that using a self-consistent approximation makes the usual renormalization procedure difficult, as discussed recently in \cite{renorm}. The argument, according to which the renormalized divergent term can be neglected, because it is independent of temperature, is really correct only when the mass is the bare mass. A self-consistent calculation involves the temperature-dependent mass, leading to the temperature-dependence of the divergent term. Renormalization thus results in temperature-dependent renormalization constants, and these should not be ignored. However, such a treatment is beyond the scope of this dissertation, and in what follows, we are going to ignore the divergent term, while still being aware of this approximation.      

\section{Pion Mass \label{sect-pionmass}}

{\it ``If there is a continuous symmetry transformation under which the Lagrangian is invariant, then either the vacuum state is also invariant under the transformation, or there must exist spinless particles of zero mass.''} This statement, written by Goldstone, Salam and Weinberg in \cite{goldi}, today is most commonly known as Goldstone's Theorem, and has to be satisfied at every order in perturbation theory. 

The spinless, massless excitations, also called  Goldstone bosons, in QCD are the pions, and are associated with the phenomenon of spontaneous chiral symmetry breaking. In the exact chiral limit we know that pions are massless at the tree level, $m_\pi=0$. In the following we verify if this stays the same to first order in $\lambda$. Let us look at the linearized field equation derived for soft pions in section \ref{sect-eom-chiral}:
\be
\partial^2\pi_s + \tilde m_\pi^2\pi_s + 2\lambda v\delta\langle\pi_f\sigma_f\rangle = 0 \,~,
\ee
where the notation introduced is
\be
\tilde m_\pi^2 &=& 2\lambda\left[\langle\pi_f^2\rangle_{eq} - \langle\sigma_f^2\rangle_{eq}\right]\nonumber\\
&=&  2\lambda\left[\int\frac{d^3p}{(2\pi)^3}\frac{1}{2E_\pi}(1+2f_\pi) - \int\frac{d^3p}{(2\pi)^3}\frac{1}{2E_\sigma}(1+2f_\sigma)\right] \,~.
\label{pionmass}
\ee
To fulfill the requirements of the theorem, expression (\ref{pionmass}) should vanish at all temperatures. However, there is no reason for this to happen, except at zero temperature, where the thermal fluctuations themselves vanish, or when the masses of the pion and the sigma are equal, which is expected only at the critical temperature. Thus, at first inspection one finds a violation of a basic theorem. 
This would mean that the presence of the heat bath alters the Goldstone boson nature of the pion, leading to the absence of massless excitations in the theory. This problem needs to be dealt with, because Goldstone's Theorem should not be violated under any circumstances. The same problem has been found also in \cite{rpa,magyar} and has been fixed using the random phase approximation in \cite{rpa}, and direct but tedious evaluation in \cite{magyar}. In the following, we present a simple and clear way to prove that this violation is only apparent. 

First, one has to recognize that (\ref{pionmass}) includes the one-loop tadpole contributions only. There is another one-loop diagram that contributes to order $\lambda$. This is incorporated in the equation of motion through the response function $\delta\langle\pi_f\sigma_f\rangle$ (see section \ref{sect-dissip1}), given by (\ref{12}). Since the response is already of the order of $\lambda v\sim\lambda^{1/2}$, and there is an overall factor of $2\lambda v\sim\lambda^{1/2}$ in front of it, the expectation values can be evaluated at lowest order, meaning that the interacting ensemble average $\langle...\rangle_{eq.}$ can be replaced by the free ensemble average $\langle...\rangle_0$. This is equivalent to the insertion of the free spectral function,
\be
\rho_{free}(p) = 2\pi\epsilon(p^0)\delta(p^2-m^2)\,~,
\ee
into expression (\ref{12}). After evaluating the frequency integrals:
\be
\delta\langle\sigma_f\pi_f\rangle = i2\lambda\bar m_\sigma^2\int d^4x'\pi_s(x')\int\frac{d^3p}{(2\pi)^3}\int\frac{d^3q}{(2\pi)^3} e^{i(\vec{p}+\vec{q})(\vec{x}-\vec{x}~')} F(\vec{p}, \vec{q}, t')\,~,
\label{sp}
\ee
with
\be
F(\vec{p}, \vec{q}, t') &=& \frac{1}{4E_\sigma E_\pi}\left[(1+f_\sigma+f_\pi)\left(e^{i(E_\sigma+E_\pi)(t-t')} -e^{-i(E_\sigma+E_\pi)(t-t')}\right)\right. \nonumber\\
&& \left. \qquad\qquad +(f_\pi-f_\sigma)\left(e^{i(E_\sigma-E_\pi)(t-t')}-e^{-i(E_\sigma-E_\pi)(t-t')}\right)\right]\,.\nonumber\\
\label{F}
\ee
Since the deviation from equilibrium is assumed to be small, one can approximate the condensate with the first term from the Taylor expansion about its equilibrium value, $\pi_s(x')\simeq \pi_s(x)$, and can be taken out of the integral. Now, the field equation has a simplied form:
\be
\partial^2\pi_s(x) + \bar m_\pi^2\pi_s (x) = 0\,~.
\ee
The total pion mass is then 
\be
\bar m_\pi^2 = \tilde{m}_\pi^2 + a_1\,~,
\label{truemass}
\ee
where 
\be
a_1 &=& i2\lambda\bar m_\sigma^2\int_0^t dt'\int\frac{d^3p}{(2\pi)^3}\int\frac{d^3q}{(2\pi)^3}\int d^3x' 
e^{i(\vec{p}+\vec{q})(\vec{x}-\vec{x}~')} F(\vec{p}, \vec{q}, t')\nonumber\\
&=& i2\lambda\bar m_\sigma^2\int_0^t dt' \int\frac{d^3p}{(2\pi)^3}\int\frac{d^3q}{(2\pi)^3}\delta^3(\vec{p}+\vec{q})F(\vec{p}, \vec{q}, t')\nonumber\\
&=& i2\lambda\bar m_\sigma^2\int\frac{d^3p}{(2\pi)^3}\int_0^t dt'F(\vec{p}, t')
\ee
and $E_1 = \sqrt{\vec{p}~^2 + \bar m_\sigma^2}$ and $E_2 = |\vec{p}~|$. Since the coherence time is finite, one can extend the time integral from zero to negative infinity\footnote{Another argument: for negative times, $t'<0$, there is no condensate, $\pi(x')=0$, because this time period corresponds to the symmetric phase, above the critical temperature.}.
\be
\int_0^t dt' e^{\pm iE(t-t')} &=& \int_{-\infty}^t dt' e^{\pm iE(t-t')} = \int_0^{\infty} dT e^{\pm iET}  \nonumber\\
&=& \int_0^{\infty} dT e^{\pm i(E\pm i\epsilon)T} = \frac{1}{\mp iE+\epsilon}
\ee
where $\epsilon\rightarrow 0^+$ is a small positive real number. Then
\be
a_1 &=& i2\lambda \bar m_\sigma^2\int\frac{d^3p}{(2\pi)^3}\frac{1}{4E_\sigma E_\pi}\left[\frac{2E_\sigma(1+2f_\pi)-2E_\pi(1+2f_\sigma)}{i(E_\pi^2-E_\sigma^2)}\right] \nonumber\\
&=&2\lambda\int\frac{d^3p}{(2\pi)^3}\left[\frac{1}{2E_\sigma}(1+2f_\sigma) - \frac{1}{2E_\pi}(1+2f_\pi)\right] \,~.
\label{a1}
\ee
Insert (\ref{a1}) and (\ref{pionmass}) in the expression for the true pion mass, (\ref{truemass}):
\be
\bar m_\pi^2 &=& 2\lambda\left[\int\frac{d^3p}{(2\pi)^3}\frac{1}{2E_\pi}(1+2f_\pi) - \int\frac{d^3p}{(2\pi)^3}\frac{1}{2E_\sigma}(1+2f_\sigma)\right] \nonumber\\
&& + 2\lambda\int\frac{d^3p}{(2\pi)^3}\left[\frac{1}{2E_\sigma}(1+2f_\sigma) - \frac{1}{2E_\pi}(1+2f_\pi)\right]\nonumber\\
&=& 0
\ee
thus proving that pions stay Goldstone bosons at one-loop level.  Another simple and straightforward proof is presented in section \ref{sect-mom} when working in frequency-momentum space.

\section{Numerical Results \label{sect-masses-results}}

\subsection{Chiral Limit}

In the theory with exact chiral symmetry, $\bar m_\pi = 0 $, the sigma meson mass is
\be
\bar m_\sigma^2 = 2\lambda\left[f_\pi^2 - 3\langle\sigma_f^2\rangle_{eq} - (N-1)\langle\pi_f^2\rangle_{eq}\right] \,~.
\label{sigmam}
\ee
The hard sigma field fluctuation,
\be
\langle\sigma_f^2\rangle = \frac{1}{2\pi^2}\int_0^\infty dp~\frac{p^2}{E}\frac{1}{e^{E/T}-1}\,~,
\ee
should be calculated with the meson field mass $\bar m_\sigma$, that is $E=\sqrt{p^2+\bar m_\sigma^2}$. For a massless pion field the fluctuation can be evaluated analytically, resulting in a simple form: 
\be
\langle\pi_f^2\rangle =  \frac{T^2}{12}\,~.
\ee
\begin{figure}[htbp] 
\begin {center}
\vspace*{-3.5cm}
\leavevmode
\hbox{%
\epsfysize=11.5cm
\epsffile{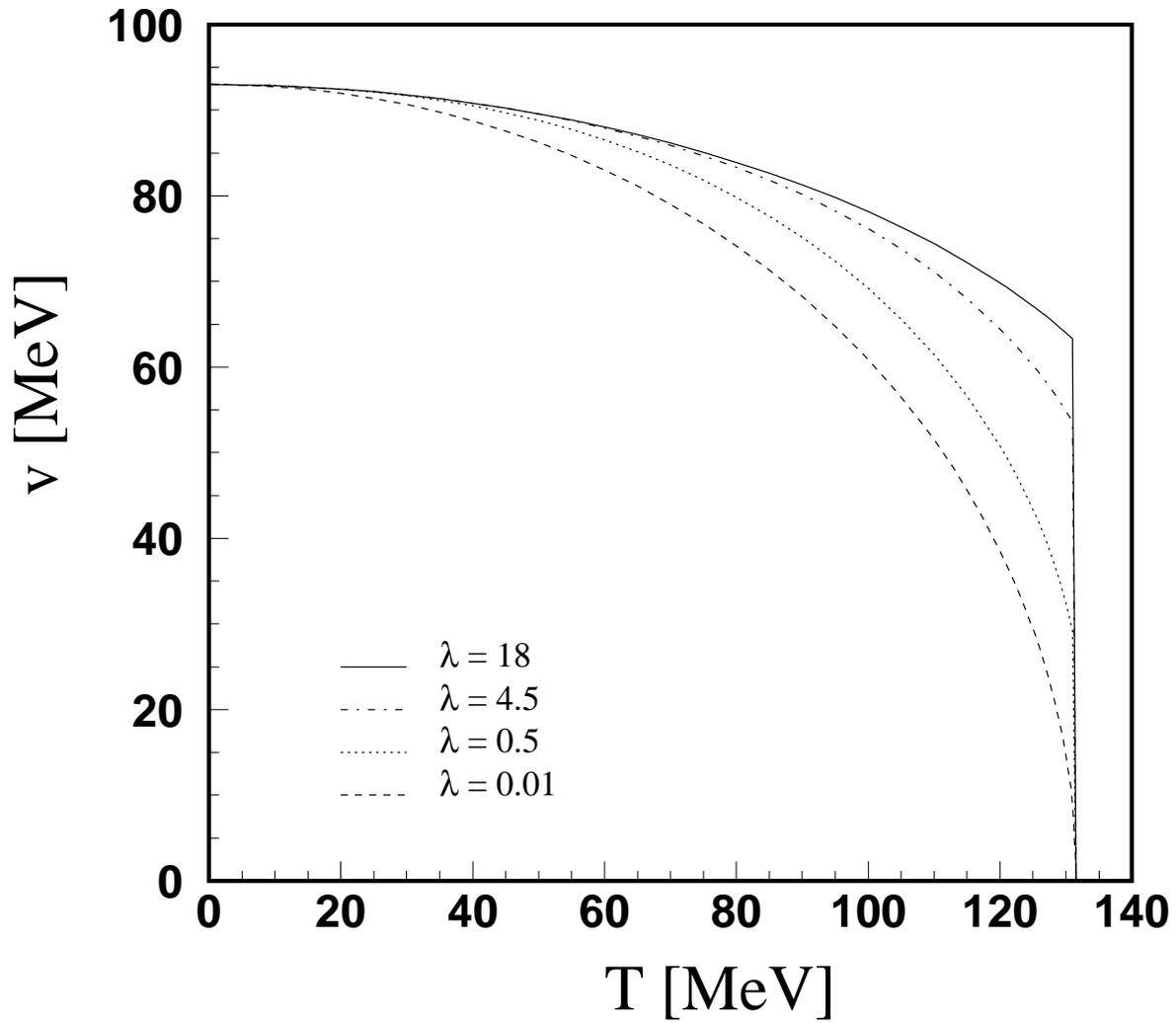}
}
\end{center}
\vspace*{6cm}
\caption{Temperature dependence of the condensate in the $O(4)$ model with exact chiral symmetry. The value of the coupling can modify the order of the phase transition.}
\label{lambda.fig}
\end{figure} 
When evaluating the field fluctuations of hard modes, the approximation $\Lambda_c\rightarrow 0$ has been made. Small momentum modes are very few inside the hard momentum bath, and including their contribution to the integrals introduces negligible small error. Physically, $\Lambda_c = 0$ means that the condensate fields contain only $k=0$ modes, meaning that the classical field configurations are homogeneous. In general, this needs not to be the case. For more discussion and analysis of $k\neq 0$ modes see Chapter \ref{sect-dispersion} on pion propagation. 

At $T=0$ thermal fluctuations obviously vanish, and the familiar $m_\sigma^2 = 2\lambda f_\pi^2$ zero temperature mass is recovered. The coupling constant is chosen to be $\lambda=18$, corresponding to a vacuum sigma mass of $m_\sigma=558~$MeV. The order parameter, which is the equilibrium condensate $v$, not only determines the mass of the sigma but also the order of the phase transition. The behavior of the condensate as the temperature changes, determined numerically in a self-consistent manner, is represented by the solid line in figure \ref{lambda.fig}. In this plot $N=4$. The effect of thermal fluctuations is to decrease the condensate with temperature from its vacuum value of $93~$MeV. At a critical temperature of $T_c\simeq 130~$MeV the condensate is completely dissolved. Above $T_c$ the fluctuations are much too large to allow the formation of any condensate. The critical temperature, defined by the vanishing of the condensate, can be estimated from the expression (\ref{sm}), and is given by $T_c^2 = \frac{12}{N+2}f_\pi^2$. In the $O(4)$ model $T_c = \sqrt{2}f_\pi=130~$MeV, which coincides with our numerically determined value. Note also the positive-definite mass at all temperatures. The tachyon problem present in mean field approximations \cite{bochkarev} is eliminated. \\
Figure \ref{lambda.fig} shows a discontinuous behavior of the condensate at $T_c$. Such a jump is characteristic of first order phase transitions. This result differs from the one in \cite{bochkarev}, where the transition was found to be second order, although at the same $T_c$. Thus, inclusion of thermal fluctuations with self-consistently resummed masses renders the phase transition first order. We also show in figure \ref{lambda.fig} how the discontinuity of the order parameter decreases with decrease of the coupling constant. For a weak enough coupling the transition is no longer first order, but a continuous second-order transition. This can be understood in the following way: the self-consistent gap equation (\ref{sigmam}) has been derived in the $O(N)$ model, which is a generalization of the sigma model for $N$ fields. In such a model, the coupling constant should be written as $\lambda/N$ \cite{largeN}, assuring the finiteness of the theory in the large-$N$ limit \cite{baym}. We can then write:
\be
\bar m_\sigma^2 = m_\sigma^2 - 2\frac{\lambda}{N}\left[3\langle\sigma_f^2\rangle_{eq} + (N-1)\langle\pi_f^2\rangle_{eq}\right] \,~.
\label{sm}
\ee
For $N\rightarrow\infty$ the contribution from the sigma field fluctuation disappears. The condensate is then
\be
v^2 = f_\pi^2-\frac{N-1}{N}\langle\pi_f^2\rangle_{eq} \simeq f_\pi^2 - \frac{T^2}{12}
\ee
In the last equality we have set $N-1\simeq N$ for large $N$. The above expression clearly shows a continuous decrease of the order parameter with increasing temperature. In the $O(4)$ model decreasing $\lambda$ by hand is equivalent to going to the large $N$ limit in the $O(N)$ model. We determined a large $N_{critical}\simeq 1800$ at which the transition is second order. For this $\lambda=18$ was held fixed, in this way assuring a correct vacuum mass for the sigma meson. This result is equivalent to having a $\lambda_{critical}\simeq 0.01$ for which the transition is second order, with $N=4$, corresponding to the correct number of degrees of freedom. Our results are consistent with the qualitative behavior found in \cite{baym}.  

Therefore, in the theory with massless pions there is a first order phase transition from a spontaneously broken symmetry phase to a symmetric phase which, for small $\lambda$ or equivalently for large $N$, becomes a second-order phase transition. The transition point is characterized by the vanishing of the sigma meson mass.

\subsection{Explicitly Broken Symmetry}

When the $\sigma$-model Lagrangian includes a symmetry breaking term the 
behavior of the equilibrium condensate is described by equation 
(\ref{equilibm}):
\be
\lambda v^3 + \lambda\left[3\langle\sigma_f^2\rangle_{eq} + 
(N-1)\langle\pi_f^2\rangle_{eq}\right]v - \lambda v_0^2v - H = 0 \,~,
\label{vh}
\ee
and the field fluctuations are given by
\be
\langle\sigma_f^2\rangle = \frac{1}{2\pi^2}\int_0^\infty dp~\frac{p^2}
{E_\sigma}\frac{1}{e^{E_\sigma/T}-1}
\label{fluc1}
\ee
and by
\be
\langle\pi_f^2\rangle =  \frac{1}{2\pi^2}\int_0^\infty dp~\frac{p^2}{E_\pi}
\frac{1}{e^{E_\pi/T}-1}\,~.
\label{fluc2}
\ee
\begin{figure}[htbp] 
\begin {center}
\leavevmode
\hbox{%
\epsfysize=11.5cm
\epsffile{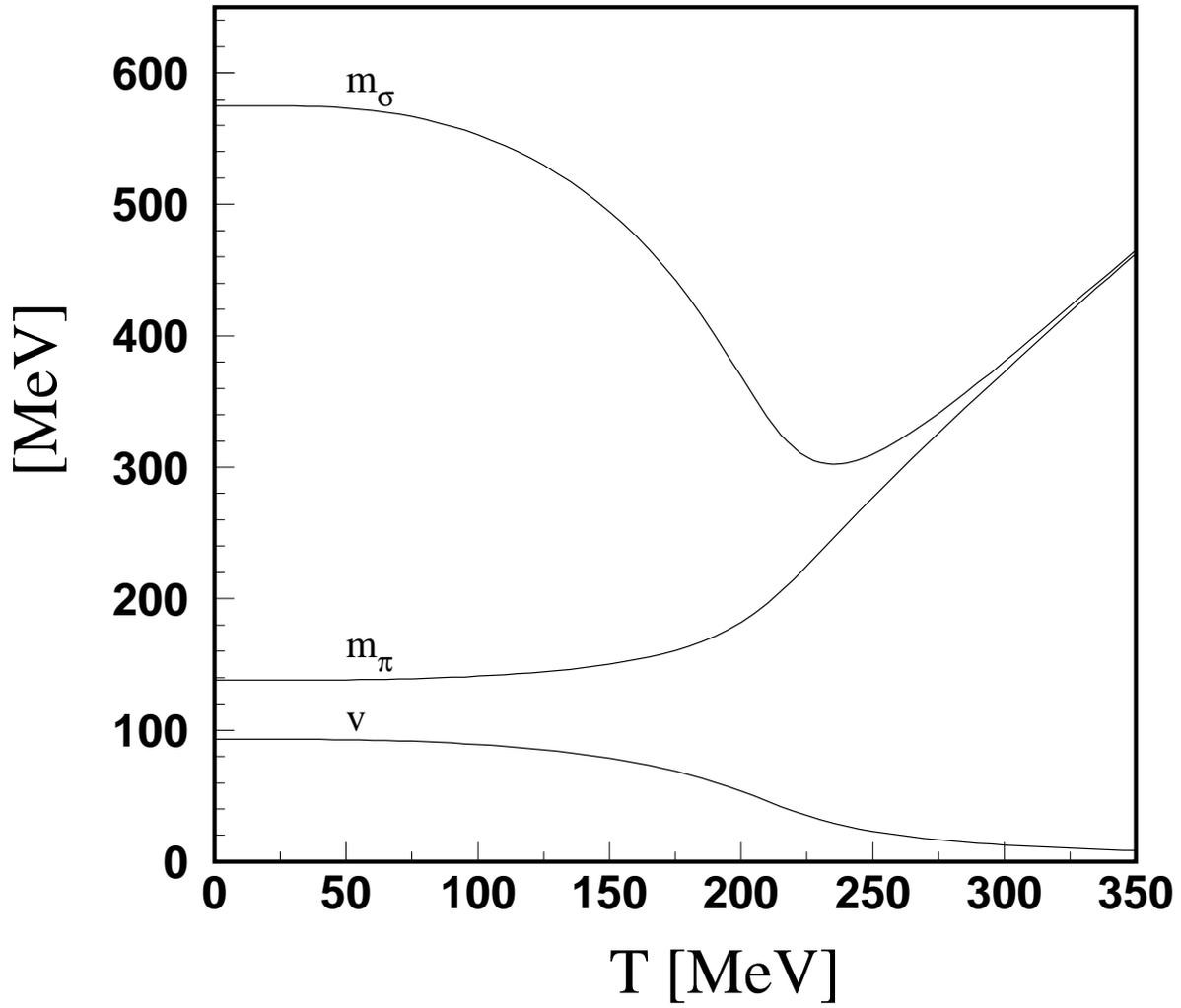}
}
\end{center}
\vspace*{6cm}
\caption{Temperature dependence of the meson masses, and of the equilibrium 
condensate in the $O(4)$ model with explicitly broken symmetry.}
\label{step.fig}
\end{figure} 
The masses in $E_\sigma=\sqrt{p^2+\bar m_\sigma^2}$ and 
$E_\pi=\sqrt{p^2+\bar m_\pi^2}$, derived in section \ref{sect-eom-nochiral} 
are
\be
\bar m_\sigma^2 = 2\lambda v^2 + \frac{H}{v}
\label{smh}
\ee
and
\be
\bar m_\pi^2 = \frac{H}{v} = \frac{f_\pi}{v}m_\pi^2   \,~,
\label{pmh}
\ee
so that (\ref{fluc1}) and (\ref{fluc2}) are the one-loop tadpoles of the 
dressed sigma and pion. Once again, determining the condensate and the meson 
masses requires a self-consistent approach: approximate the value of the 
condensate $v$, then solve for the masses $\bar m_\sigma$ and $\bar m_\pi$, 
thus obtaining $\langle\sigma_f^2\rangle_{eq}$ and 
$\langle\pi_f^2\rangle_{eq}$, which in turn is used to determine $v$ by 
solving equation (\ref{vh}). Keep repeating these steps with this newly 
obtained $v$ as long as the 
difference between the old and new values is below a required accuracy. The 
degree of symmetry breaking is given by $H=f_\pi m_\pi^2$, where 
$m_\pi=138~$MeV is the zero temperature pion mass, and 
$v_0^2=f_\pi^2-m_\pi^2/\lambda$, with $\lambda=18$, $f_\pi=93~$MeV. With 
these choice of constants the zero temperature sigma meson mass is 
$m_\sigma=575~$MeV.\\
The numerically determined self-consistent solutions for the condensate and 
the meson masses are displayed in figure \ref{step.fig}. In this case a 
qualitatively different behavior is observed. There is no phase transition. 
The equilibrium condensate monotonically decreases with increasing 
temperature. Even though one cannot define a critical temperature, there is a 
crossover region where the sigma and pion mass starts to approach degeneracy. 
This region, $160\leq T\leq 225~$MeV, corresponds to much higher temperatures 
than the critical temperature in the exact chiral limit, $T_c\simeq 130~$MeV. 
The reason is that in the present case fluctuations are smaller because the 
masses are larger. The temperature must be at least of the order of the pion 
mass in order to have significant fluctuations which can start destroying the 
condensate.
\chapter{Pion Propagation in Medium \label{sect-dispersion}}

When particles go through a medium, not only their properties are modified but also their dynamics: for example, their velocity is changed. Sigma mesons are heavy and their velocity is not altered in any significant manner \cite{ayala}. Pions, on the other hand, massless or massive, have their speed altered by the hard modes. In this chapter the change in velocity is determined. First, Goldstone modes are analyzed in coordinate space. Then the modification of the pion dispersion relation compared to the vacuum relation is discussed, taking into account the resummed meson masses. Finally, the dispersion relation of massive pion waves is determined for different temperatures. 

\section{Goldstone Bosons \label{sect-goldi}}

In the limit of exact chiral symmetry pions are massless. At zero temperature Lorentz invariance requires that massless bosons travel with the speed of light. However, at finite temperature, the existence of the thermal medium singles out a reference frame, breaking the Lorentz invariance. A direct implication of this is the modification of the speed of the Goldstone bosons. This idea has been previously looked at by Pisarski and Tytgat in \cite{rob}. They studied the propagation of cool pions through a thermal medium. Cool means that their discussion is based on the low temperature expansion of the self-energy. The analysis presented here is different. We first evaluate the physical response functions directly in coordinate space and show that our results are consistent with the results of \cite{rob}. Next, the self-energy of the soft pion field is expanded in frequency and momentum to up to fourth order. The starting point of our analysis is, once again, the pion field equation (\ref{pioneom}): 
\be
\partial^2\pi_s + 2\lambda v\delta\langle\sigma_f\pi_f\rangle + \tilde{m}_\pi^2\pi_s= 0\,~ .
\label{peom}
\ee
%

\subsection{Coordinate space: expansion of the field about its equilibrium value \label{sect-coord}}

The effect of the hard modes on the velocity of the soft modes may be seen by renormalizing the equation of motion. The idea is to expand the slow field $\pi_s(x')$ within the response function, $\delta\langle\sigma_f\pi_f\rangle$, in a Taylor series about the point $x$. This can be done since we assumed small deviations from equilibrium. The expansion is in all four coordinates. The general form of a four-dimensional Taylor expansion is
\be
f(\vec a+\vec x, \tau +t) &=&  f(\vec a,\tau) + t\dot f(\vec a,\tau) + \frac{t^2}{2}\ddot f(\vec a,\tau) + t\vec x \cdot\nabla\dot f(\vec a,\tau) + \frac{t^2}{2}\vec x \cdot\nabla\ddot f(\vec a,\tau) \nonumber \\
&+& \frac{1}{2}(\vec x \cdot\nabla)^2f(\vec a,\tau) +\frac{t}{2}(\vec x\cdot\nabla)^2\dot f(\vec a,\tau) + ...
\label{taylor1}
\ee
For now, let us disregard possible dissipation and keep only terms that are relevant for renormalization of the equation of motion:
\be
\pi_s(t',\vec x~')\simeq \pi_s(t,\vec x) + \frac{(t-t')^2}{2}\ddot\pi_s(t,\vec x) + \frac{1}{2}\left[(\vec x - \vec x~')\cdot\nabla\right]^2\pi_s(t, \vec x)
\label{taylor2}
\ee
The result of inserting the expansion (\ref{taylor2}) into the response function (\ref{sp}) and this into (\ref{peom}) is the renormalized equation of motion:
\be
\ddot\pi_s(x) - \nabla^2\pi_s(x) + \tilde m_\pi^2\pi_s(x) + a_1\pi_s(x) + a_2\ddot\pi_s(x) + a_3 \nabla^2\pi_s(x) = 0\nonumber
\ee
\be
(1+a_2)\ddot\pi_s(x) - (1-a_3) \nabla^2\pi_s(x) + (\tilde{m}_\pi^2+a_1)\pi_s(x) = 0 \nonumber
\ee
\be
\ddot\pi_s(x) - v_\pi^2\nabla^2\pi_s(x) + \bar m_\pi ^2 \pi_s(x) = 0 \,~. \nonumber
\ee
It has been proven in section \ref{sect-pionmass} that the pion mass, which includes one-loop corrections, is zero (Goldstone's Theorem),
\be
\bar m_\pi ^2 = \frac{\tilde{m}_\pi ^2+a_1}{1+a_2} = 0 \,~.
\ee
Thus the equation describing the dynamics of Goldstone modes is a familiar wave-equation:
\be
\ddot\pi_s(x) - v_\pi^2\nabla^2\pi_s(x) = 0 \,~,
\ee
where the speed of propagation of the pions is given by
\be
v_\pi^2 =  \frac{1-a_3}{1+a_2}\simeq 1-a_3-a_2\,~.
\label{speed}
\ee
In the following we evaluate $a_2$ and $a_3$ explicitly\footnote{The lower limit of the time integration is extended to $-\infty$. The arguments for doing so are presented in section \ref{sect-pionmass}.}. 
\be
a_2 &=& i\lambda \bar m_\sigma^2\int_{-\infty}^t~dt'~(t-t')^2\int\frac{d^3p}{(2\pi)^3}\int\frac{d^3q}{(2\pi)^3}\int d^3x'e^{i(\vec{p}+\vec{q})(\vec{x}-\vec{x}~')}F(\vec{p},\vec{q},t')\nonumber\\
&=& \lambda\bar m_\sigma^2\int\frac{d^3p}{(2\pi)^3}\frac{1}{E_\sigma E_\pi}\frac{(1+f_\sigma+f_\pi)(E_\sigma-E_\pi)^3 + (f_\pi-f_\sigma)(E_\sigma+E_\pi)^3}{(E_\sigma^2-E_\pi^2)^3}\nonumber\\
&=&\frac{\lambda}{\bar m_\sigma^4}\int\frac{d^3p}{(2\pi)^3}\left[(1+2f_\pi)\frac{E_\sigma^2+3E_\pi^2}{E_\pi} - (1+2f_\sigma)\frac{E_\pi^2+3E_\sigma^2}{E_\sigma}\right]\,~.
\label{a2}
\ee
Here $E_\sigma=\sqrt{p^2+\bar m_\sigma^2}$ and $E_\pi=p$. Determining $a_3$ is more cumbersome because it requires the extraction and evaluation of the $\nabla^2\pi_s$ term from $\left[(\vec x - \vec x')\cdot\nabla\right]^2\pi_s$ of the Taylor expansion (\ref{taylor2}). In Cartesian coordinates
\be
(\vec{r}\cdot\nabla)^2 &=& \left(x\frac{\partial}{\partial x} + y\frac{\partial}{\partial y} + z\frac{\partial}{\partial z}\right)^2\nonumber\\
&=& x^2\frac{\partial^2}{\partial x^2} + y^2\frac{\partial^2}{\partial y^2} + z^2\frac{\partial^2}{\partial z^2} + 2xy\frac{\partial^2}{\partial x\partial y} + 2xz\frac{\partial^2}{\partial x\partial z} + 2yz\frac{\partial^2}{\partial y\partial z}\nonumber\,~.
\ee
One can neglect the cross-terms, since these do not contribute to $\nabla^2~$. The retained expression is
\be
&&i\lambda m_\sigma^2\int_0^t dt' \int d^3x'\int\frac{d^3p}{(2\pi)^3}\int\frac{d^3q}{(2\pi)^3} e^{i(\vec{p}+\vec{q})(\vec{x}-\vec{x}~')}F(\vec{p}, \vec{q}, t')\nonumber\\
&& \qquad\quad\times\left((\vec{x}-\vec{x}~')^2\frac{\partial^2\pi_s}{\partial x^2} + (\vec{y}-\vec{y}~')^2\frac{\partial^2\pi_s}{\partial y^2} + (\vec{z}-\vec{z}~')^2\frac{\partial^2\pi_s}{\partial z^2}\right)\,~.
\label{extract}
\ee
Then\protect\footnote{
\be
&& \int_{-\infty}^{\infty} dx'\int_{-\infty}^{\infty} dy'\int_{-\infty}^{\infty} dz' e^{i(k_xx'+k_yy'+k_zz')}\left(x'^2\frac{\partial^2\phi}{\partial x^2} + y'^2\frac{\partial^2\phi}{\partial y^2} + z'^2\frac{\partial^2\phi}{\partial z^2}\right)\nonumber\\
&& =\frac{\partial^2\phi}{\partial x^2} \int dx'x'^2e^{ik_xx'}\int dy'e^{ik_yy'}\int dz'e^{ik_zz'} + \frac{\partial^2\phi}{\partial y^2}\int dx'e^{ik_xx'}\int dy'y'^2e^{ik_yy'}\int dz'e^{ik_zz'} \nonumber\\
&& \qquad+ ~\frac{\partial^2\phi}{\partial z^2}\int dx'e^{ik_xx'}\int dy'e^{ik_yy'}\int dz'z'^2e^{ik_zz'} \nonumber\\
&& = -(2\pi)^3\left[\frac{\partial^2\phi}{\partial x^2}\frac{d^2\delta(k_x)}{dk_x^2}\delta(k_y)\delta(k_z) + \frac{\partial^2\phi}{\partial y^2}\delta(k_x)\frac{d^2\delta(k_y)}{dk_y^2}\delta(k_z) + \frac{\partial^2\phi}{\partial z^2}\delta(k_x)\delta(k_y)\frac{d^2\delta(k_z)}{dk_z^2}\right] \nonumber
\ee
} separate the integrand as $F = f_\sigma F_\sigma +f_\pi F_\pi+F_0$, with
\be
F_\sigma &=& \frac{1}{4E_\sigma E_\pi}\left( e^{i(E_\sigma+E_\pi)(t-t')} -e^{-i(E_\sigma+E_\pi)(t-t')}-e^{i(E_\sigma-E_\pi)(t-t')}+e^{-i(E_\sigma-E_\pi)(t-t')}\right)\,~,\nonumber\\
F_\pi &=& \frac{1}{4E_\sigma E_\pi}\left(e^{i(E_\sigma+E_\pi)(t-t')} -e^{-i(E_\sigma+E_\pi)(t-t')}+e^{i(E_\sigma-E_\pi)(t-t')}-e^{-i(E_\sigma-E_\pi)(t-t')}\right)\,~,\nonumber\\
F_0 &=& \frac{1}{4E_\sigma E_\pi}\left(e^{i(E_\sigma+E_\pi)(t-t')} -e^{-i(E_\sigma+E_\pi)(t-t')}\right)\,~,\nonumber
\ee
and use the following formulas, when evaluating the integrals:
\be
\int_{-\infty}^\infty dx f(x)\delta(x-a) = f(a)\,~,
\ee
\be
\int_{-\infty}^\infty dx f(x)\frac{d^m\delta(x-a)}{dx^m} = (-1)^m\frac{d^mf(a)}{dx^m}\,~.
\ee
Expression (\ref{extract}) is rewritten as:
\be
&&-i\lambda m_\sigma^2\int_0^t dt'\left[\int\frac{d^3p}{(2\pi)^3}f_\sigma\left(\frac{\partial^2\phi}{\partial x^2}\frac{\partial^2 F_\sigma}{\partial q_x^2} + \frac{\partial^2\phi}{\partial y^2}\frac{\partial^2 F_\sigma}{\partial q_y^2} + \frac{\partial^2\phi}{\partial z^2}\frac{\partial^2 F_\sigma}{\partial q_z^2}\right)_{q=-p}\right.\nonumber\\
&&\qquad\qquad\qquad + \left. \int\frac{d^3q}{(2\pi)^3}f_\pi\left(\frac{\partial^2\phi}{\partial x^2}\frac{\partial^2 F_\pi}{\partial p_x^2} + \frac{\partial^2\phi}{\partial y^2}\frac{\partial^2 F_\pi}{\partial p_y^2} + \frac{\partial^2\phi}{\partial z^2}\frac{\partial^2 F_\pi}{\partial p_z^2}\right)_{p=-q}\right.\nonumber\\
&& \qquad\qquad\qquad\quad + \left.\int\frac{d^3p}{(2\pi)^3}\left(\frac{\partial^2\phi}{\partial x^2}\frac{\partial^2 F_0}{\partial q_x^2} + \frac{\partial^2\phi}{\partial y^2}\frac{\partial^2 F_0}{\partial q_y^2} + \frac{\partial^2\phi}{\partial z^2}\frac{\partial^2 F_0}{\partial q_z^2}\right)_{q=-p}\right]\,~.\nonumber
\ee
Notice that $F_0$, $F_\sigma$ and $F_\pi$ are quadratically dependent of the momentum, and so these are symmetric in $p_x$, $p_y$ and $p_z$. Therefore, the derivatives with respect to the different components of the momentum are equal, and one can factor these out. The expression of interest is then
\be
a_3 &=& -i\lambda \bar m_\sigma^2\int_{-\infty}^t dt'\left[\int\frac{d^3p}{(2\pi)^3}f_\sigma\frac{\partial^2 F_\sigma}{\partial q_x^2} + \int\frac{d^3q}{(2\pi)^3}f_\pi\frac{\partial^2 F_\pi}{\partial p_x^2} + \int\frac{d^3p}{(2\pi)^3}\frac{\partial^2 F_0}{\partial q_x^2}\right]_{q=-p}\nonumber\\
&=& -\lambda \bar m_\sigma^2\left[\int\frac{d^3p}{(2\pi)^3}\frac{f_\sigma}{E_\sigma}\frac{\partial^2}{dq_x^2}\left(\frac{1}{E_\sigma^2-E_\pi^2}\right) - \int\frac{d^3q}{(2\pi)^3}\frac{f_\pi}{E_\pi}\frac{\partial^2}{dp_x^2}\left(\frac{1}{E_\sigma^2-E_\pi^2}\right) \right.\nonumber\\
&& \qquad\quad \left. -\int\frac{d^3p}{(2\pi)^3}\frac{1}{2E_\sigma}\frac{\partial^2}{dq_x^2}\left(\frac{1}{E_\pi(E_\sigma+E_\pi)}\right)\right]_{q=-p}\,~.
\ee
Here $E_\sigma=\sqrt{p^2+\bar m_\sigma^2}$ and $E_\pi = q$. The last term is the vacuum contribution and is divergent for large momenta. So is the term proportional to 1 in (\ref{a2}). These divergences are removed with a usual vacuum renormalization procedure. For the purpose of our study here, let us assume that this has been done and ignore all these terms, focusing on the finite temperature contributions only.  
\be
a_3 &=& -\lambda\bar m_\sigma^2\left[\int\frac{d^3p}{(2\pi)^3}\frac{f_\sigma}{E_\sigma}\left(\frac{2}{\bar m_\sigma^4}+\frac{8p_x^2}{\bar m_\sigma^6}\right) - \int\frac{d^3q}{(2\pi)^3}\frac{f_\pi}{E_\pi}\left(-\frac{2}{\bar m_\sigma^4}+\frac{8q_x^2}{\bar m_\sigma^6}\right)\right]\nonumber\\
&=& -\frac{2\lambda}{\bar m_\sigma^2}\int\frac{d^3p}{(2\pi)^3}\left[\frac{f_\sigma}{E_\sigma}\left(1+\frac{4}{3}\frac{p^2}{\bar m_\sigma^2}\right) + \frac{f_\pi}{E_\pi}\left(1-\frac{4}{3}\frac{p^2}{\bar m_\sigma^2}\right)\right]
\label{a3}
\ee
Putting the ingredients (\ref{a2}) and (\ref{a3}) into (\ref{speed}) one obtains:
\be
v_\pi^2 &=& 1 - a_2 - a_3\nonumber\\
&=& 1 - \frac{2\lambda}{\bar m_\sigma^4}\int\frac{d^3p}{(2\pi)^3}\left[f_\pi\frac{E_\sigma^2+3E_\pi^2}{E_\pi} - f_\sigma\frac{E_\pi^2+3E_\sigma^2}{E_\sigma}\right] \nonumber\\
&& \quad + \frac{2\lambda}{\bar m_\sigma^2}\int\frac{d^3p}{(2\pi)^3}\left[\frac{f_\sigma}{E_\sigma}\left(1+\frac{4}{3}\frac{p^2}{\bar m_\sigma^2}\right) + \frac{f_\pi}{E_\pi}\left(1-\frac{4}{3}\frac{p^2}{\bar m_\sigma^2}\right)\right]\nonumber\\
&=& 1 - \frac{2\lambda}{\bar m_\sigma^2}\int\frac{d^3p}{(2\pi)^3}\left[\frac{f_\pi}{E_\pi}\frac{16}{3}\frac{p^2}{\bar m_\sigma^2} - \frac{f_\sigma}{E_\sigma}\left(4+\frac{16}{3}\frac{p^2}{\bar m_\sigma^2}\right)\right]\nonumber\\
&=& 1 - \frac{\lambda}{\pi^2\bar m_\sigma^2}\int_0^{\infty}dp p^2\left[\frac{f_\pi}{E_\pi}\frac{16}{3}\frac{p^2}{\bar m_\sigma^2} - \frac{f_\sigma}{E_\sigma}\left(4+\frac{16}{3}\frac{p^2}{\bar m_\sigma^2}\right)\right]
\label{speedgen}
\ee
At low temperatures the large sigma mass leads to exponential suppression of the term involving $f_\sigma=1/(e^{E_\sigma/T}-1)$ in (\ref{speedgen}). The solid line in figure \ref{lambda.fig} shows that the sigma mass is large up to quite high temperatures. Therefore, this approximation breaks down only very close to the critical temperature. The rest of (\ref{speedgen}) can be done analytically since the pion mass is zero:
\be
\int_0^{\infty}dp p^2\frac{f_\pi}{E_\pi}\frac{16}{3}\frac{p^2}{\bar m_\sigma^2} =\frac{16}{3\bar m_\sigma^2}\int_0^{\infty}dp\frac{p^3}{e^{\frac{p}{T}}-1} = \frac{16}{3\bar m_\sigma^2}\frac{\pi^4T^4}{15}\,~.
\ee
The final expression for the corrected pion velocity is
\be
v_\pi^2 = 1 - \frac{16}{45}\lambda\pi^2\frac{T^4}{\bar m_\sigma^4} \,~.
\label{pispeed} 
\ee
Equation (\ref{pispeed}) describes the expected result: Goldstone waves travel slower than the speed of light when propagating in a thermal medium. Our result is consistent with the result in \cite{rob}; however, it should be emphasized that our only assumption is that the condensate is only slightly out of equilibrium. 

\subsection{Momentum space: expansion of the self-energy about zero frequency 
and momentum \label{sect-mom}}

\subsubsection{Second Order Corrections}

In this section we present an alternative way for obtaining expression (\ref{pispeed}) for the velocity of soft pion modes. The propagation properties of particles in a thermal medium are governed, in general, by the position of the pole of the propagator. This is determined by Fourier transforming the equation of motion (\ref{pioneom}) into frequency-momentum space. One readily obtains
\be
k_0^2 = \vec{k}~^2 + \tilde{m_\pi}^2 + \Pi_\pi(k_0, \vec{k})\,~.
\label{Ft}
\ee
Here $(k_0,\vec{k})$ is the pion four-momentum. The pion self-energy, $\Pi_\pi(k_0,\vec{k})$, has a real and an imaginary part. Taking $\vec{k}$ real means that the frequency has an imaginary part, $k_0 = \omega - i\Gamma$, too. We assume that the imaginary part, $\Gamma$, which determines the width, is small relative to the real part, $\omega$, which is the energy. Thus, the pions are only slightly damped. The question of damping is addressed in detail in chapters \ref{sect-dissip1} and \ref{sect-dissip2}. For now, let us keep our focus on the real part of equation (\ref{Ft}). This is the function relating the energy of the soft pion, $\omega$, moving through the heat bath, to its momentum, $k$, known as the pion dispersion relation: 
\be
\omega^2 =  \vec{k}~^2 + \tilde m_\pi^2 + \mbox{Re}\Pi_\pi(\omega, \vec{k})\,~.
\label{disp}
\ee
The pions in the condensate, being soft, allow one to expand the real part of the self-energy about zero frequency $\omega=0$ and momentum $k=|\vec{k}~|=0$. The dominant non-vanishing terms are
\be
\mbox{Re}~\Pi_\pi(\omega, \vec{k}) &\simeq& \mbox{Re}~\Pi_\pi(0,0) + \frac{1}{2}\frac{d^2\mbox{Re}~\Pi_\pi(0,0)}{d\omega^2}\omega^2 + \frac{1}{2}\frac{d^2\mbox{Re}~\Pi_\pi(0,0)}{dk^2}k^2\nonumber\\
&=& 2\lambda \bar m_\sigma^2\int\frac{d^3p}{(2\pi)^3}G(\omega,k)\nonumber\\
&=& b_1 + b_2\omega^2 + b_3k^2 \,~,
\label{rep}
\ee
where
\be
G(\omega,k) &=& \frac{1}{4E_\sigma E_\pi}\left[(1+f_\sigma+f_\pi)\left(\frac{1}{\omega-E_\sigma-E_\pi} - \frac{1}{\omega+E_\sigma+E_\pi}\right)\right. \nonumber\\
&& \left. \qquad\qquad +(f_\pi-f_\sigma)\left(\frac{1}{\omega-E_\sigma+E_\pi} - \frac{1}{\omega+E_\sigma-E_\pi}\right)\right] \nonumber\\
 &\equiv & G_0 + f_\sigma G_\sigma + f_\pi G_\pi
\label{G}
\ee
Equations (\ref{rep}) and (\ref{disp}) together lead to 
\be
\omega^2(1-b_2) = k^2(1+b_3) +  \tilde{m_\pi}^2 + b_1\nonumber
\ee
and so
\be
\omega^2 = v_\pi^2k^2+ \bar m_\pi^2 \,~,
\label{pidisp1}
\ee
where we defined 
\be
v_\pi^2 \simeq 1 + b_2 + b_3 \,~,
\ee
and
\be
\bar m_\pi^2 = \tilde m_\pi^2 + b_1 \,~.
\label{pmb}
\ee
In the following let us determine $b_1$, $b_2$ and $b_3$ explicitly from their definitions. The mass term is
\be
b_1 &=&  \mbox{Re}\Pi_\pi(0,0) = 2\lambda\bar m_\sigma^2\int\frac{d^3p}{(2\pi)^3}G(0,0)\nonumber\\
&=& 2\lambda \bar m_\sigma^2\int\frac{d^3p}{(2\pi)^3}\frac{1}{4E_\sigma E_\pi}\left(f_\sigma\frac{4E_\pi}{E_\sigma^2-E_\pi^2}+ f_\pi\frac{-4E_\sigma}{E_\sigma^2-E_\pi^2}\right)\nonumber\\
&=& 2\lambda\int\frac{d^3p}{(2\pi)^3}\left[\frac{f_\sigma}{E_\sigma} - \frac{f_\pi}{E_\pi}\right]\,~.
\label{b1}
\ee
The result (\ref{b1}) inserted into (\ref{pmb}) together with expression (\ref{pionmass}) for $ \tilde m_\pi$ yields another proof of Goldstone's Theorem at one-loop order
\be
\bar m_\pi = 0\,~,
\ee
Relation (\ref{pidisp1}) shows the existence of zero frequency excitations at zero momentum.\\
Straightforward calculation leads to 
\be
b_2 &=& \frac{1}{2}\frac{d^2\mbox{Re}\Pi_\pi(0,0)}{d\omega^2} = \lambda \bar m_\sigma^2\int\frac{d^3p}{(2\pi)^3}\frac{d^2G(0,0)}{d\omega^2}\nonumber\\
&=&\lambda \bar m_\sigma^2\int\frac{d^3p}{(2\pi)^3}\left[2f_\sigma\frac{E_\pi^2+3E_\sigma^2}{E_\sigma}-2f_\pi\frac{E_\sigma^2+3E_\pi^2}{E_\pi}\right]\frac{1}{\bar m_\sigma^6}\nonumber\\
&=& - \frac{2\lambda}{\bar m_\sigma^2}\int\frac{d^3p}{(2\pi)^3}\left[\frac{f_\pi}{E_\pi}\left(1+4\frac{p^2}{\bar m_\sigma^2}\right)-\frac{f_\sigma}{E_\sigma}\left(3+4\frac{p^2}{\bar m_\sigma^2}\right)\right]\,~.
\label{b2}
\ee
When determining $b_3$ we use the separation (\ref{G}):
\be
b_3 &=& \frac{1}{2}\frac{d^2\mbox{Re}\Pi_\pi(0,0)}{dk^2} = \lambda \bar m_\sigma^2\int\frac{d^3p}{(2\pi)^3}\frac{d^2G(0,0)}{dk^2}\nonumber\\
&=& \lambda\bar m_\sigma^2\int\frac{d^3p}{(2\pi)^3}\left[f_\sigma\frac{d^2G_\sigma(0,0)}{dk^2} + f_\pi\frac{d^2G_\pi(0,0)}{dk^2}\right]\,~.
\ee
In the first term $E_\sigma=\sqrt{p^2+\bar m_\sigma^2}$ and $E_\pi=\sqrt{p^2+k^2-\vec{k}\cdot\vec{p}}$, whereas for the second term a change of variables has been done, resulting in $E_\sigma=\sqrt{p^2+\bar m_\sigma^2+k^2+\vec{k}\cdot\vec{p}}$ and $E_\pi=p$. Then
\be
b_3 &=& \frac{2\lambda}{\bar m_\sigma^2}\int\frac{d^3p}{(2\pi)^3}\left[\frac{f_\sigma}{E_\sigma}\left(1+\frac{4}{3}\frac{p^2}{\bar m_\sigma^2}\right) + \frac{f_\pi}{E_\pi}\left(1-\frac{4}{3}\frac{p^2}{\bar m_\sigma^2}\right)\right]\,~.
\label{b3}
\ee
Notice that $b_2=-a_2$  and $b_3=-a_3$, given by expressions (\ref{a2}) and (\ref{a3}) respectively, resulting in the same dispersion relation for the Goldstone modes, as obtained in configuration space,
\be
v_\pi^2 =  1 - \frac{\lambda}{\pi^2\bar m_\sigma^2}\int_0^{\infty}dp p^2\left[\frac{f_\pi}{E_\pi}\frac{16}{3}\frac{p^2}{\bar m_\sigma^2} - \frac{f_\sigma}{E_\sigma}\left(4+\frac{16}{3}\frac{p^2}{\bar m_\sigma^2}\right)\right]\,~.
\label{speed2gen}
\ee

The vacuum dispersion relation for massless pions, which are actually on the light cone, is
\be
\omega = k\,~.
\label{vac}
\ee
In sections \ref{sect-coord} and \ref{sect-mom} the second order correction to this relation has been determined. We call the correction second order, because it is of the order of $k^2$. The corrected dispersion relation is 
\be
\omega = v_\pi k\,~,
\label{2nd}
\ee
where $v_\pi$ is the solution of
\be
v_\pi^2 &=& 1 - \frac{\lambda}{\pi^2}\left[\frac{16}{3\bar m_\sigma^4}\int_0^{\infty}dp p^4\frac{f_\pi}{E_\pi} -\frac{4}{\bar m_\sigma^2}\int_0^{\infty}dp p^2\frac{f_\sigma}{E_\sigma} - \frac{16}{3\bar m_\sigma^4}\int_0^{\infty}dp p^4\frac{f_\sigma}{E_\sigma}\right]\nonumber\\
&\simeq& 1 - \frac{16}{45}\lambda\pi^2\frac{T^4}{\bar m_\sigma^4} \,~.
\label{speed2}
\ee
This result coincides with the result obtained in coordinate-space, expression (\ref{pispeed}). Figure \ref{lambda.fig} shows that the sigma is heavy, $T\ll \bar m_\sigma$, right up to the critical temperature, which justifies the second line of (\ref{speed2}). This takes into account that sigma terms are exponentially suppressed. The dispersion relation (\ref{2nd}) is the same as obtained for sound waves propagating with velocity $v_\pi$ in a non-viscous fluid (see for example \cite{fetter}). Therefore, up to second order the phase velocity, 
\be
v_\pi^{ph} = \frac{\omega}{k} = v_\pi\,~,
\ee
and the group velocity, 
\be
v_\pi^{gr} = \frac{d\omega}{dk} = v_\pi\,~,
\ee
of a Goldstone wave are the same, and are decreasing as the temperature is increasing. This can be attributed to the fact that the higher the temperature, the more are the interactions between the different modes, and these can cause significant slowing down. 

\subsubsection{Fourth Order Corrections}

One can analyze the dispersion relation further by including higher order corrections to the velocity. The next-to-leading order non-vanishing terms in the self-energy expansion about small energy and momenta are of fourth-order.
\be
\mbox{Re}~\Pi_\pi(\omega, \vec{k}) \simeq  b_1 + b_2\omega^2 + b_3k^2 + b_4 \omega^2 k^2 + b_5k^4 + b_6\omega^4\,~.
\label{selfour}
\ee
To determine the energy in function of momentum one needs to solve a quadratic equation in $\omega^2$, equation that results from inserting (\ref{selfour}) in (\ref{disp}). The result is
\be
\omega^2 &\simeq  & k^2\left[1 + b_2 + b_3 + k^2(b_4 +b_5)\right]\nonumber\\
&=& v_\pi^2(k)k^2\,~.
\ee
An analytic result for $v_\pi$ can be obtained when $m_\sigma\gg T$. The result is
\be
v_\pi^2(k) = 1 - \frac{16}{45}\lambda\pi^2\frac{T^4}{\bar m_\sigma^4} + \frac{k^2}{\bar m_\sigma^2}\left(\frac{\lambda}{6}\frac{T^2}{\bar m_\sigma^2} + \frac{4\lambda\pi^2}{5}\frac{T^4}{\bar m_\sigma^4}\right)\,~.
\label{pi4thspeed}
\ee
When including fourth order terms, the velocity depends not only on temperature, but also on momentum. Therefore, the group and phase velocities are different.

\subsection{Numerical Results}

\begin{figure}[htbp] 
\begin {center}
\vspace*{-4cm}
\leavevmode
\hbox{%
\epsfysize=11.5cm
\epsffile{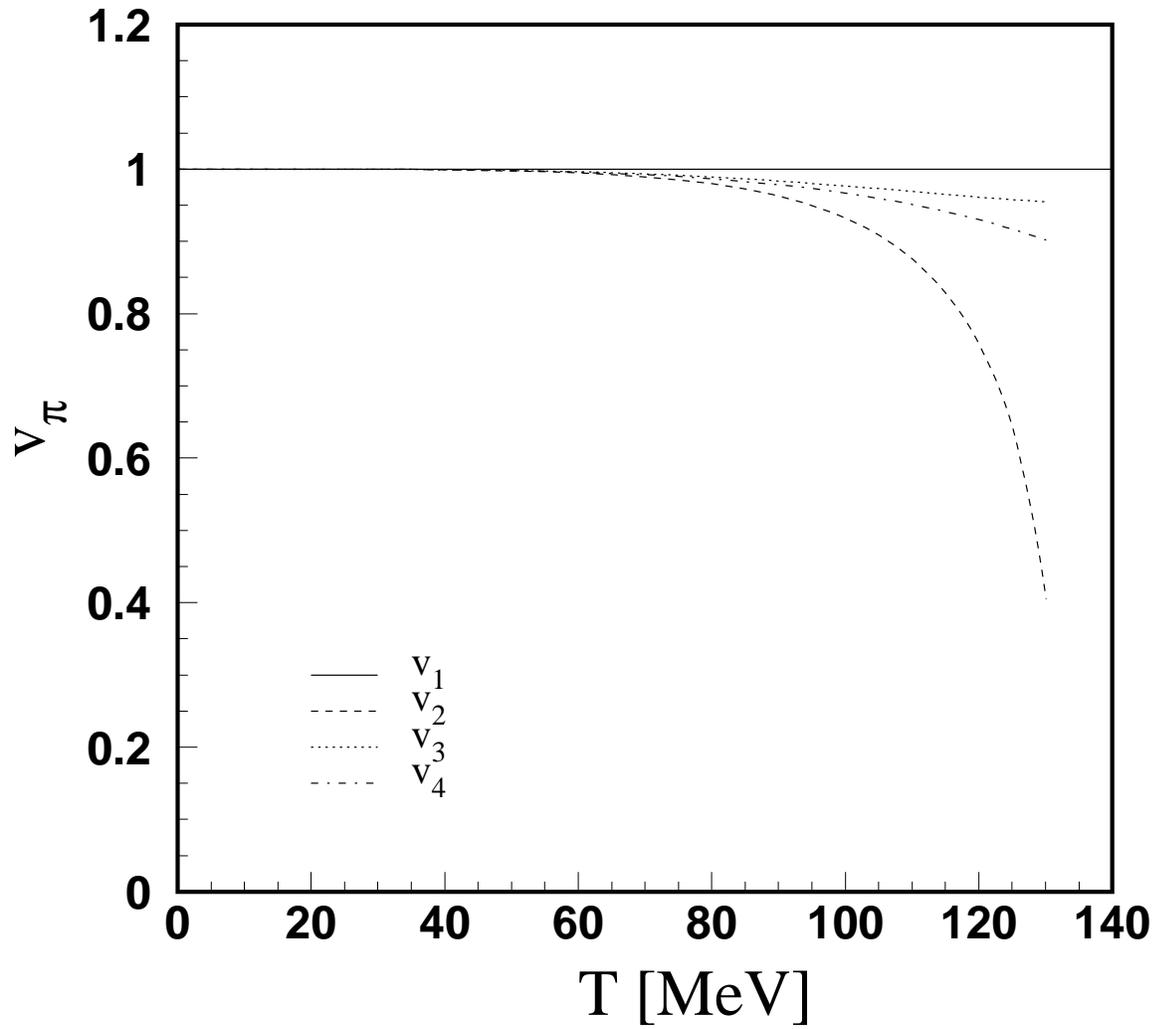}
}
\end{center}
\vspace*{5cm}
\caption{Temperature dependence of the speed of massless pions. See text for 
definition of $v_1,v_2,v_3$ and $v_4$.}
\label{speed2self.fig}
\end{figure}

The temperature dependence of the speed of massless pions is presented in 
figure \ref{speed2self.fig}, where four different expressions for this are 
compared. The following notation is used:
\be
v_1=1
\ee
is the speed of light (solid line); $v_2$ is the result of our self-consistent
 calculation given by equation (\ref{speed2}), using temperature-dependent 
sigma mass, $\bar m_\sigma^2=2\lambda v(T)^2$, determined in Chapter 
\ref{sect-masses}, (dashed line)
\be
v_2^2 = 1 - \frac{16}{45}\lambda\pi^2\frac{T^4}{\bar m_\sigma^4} \,~;
\ee
then 
\be
v_3^2 = 1 - \frac{16}{45}\lambda\pi^2\frac{T^4\bar m_\sigma^2}{m_\sigma^6} 
\ee
takes into account the temperature-dependence of the coupling $\lambda v(T)$, 
but assumes the $T=0$ mass for sigmas in the heat bath, 
$\bar m_\sigma=m_\sigma\sim 600~$MeV (dotted line); finally 
\be
v_4^2 = 1 - \frac{8\pi^2}{45}\frac{T^4}{f_\pi^2m_\sigma^2} 
\ee
is equation (24) from \cite{rob}, where both the coupling and the sigma mass 
is set to their $T=0$ values of $v=f_\pi$ and $\bar m_\sigma=m_\sigma$ 
(dotted-dashed line). 
For all of these possibilities one can clearly see a deviation of the pion 
velocity from the speed of light at finite temperatures. 

Let us look at $\lambda/\bar m_\sigma^4$ in $v_2$ more closely. Here the 
resummed sigma mass $\bar m_\sigma^2=2\lambda v^2$ introduces not only 
temperature-dependence but also coupling constant dependence. For small 
coupling, $\lambda\rightarrow 0$, $v_2$ is blown up by 
$\lambda/\bar m_\sigma^4\sim 1/\lambda$. Such behavior is not encountered in 
$v_4$. Does this mean that the effect of the medium cannot be included in a 
perturbative manner and a non-perturbative approach is needed? One way to 
reconsider this situation is by looking at $v_3$. Here the 
temperature-dependence only of the condensate is taken into account, leading 
to a $\lambda^2$ correction to the velocity. This does not introduce any bad 
behavior, since with no interactions, $\lambda\rightarrow 0$, the massless 
pions travel as they should, with the speed of light. 

\begin{figure}[htbp] 
\begin {center}
\vspace*{-2cm}
\leavevmode
\hbox{%
\epsfysize=11.5cm
\epsffile{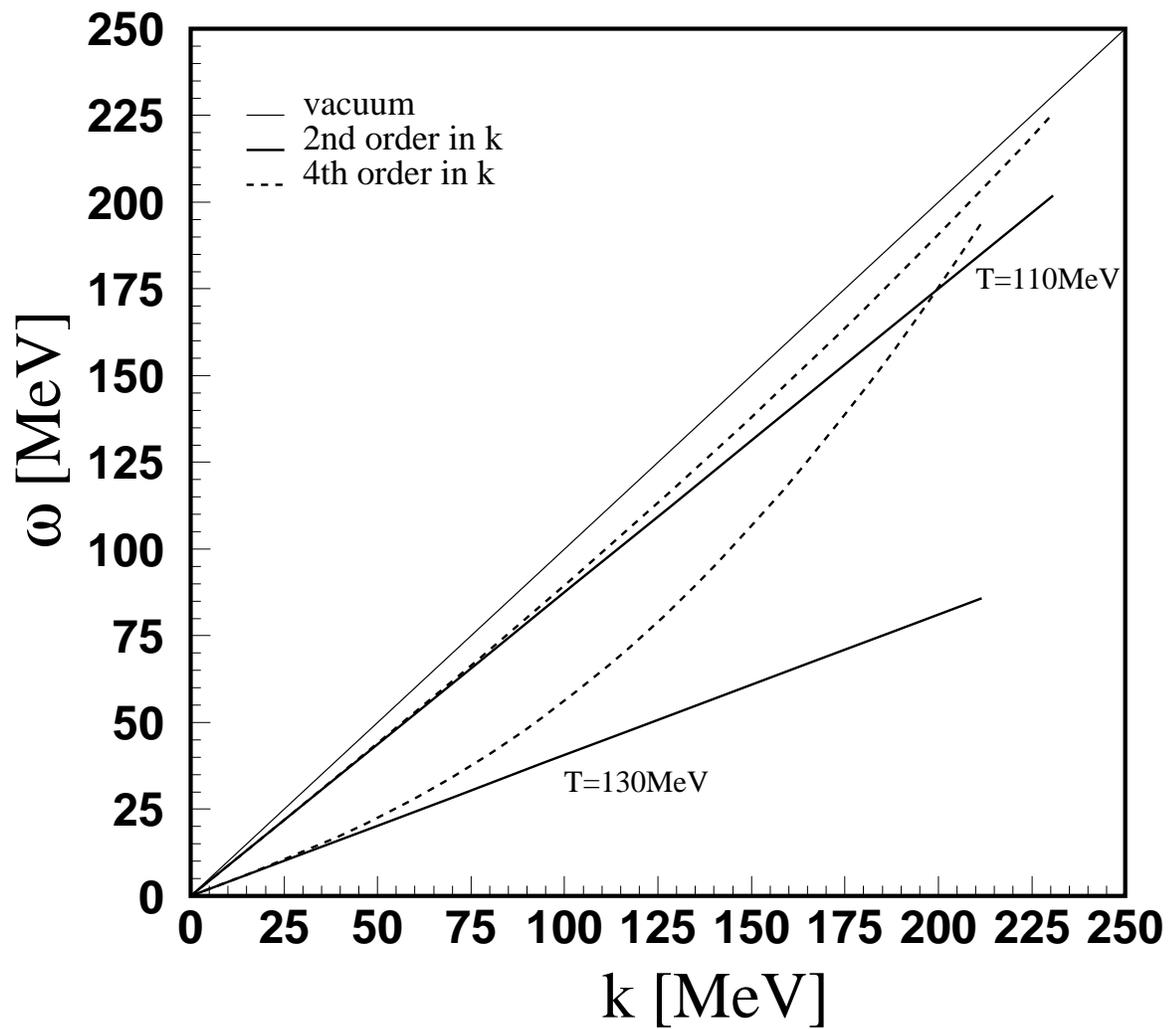}
}
\end{center}
\vspace*{5cm}
\caption{Pion dispersion curve in exact chiral limit for $T=110~$MeV (upper 
curves) and $T=130~$MeV (lower curves).}
\label{speed4th.fig}
\end{figure}

The temperature dependence of the dispersion relation of Goldstone waves is 
presented in figure \ref{speed4th.fig}. 
All the terms of the 2nd-order (\ref{speed2}) (solid curves), and the 4th-order (\ref{pi4thspeed}) (dashed curves) velocity are included in the numerics. Also, the resummed sigma mass determined in section \ref{sect-masses-results} is used. This figure shows a softening of the dispersion due to second-order corrections compared to the dispersion relation in vacuum. This softening becomes significant at higher temperatures, and is more accentuated as the temperature increases. The 2nd-order curves are straight lines whose slope gives the velocity of the massless pions. The deviation of this from the speed of light is clearly noticeable. 
\begin{figure}[htbp] 
\begin {center}
\vspace*{-2cm}
\leavevmode
\hbox{%
\epsfysize=11.5cm
\epsffile{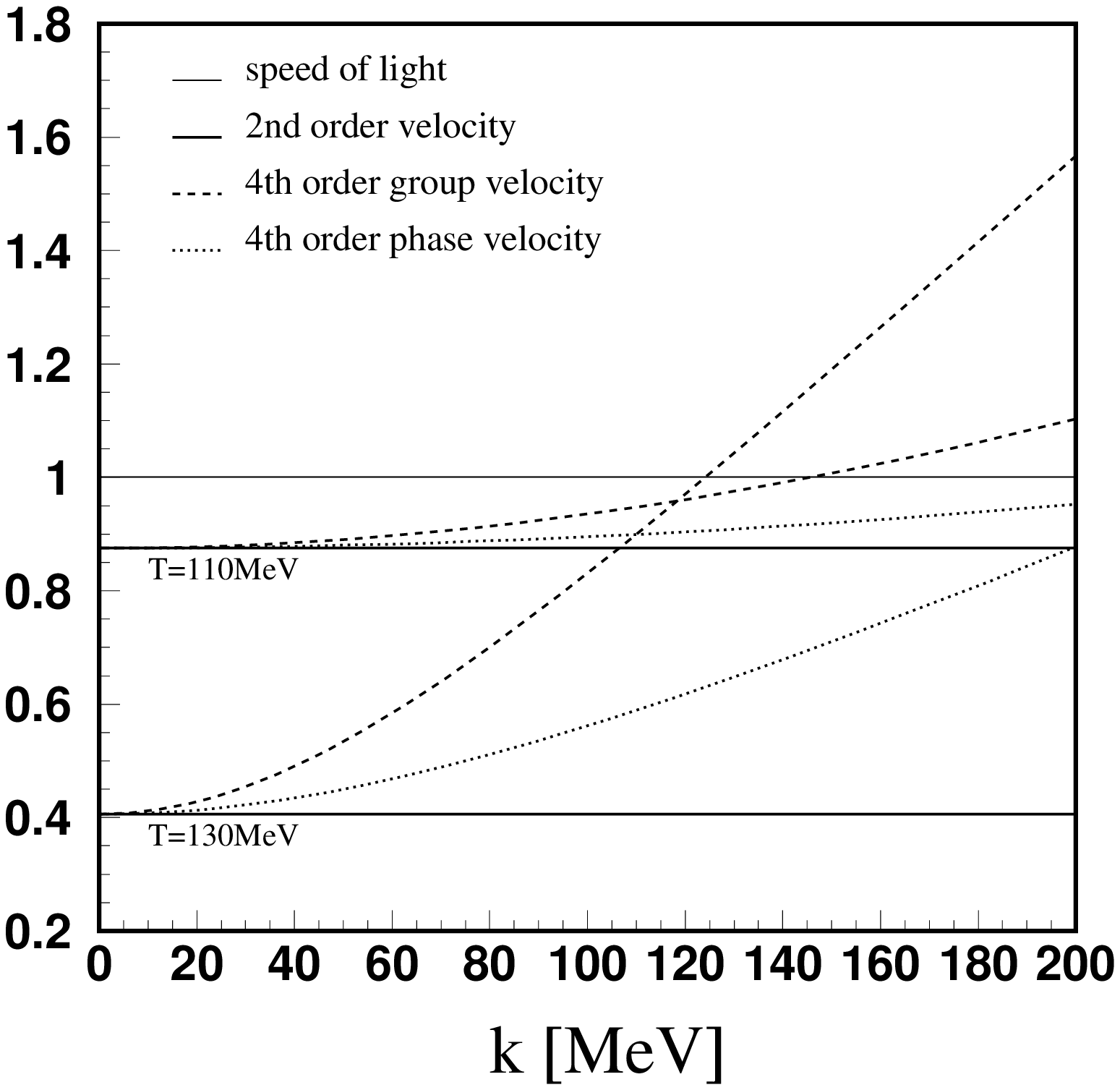}
}
\end{center}
\vspace*{5cm}
\caption{Group velocity and phase velocity of Goldstone-waves versus momentum 
for $T=110~$MeV (upper curves) and $T=130~$MeV (lower curves).}
\label{group.fig}
\end{figure} 
The 4th-order contribution to the velocity is opposite in sign 
(see (\ref{pi4thspeed})), so its inclusion results in a harder dispersion 
relation compared to the 2nd-order results, but still softer than the vacuum 
dispersion. With increasing momentum a monotonically increasing behavior of 
the 4th-order contribution with respect to the 2nd order one is found. The 
slope of these curves determines the group velocity. 

The behavior of the group velocity in terms of momentum for different 
temperatures is shown in figure \ref{group.fig}. One can see that 4th-order 
contributions introduce a momentum-dependent change in the velocity compared 
to the 2nd-order results. The deviation from the speed of light is greatest 
for true Goldstone bosons (modes with $k=0$). Looking at harder excitations, 
but still in the soft sector, one finds a decreasing deviation from the speed 
of light. Notice that at a certain momentum the group velocity becomes 
superluminous, setting an upper bound for the validity of our model. This 
limiting momentum, denote it by $\Lambda_c$, is shifted to smaller values for 
higher temperatures. The scale $\Lambda_c$ is controlled by the coefficient 
of the $k^4$ term in (\ref{pi4thspeed}). The overall validity of the 
self-energy expansion is limited to momenta as large as $125~$MeV. This 
$\Lambda_c$ can be thought of as the momentum scale defined in section 
\ref{sect-eom}, which separates hard modes from soft modes. One can claim, 
then, that only modes with $\mid\vec{k}\mid<\Lambda_c$ can be treated 
classically, as part of the condensate, and this is consistent with the 
original premise of our model.

In figure \ref{group.fig} the momentum dependence of the phase velocity is 
also plotted. This is related to the index of refraction of the medium, 
$n=c/v$, showing that the heat bath is less dispersive for Goldstone waves 
of larger $k$. 

\section{Massive Pions}

For the theory with explicitly broken chiral symmetry the dispersion relation,
 determined by Fourier transforming equation (\ref{pioneomm}), is given by
\be
\omega^2 = k^2 + \bar m_\pi^2 + \tilde m_\pi^2 + \mbox{Re}~\Pi_\pi(\omega, 
\vec{k})\,~.
\ee
Here $\bar m_\pi$ is the pion mass given by expression (\ref{pionmassh}) in 
terms of the $T=0$ pion mass, $m_\pi$,
\be
\bar m_\pi^2 = \frac{f_\pi}{v}m_\pi^2   \,~,
\ee
and $\tilde m_\pi$ is an apparent contribution which cancels out later. 
Knowledge of a more accurate pion dispersion relation is expected to lead for 
instance, to more accurate description of the dilepton spectrum in heavy ion 
collisions \cite{gale}. The method for determining an explicit relation 
between the energy $\omega$ and momentum $k$ is the same as in the previous 
section: Taylor expand the self-energy about small energy and momentum, 
retaining the nonzero leading (2nd) order and next to leading (4th) order 
terms, as in (\ref{selfour}). Because it is lengthy, the evaluation is 
omitted, and only the final expression for the dispersion relation is 
stated:   
\be
\omega = \sqrt{v_\pi^2(k)k^2 + \bar m_\pi^2} \,~,
\label{disp-m}
\ee
with the function
\be
v_\pi^2(k) = 1 &+& \frac{4\lambda}{3\pi^2}\frac{1}{(\bar m_\sigma^2-\bar 
m_\pi^2)^2}\left[3\bar m_\sigma^2I_\sigma + 4I_2 - 3\bar m_\pi^2I_3 - 4I_4
\right]\nonumber\\
&+&\frac{k^2}{\bar m_\sigma^2-\bar m_\pi^2}\frac{\lambda}{\pi^2}\frac{1}
{(\bar m_\sigma^2-\bar m_\pi^2)^2}\left[(11\bar m_\sigma^2+\bar m_\pi^2)I_1 + 
4\frac{11\bar m_\sigma^2-3\bar m_\pi^2}{\bar m_1^2-\bar m_\pi^2}I_2 \right.
\nonumber\\
&& \qquad\qquad\qquad\quad\left. + (11\bar m_\pi^2+\bar m_\sigma^2)I_3 +  
4\frac{3\bar m_\sigma^2-11\bar m_\pi^2}{\bar m_\sigma^2-\bar m_\pi^2}
I_4\right] \,~,
\ee
\newpage

\enlargethispage*{8in}
where 
\be
I_1 = \int_0^\infty\frac{p^2}{E_\sigma}f_\sigma, ~~I_2 = \int_0^\infty
\frac{p^4}{E_\sigma}f_\sigma, ~~I_3 = \int_0^\infty\frac{p^2}{E_\pi}f_\pi,
~~I_4 = \int_0^\infty\frac{p^4}{E_\pi}f_\pi \,~.
\ee
The energies, $E_\sigma=\sqrt{p^2+\bar m_\sigma^2}$ and $E_\pi=\sqrt{p^2+\bar 
m_\pi^2}$, are evaluated with temperature-dependent masses determined in 
section \ref{sect-masses-results}. One should be careful not to identify 
$v_\pi$ as a velocity. Due to the nontrivial nature of the dispersion 
relation, finding the group velocity requires the evaluation of $d\omega/dk$, 
or looking at the slope of curves.

\begin{figure}[htbp] 
\begin {center}
\vspace{-2cm}
\leavevmode
\hbox{%
\epsfysize=11.5cm
\epsffile{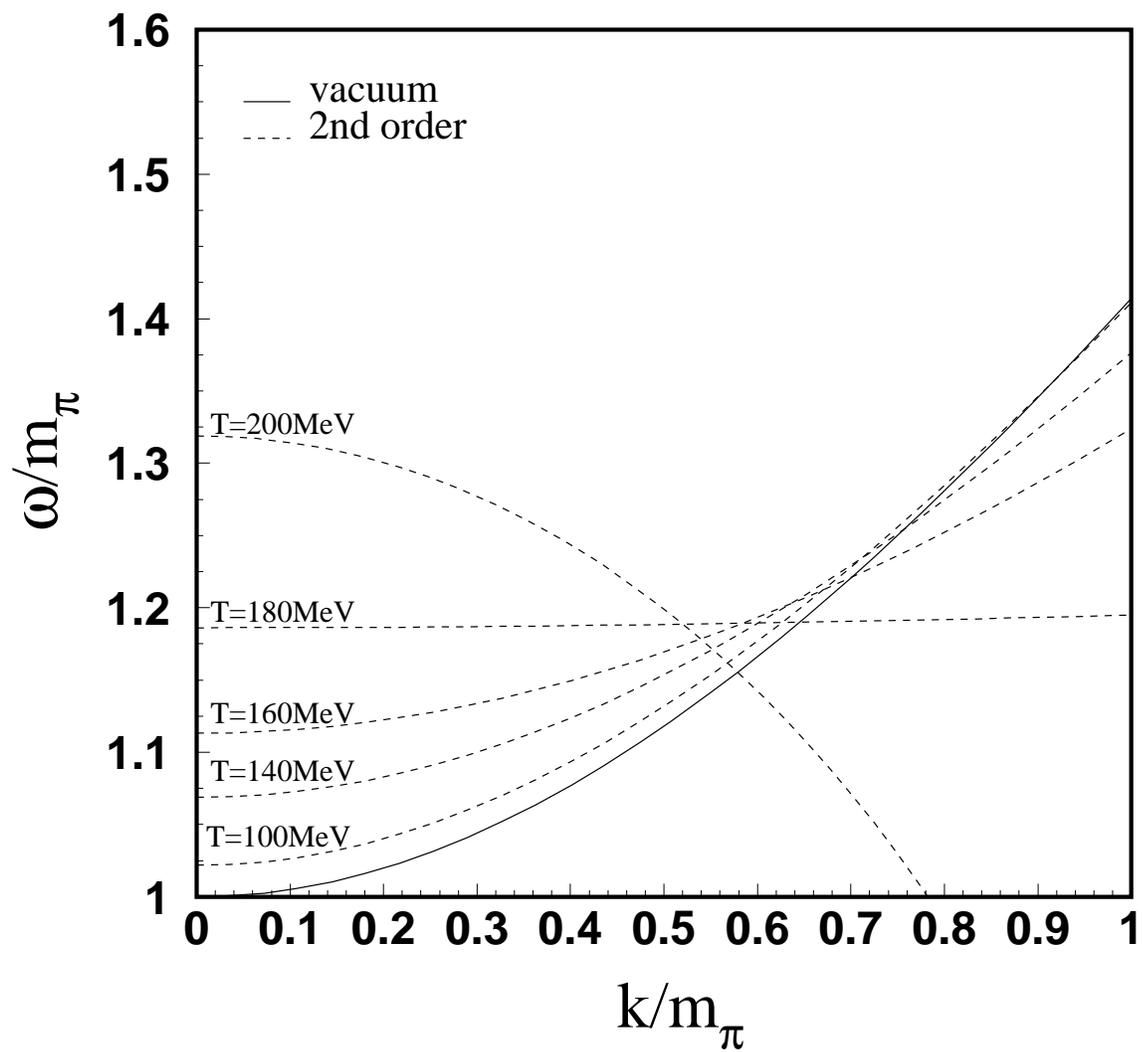}
}
\end{center}
\vspace*{5cm}
\caption{Dispersion curves for massive pions corrected to 2nd-order for 
different temperatures.}
\label{new2.fig}
\end{figure}

\begin{figure}[htbp] 
\begin {center}
\leavevmode
\hbox{%
\epsfysize=11.5cm
\epsffile{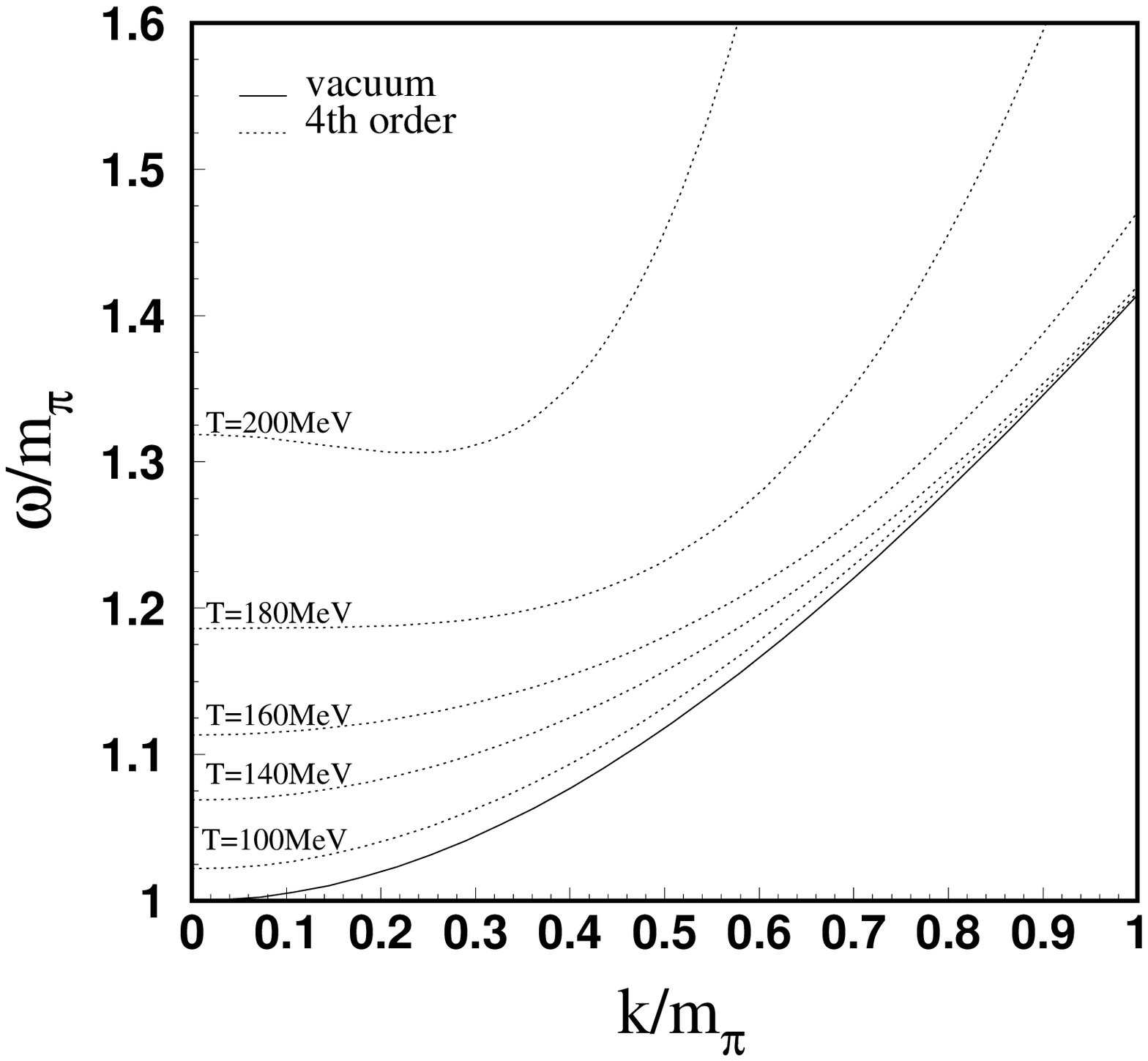}
}
\end{center}
\vspace*{5cm}
\caption{Dispersion curves for massive pions corrected to 4th-order for 
different temperatures.}
\label{new4.fig}
\end{figure}

\begin{figure}[htbp] 
\begin {center}
\hspace*{-1cm}
\leavevmode
\hbox{%
\epsfysize=11cm
\epsffile{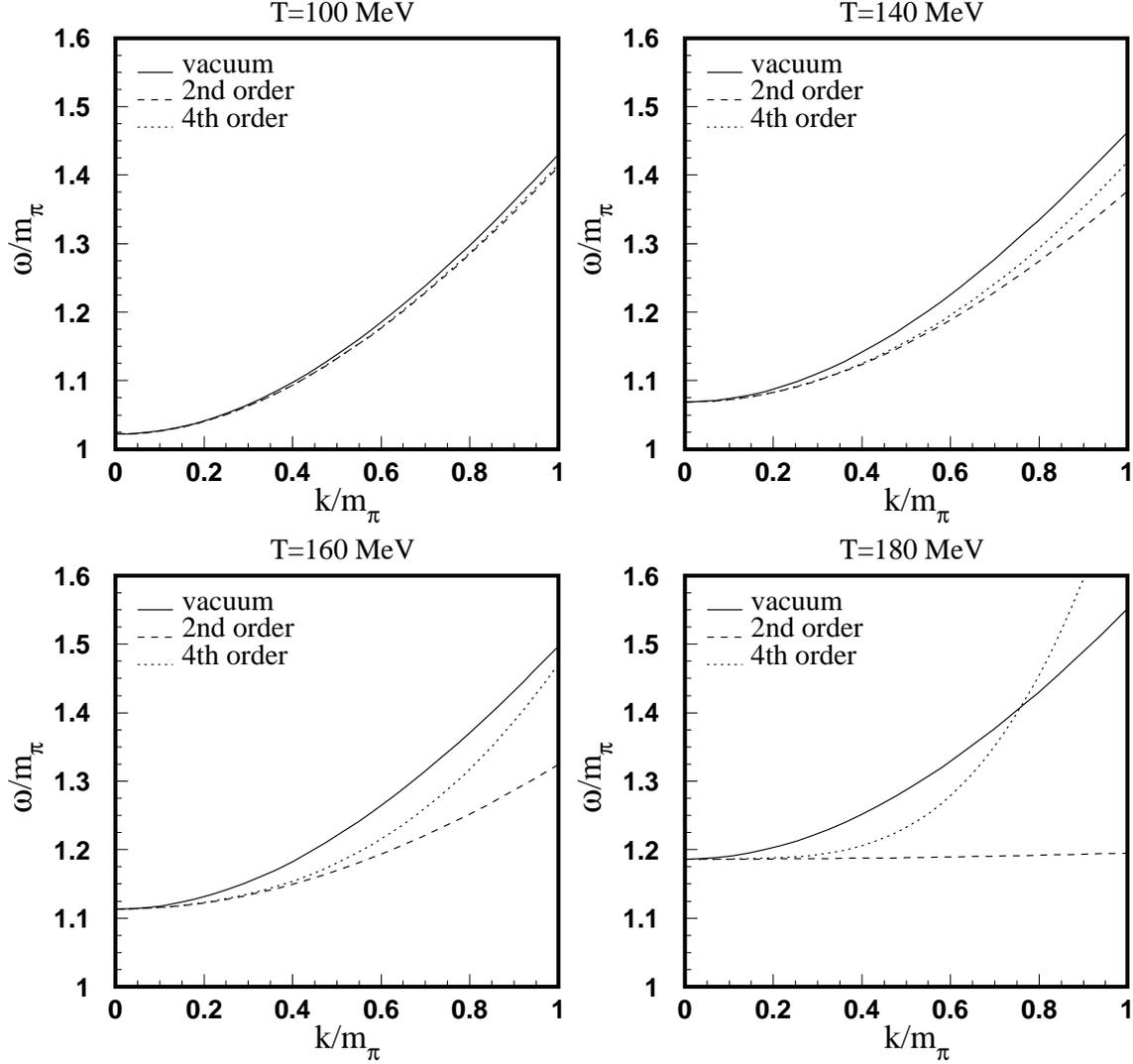}
}
\end{center}
\vspace*{5cm}
\caption{Pion dispersion curve $\omega^2=v_\pi^2(k)k^2+\bar m_\pi^2$ for 
different temperatures in the theory with explicitly broken chiral symmetry. 
Results with $v_\pi=1$ (solid), $v_\pi^2=a$ (dashed) and $v_\pi^2=a+bk^2$ 
(dotted) are shown.}
\label{hspeed4th.fig}
\end{figure}

The temperature dependence of the dispersion relation of massive pions can be 
obtained only by performing numerical computations. Figure \ref{new2.fig} 
shows dispersion curves corrected to 2nd-order in k for different 
temperatures. We found that with increasing temperature there is a f
lattening of the curves, which corresponds to decreasing group velocity.
At around $T=180~$MeV the velocity of all the excitations is vanishing. Going 
to even higher temperatures results in a negative slope and so this is a 
clearly unphysical region. Inclusion of higher order contributions changes 
the qualitative behavior, as shown in figure \ref{new4.fig}. 
At temperatures as high as 200 MeV there is not only negative slope at low 
momenta, but also a very steep rise at higher momenta, leading to speeds that 
exceed the speed of light. As before, at temperatures this high our results 
are unphysical. This temperature range is the same as the temperature range 
where the difference between the sigma and pion mass is diminishing, and so 
the pion starts becoming degenerate with the sigma (see figure 
\ref{step.fig}): symmetry is approximately restored. Our result could be 
signalling that at temperatures above $180~$MeV pions are not good degrees of 
freedom anymore. 
A comparison of results with 2nd- and 4th-order contributions are presented 
in figure \ref{hspeed4th.fig}. These are compared to the vacuum dispersion 
relation, too. Here vacuum dispersion means $v_\pi=1$, and so 
$\omega = \sqrt{k^2+\bar m_\pi^2}$. Deviation from the vacuum dispersion 
relation is significant above $T=100~$MeV. In the very soft sector there is 
no noticeable difference between the two corrected curves. Their behavior is 
both in qualitative and quantitative agreement with corresponding results 
obtained in a recent publication \cite{ayala}.
 The ground for comparison is limited, though, because Ayala {\it et al.}~studied the kinematic region with $k\leq 0.2m_\pi$ and temperatures of the order of the pion mass only. Significant difference between the 2nd- and 4th-order curves becomes noticeable in the kinematic region with $k\geq 0.3m_\pi$.
\newpage
%
\chapter{Dissipation at One-Loop Order\label{sect-dissip1}}

Dissipation of the condensate occurs because energy can be transferred between the condensate and the heat bath through the interactions of slow and fast modes. The dissipation can be determined by analyzing how interactions change the width of particles or collective modes, since this is related to their damping. In the present chapter the effect at one-loop order is calculated by evaluating the response functions. We show the direct connection to the physical processes responsible for dissipation. After a general discussion, damping of the long-wavelength sigma and pion fields is determined, and numerical results are presented.  

\section{General Formalism \label{sect-genform}}

The equations of motion for long-wavelength sigma and pion fields (eqs.~(\ref{sigmaeom})and (\ref{pioneom}) in the exact chiral limit, and eqs.~(\ref{sigmaeomm}) and (\ref{pioneomm}) for explicitly broken symmetry) take the form
\be
\partial^2\phi(x) + m^2 \phi(x) + F(x) = 0 \,~,
\label{geneom}
\ee
where $\phi=\sigma_s$ or $\pi_s$. Based on the response functions (see (\ref{1^2}), (\ref{2^2}) and (\ref{12})) one can write 
\be
F(x) = \int d^4x'\phi(x')\Pi(x-x')\,~.
\ee
Transformed into frequency-momentum space equation (\ref{geneom}) reads
\be
-k^2 + m^2 + \Pi(k) = 0\,~.
\label{keom}
\ee
The function $\Pi(k)$ is given by
\be
\Pi(k) &=&- ig^2\int d^4x'\int\frac{d^4p}{(2\pi)^4}\int\frac{d^4q}{(2\pi)^4}e^{i(k-p-q)(x-x')}\rho_1(p)\rho_2(q)(1+f(p^0)+f(q^0))\nonumber\\
&=& \!\!\!\!g^2\int\frac{d^3p}{(2\pi)^3} \frac{1}{4E_1E_2}\left[(1+f_1+f_2)\left(\frac{1}{\omega-E_1-E_2+i\epsilon}-\frac{1}{\omega+E_1+E_2 + i\epsilon}\right)\right.\nonumber\\
&& \quad \left. + (f_2-f_1)\left(\frac{1}{\omega-E_1+E_2+i\epsilon} - \frac{1}{\omega+E_1-E_2+i\epsilon}\right)\right]\,~,
\label{integral}
\ee
where $g$ is the coupling constant of cubic interactions, and the indices $1$ and $2$ refer to either of the hard modes, $\sigma_f$, and $\pi_f$, respectively. Also $E_1=\sqrt{m_1^2+(\vec{p}+\vec{k})^2}$ and $E_1=\sqrt{m_2^2+\vec{p}~^2}$, and $k=(k^0,\vec{k})$ is the 4-momentum of the soft sigma meson or pion. The frequency has a real and an imaginary part, $k^0=\omega-i\Gamma$, assuming that $\vec{k}$ is real. The last equality in (\ref{integral}) was obtained with the insertion of the free spectral functions (see discussion in section \ref{sect-pionmass}). Equation (\ref{integral}) coincides with (\ref{general}). This is a clear proof that the effect of the medium as described by $\Pi(k)$ can be identified with the self-energy. 

The real part of the self-energy participates in the dispersion relation 
\be
\omega^2 = \vec{k}~^2+m^2+\mbox{Re}~\Pi(\omega,\vec{k})\,~,
\ee
which has been analyzed in previous chapters. With the assumption of weak damping, $\Gamma\ll\omega$, the imaginary part of the self-energy completely determines the width of excitations: 
\be
\Gamma = -\frac{\mbox{Im}~\Pi(\omega,\vec{k})}{2\omega}\,~.
\label{Gamma/2}
\ee
$\Gamma$ is the rate at which mesonic modes with energy $\omega$ and momentum $\vec{k}$ approach equilibrium. This rate is determined by physical processes which can be identified from the imaginary part of the self-energy. 

There are several diagrams contributing to the self-energy at one-loop order. The tadpoles are real and their effect is only to modify the mass (see chapter \ref{sect-masses}). Dissipative effects come from non-local diagrams like the one shown in figure \ref{general.fig}. The imaginary part of this Feynman diagram is given by equation (\ref{imgen}). There are eight processes embedded in (\ref{imgen}). For details the reader is referred to \cite{weldon}. 
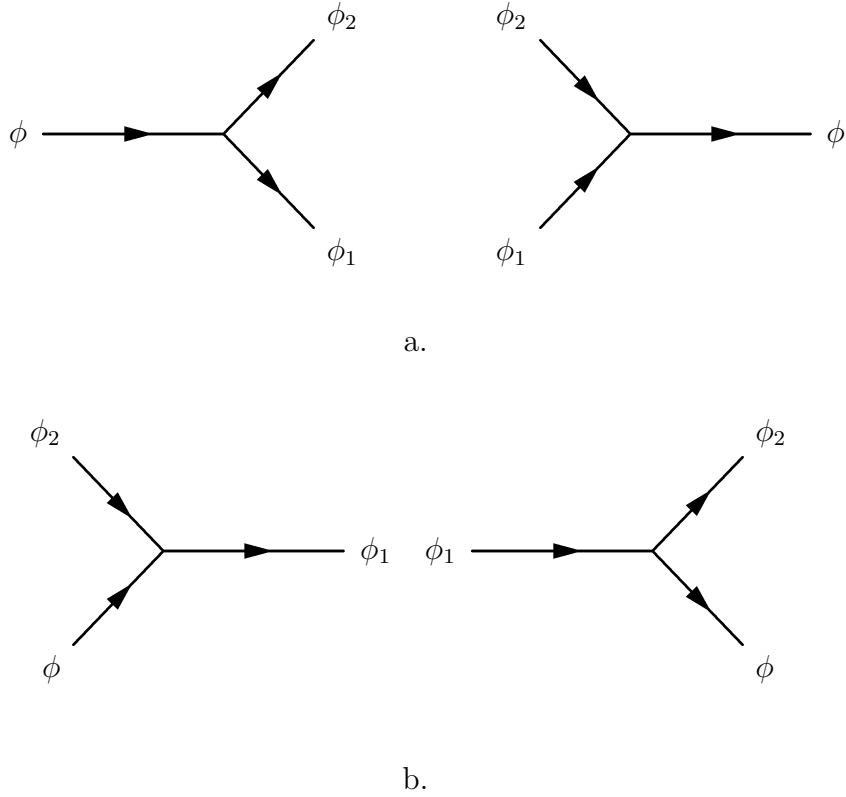
\begin{figure}[htbp]
\bc
\setlength{\unitlength}{1mm}
\be
\parbox{30mm}{
\begin{fmfgraph*}(40,25)
  \fmfleft{i}
  \fmfright{o1,o2}
  \fmf{plain_arrow}{i,v1}
  \fmf{plain_arrow}{v1,o1}
  \fmf{plain_arrow}{v1,o2}
\fmflabel{$\phi$}{i}
\fmflabel{$\phi_1$}{o1}
\fmflabel{$\phi_2$}{o2}
\end{fmfgraph*}}
\hspace*{2.2cm} 
\hspace*{1cm}
\parbox{30mm}{
\begin{fmfgraph*}(40,25)
  \fmfleft{i1,i2}
  \fmfright{o}
  \fmf{plain_arrow}{i1,v1}
  \fmf{plain_arrow}{i2,v1}
  \fmf{plain_arrow}{v1,o}
\fmflabel{$\phi$}{o}
\fmflabel{$\phi_1$}{i1}
\fmflabel{$\phi_2$}{i2}
\end{fmfgraph*}
}\nonumber
\ee
\ec
\vspace*{.5cm}
\bc
a.
\ec
\vspace*{.5cm}
\be
\parbox{35mm}{
\begin{fmfgraph*}(40,25)
  \fmfleft{i1,i2}
  \fmfright{o}
  \fmf{plain_arrow}{i1,v1}
  \fmf{plain_arrow}{i2,v1}
  \fmf{plain_arrow}{v1,o}
\fmflabel{$\phi_1$}{o}
\fmflabel{$\phi$}{i1}
\fmflabel{$\phi_2$}{i2}
\end{fmfgraph*}}
\hspace*{2.2cm} 
\parbox{35mm}{
\begin{fmfgraph*}(40,25)  
  \fmfleft{i}
  \fmfright{o1,o2}
  \fmf{plain_arrow}{i,v1}
  \fmf{plain_arrow}{v1,o1}
  \fmf{plain_arrow}{v1,o2}
\fmflabel{$\phi_1$}{i}
\fmflabel{$\phi$}{o1}
\fmflabel{$\phi_2$}{o2}
\end{fmfgraph*}
}\nonumber
\ee
\vspace*{.5cm}
\bc
b.
\ec

\bc
\parbox{11cm}{\caption{\small Reactions with $\omega\geq 0$ and $m_1\geq m_2$. (a) requires $\omega^2-k^2\geq (m_1+m_2)^2$, and (b) possible for $m_2^2-m_1^2\leq \omega^2-k^2\leq (m_1-m_2)^2$.}
\label{decay.fig}}
\ec
\end{figure} 

Figure \ref{decay.fig} shows the two pairs of diagrams that are possible for positive energies, $\omega \geq 0$. In the following we focus on these processes, since such processes are responsible for the appearance and disappearance of pions and sigma mesons from the condensate. For $\omega \geq 0$ the contributions to the imaginary part of the self-energy are limited to 
\be
&&\!\!\!\!\!\!\!\!\!\!\!\mbox{Im}~\Pi(\omega,\vec{k}) = -\pi g^2\int\frac{d^3p}{(2\pi)^3}\frac{1}{4E_1E_2}\left[(1+f_1+f_2)\delta(\omega-E_1-E_2) \right.\nonumber\\
&&\left. \qquad\qquad \qquad\qquad\qquad\qquad\qquad\qquad+(f_2-f_1)\delta(\omega-E_1+E_2) \right]\,\, .
\label{relevantgen}
\ee
For fixed masses and given values of the energy and momentum only one of the $\delta$-function constraints can be satisfied. For the sake of definiteness, assume $m_1\geq m_2$. The process in fig.~\ref{decay.fig}(a) with $\omega=E_1+E_2$ is possible for $\omega^2-k^2\geq (m_1+m_2)^2$, and the pair in fig.~\ref{decay.fig}(b) with $\omega+E_2=E_1$ is possible for $m_2^2-m_1^2\leq \omega^2-k^2 \leq (m_1-m_2)^2$. Rewriting $1+f_1+f_2=(1+f_1)(1+f_2)-f_1f_2$ allows for identification of the physical processes directly. These include the decay $\phi\rightarrow\phi_1\phi_2$ with statistical weight $(1+f_1)(1+f_2)$ minus its inverse, which is the fusion $\phi_1\phi_2\rightarrow\phi$, with statistical weight $f_1f_2$. Note that fusion is an exclusively finite temperature process. Similarly, write $f_2-f_1=f_2(1+f_1)-f_1(1+f_2)$, and identify the corresponding reactions as creation $\phi\phi_2\rightarrow\phi_1$ and decay $\phi_1\rightarrow\phi\phi_2$. 

In the following we will find explicit expressions for the imaginary part of the self-energy and discuss the two contributions separately.

\subsection{Processes Above the Two-Particle Threshold}

In the kinematic region where 
\be
\omega^2-\vec{k}~^2\geq (m_1+m_2)^2
\label{thres-cond}
\ee
the possible reactions are decay and fusion as displayed in fig.~\ref{decay.fig}(a) and described by 
\be
\mbox{Im}~\Pi(\omega,\vec{k}) = -\pi g^2\int\frac{d^3p}{(2\pi)^3}\frac{1}{4E_1E_2}(1+f_1+f_2)\delta(\omega-E_1-E_2)\,\, ,
\label{imthres}
\ee
where $E_1 = \sqrt{m_1^2 +({\vec k}+{\vec p}~)^2}$ and $E_2 = \sqrt{m_2^2 + \vec{p}~^2}$. Evaluating (\ref{imthres}) is straightforward but tedious. To simplify our task we only consider two cases that are of particular interest for our model.

\subsubsection{Case I: $\bf m_1=m_2=m$.}

When evaluating $\mbox{Im}~\Pi(\omega,\vec{k})$ we write the integration measure in the usual way
\be
\int d^3p=2\pi\int_0^\infty p^2dp\int_{-1}^1d\cos\theta\,\, ,
\label{measure}
\ee
and perform the integral over angles first. This requires a change of variables in the delta function. Based on the formula
\be
\delta(f(x)) = \sum_i\frac{\delta(x-x_i)}{\mid\frac{df}{dx}\mid_{x=x_i}}, ~~~~f(x_i)=0\,\, ,
\ee
one can write
\be
\delta(\omega-E_1-E_2) &=& \delta\left(\omega - \sqrt{m^2+({\vec k}+{\vec p}})^2 - \sqrt{m^2+\vec{p}^2}\right)\nonumber\\
&=&  \delta\left(\omega -  \sqrt{m^2+k^2+p^2+2pk\cos{\theta}} - \sqrt{m^2+p^2}\right)\nonumber\\
&=& \frac{E_1}{kp}\delta(\cos{\theta}-\cos{\theta}_0)
\ee
where
\be
\cos{\theta}_0= \frac{\omega^2 - k^2 - 2\omega\sqrt{m^2+p^2}}{2pk}\,~.
\ee
Here $k$ and $p$ are the magnitudes of the 3-momenta. There is a lower limit, $\Lambda_c$, for the hard loop momentum integration. However, the inequality $-1\leq \cos{\theta}_0\leq 1$ must hold, and this sets natural restrictions on the limits of integration:
\be
\mbox{Im}~\Pi(\omega,\vec{k}) &=&  -\frac{g^2}{16\pi k}\int_{p_{-}}^{p_{+}} dp \frac{p}{E_2}\left[1 + f(\omega-E_2) + f(E_2)\right]
\label{mom}
\ee
with 
\be
p_{+}= \frac{k}{2}+\frac{\omega}{2}\sqrt{1-\frac{4m^2}{\omega^2-k^2}} ~~\mbox{and} ~~~
p_{-}= \left|~\frac{k}{2}-\frac{\omega}{2}\sqrt{1-\frac{4m^2}{\omega^2-k^2}}~\right|\,\, .\nonumber
\ee
After a change of variables: 
\be
\mbox{Im}~\Pi(\omega,\vec{k}) =  -\frac{g^2}{16\pi k}\int_{\omega_{-}}^{\omega_{+}} dE \left[1 + 2 f(E)\right]
\label{imgenthres}
\ee
where
\be
\omega_{\pm}=\sqrt{p_{\pm}^2+m^2} = \frac{\omega}{2}\pm \frac{k}{2}
\sqrt{1-\frac{4m^2}{\omega^2-k^2}}\,\, .
\ee
Provided that $\omega^2-k^2\geq 4m^2$, where $m$ is the mass of the particle in the loop, one gets 
\be
\mbox{Im}~\Pi(\omega,\vec{k}) = -\frac{g^2}{16\pi}\left[\sqrt{1-\frac{4m^2}{\omega^2-k^2}} + 2\frac{T}{k}\log\left(\frac{1-e^{-\frac{\omega_{+}}{T}}}{1-e^{-\frac{\omega_{-}}{T}}}\right)\right]\,\, .
\label{threshold}
\ee
Notice that the $T=0$ and $T\neq 0$ contributions are clearly separated. The vacuum contribution is
\be
\mbox{Im}~\Pi(\omega,\vec{k},T=0) = -\frac{g^2}{16\pi}\sqrt{1-\frac{4m^2}{\omega^2-k^2}}\,\, .
\label{zerodecay}
\ee
The decay of a heavy boson into two lighter bosons (left diagram in figure \ref{decay.fig}(a)) is possible even at zero temperature. This is textbook material. 

The decay and formation processes are obtained by setting $\omega^2-k^2=M^2$ in (\ref{threshold}). Here $M$ denotes the mass of the external particle.
\be
\mbox{Im}~\Pi(M,k) =  -\frac{g^2}{16\pi}\left[\sqrt{1-\frac{4m^2}{M^2}} + 2\frac{T}{k}\log\left(\frac{1-\exp{(-\frac{\omega}{2T}-\frac{k}{2T}\sqrt{1-\frac{4m^2}{M^2}})}}{1-\exp{(-\frac{\omega}{2T}+\frac{k}{2T}\sqrt{1-\frac{4m^2}{M^2}})}}\right)\right]\nonumber\,\, .
\ee
It should be emphasized that this expression is different than zero only when the threshold condition $M^2\geq 4m^2$ is satisfied; that is, the two-particle production threshold is below the pole of the propagator of the boson. Recently, contributions to this process from multi-loop self-energy diagrams have been resummed \cite{kapwong}. However, inclusion of these is beyond the scope of the present study. 

Since our interest is in the dissipation of soft excitations from the condensate, we evaluate (\ref{imgenthres}) for small momenta, $k\ll T$
\be
\mbox{Im}~\Pi(\omega,\vec{k}) &=& -\frac{g^2}{16\pi k}\left[(\omega_+-\omega_-)(1+2f(\omega_-))\right]\nonumber\\
&=& -\frac{g^2}{16\pi}\sqrt{1-\frac{4m^2}{\omega^2-k^2}}(1+2f(\omega_-))\nonumber\\
&\simeq& \mbox{Im}\Pi(\omega,\vec{k},T=0)\left[1+2f\left(\frac{\omega}{2T}\right)\right]\,\, .
\label{smallkthres}
\ee
In case the external boson is much heavier than the bosons inside the loop, these can be thought of as massless, $m\simeq0$, and then (\ref{threshold}) reduces to
\be
\mbox{Im}~\Pi(\omega,\vec{k}) =  -\frac{g^2}{16\pi}\left[1+2\frac{T}{k}\log{\left(\frac{1-e^{-\frac{\omega_{+}}{T}}}{1-e^{-\frac{\omega_{-}}{T}}}\right)}\right]\,\, ,
\ee
where
\be
\omega_\pm = \frac{1}{2}(\omega\pm k)\,\, .
\ee
For $k\ll T$ this reads  
\be
\mbox{Im}\Pi(\omega,\vec{k})\simeq -\frac{g^2}{16\pi}\left[1+2\frac{T}{k}\log{\frac{\omega+k}{\omega-k}}
\right]\,\, .
\ee
%

\subsubsection{Case II: $\bf m_1=m$ and $\bf m_2=0$}  

In the scenario with one massless and one massive boson inside the loop (see figure \ref{general.fig}), the evaluation of (\ref{imthres}) follows the same train of analysis as above, resulting in 
\be
\mbox{Im}~\Pi(\omega,\vec{k}) = -\frac{g^2}{16\pi k}\int_{p_1}^{p_2} dp \left[1+f(p)+f(\omega-p)\right] \,\, ,
\ee
with 
\be
p_1 = \frac{\omega^2-k^2-m^2}{2(\omega+k)}~~\mbox{and}~~~p_2 = \frac{\omega^2-k^2-m^2}{2(\omega-k)}\,\, .
\ee
The threshold condition that must be satisfied is $\omega^2-k^2\geq m^2$. The final expression is given by
\be
\mbox{Im}~\Pi(\omega,\vec{k}) &=& -\frac{g^2}{16\pi}\left[1-\frac{m^2}{\omega^2-k^2} + \frac{T}{k}\log{\frac{\left(1-e^{-\frac{p_2}{T}}\right)\left(1-e^{\frac{p_1-\omega}{T}}\right)}{\left(1-e^{-\frac{p_1}{T}}\right)\left(1-e^{\frac{p_2-\omega}{T}}\right)}}\right]\,~.\nonumber\\
\ee
Once again, a non-vanishing zero temperature contribution can be identified, 
\be
\mbox{Im}~\Pi(\omega,\vec{k},T=0) =  -\frac{g^2}{16\pi}\left[1-\frac{m^2}{\omega^2-k^2}\right]\,\, .
\ee
Note that the decay of an on-shell massless boson is kinematically forbidden. 

\subsection{Landau damping}

Landau damping is a finite temperature mechanism in which particles are created and/or destroyed through scattering with particles from the heat bath. Such processes are displayed in figure \ref{decay.fig}.b and are described mathematically by 
\be
\mbox{Im}~\Pi(\omega,\vec{k}) = -\pi g^2\int\frac{d^3p}{(2\pi)^3}\frac{1}{4E_1E_2}(f_2-f_1)\delta(\omega-E_1+E_2)\,\, ,
\label{imland}
\ee
provided the condition 
\be
\omega^2-k^2\leq (m_1-m_2)^2
\label{land-cond}
\ee
is satisfied. The integrals in (\ref{imland}) can be completed resulting in 
\be
\mbox{Im}~\Pi(\omega,\vec{k}) &=& -\frac{g^2}{16\pi k}\int_{p_-}^{p_+} dp\frac{p}{E_2}(f(E_2)-f(E_1))\nonumber\\
&=&  -\frac{g^2}{16\pi k}\int_{\omega_-}^{\omega_+} dE(f(E)-f(E+\omega))\,\, .
\ee
The final expression is 
\be
\mbox{Im}~\Pi(\omega,\vec{k}) =  -\frac{g^2}{16\pi}\frac{T}{k}\left[\log{\left(\frac{e^{\frac{\omega_+}{T}}-1}{e^{\frac{\omega_-}{T}}-1}\right)} + \log{\left(\frac{e^{\frac{\omega_+ +\omega}{T}}-1}{e^{\frac{\omega_- +\omega}{T}}-1}\right)}\right]\, ,
\label{landaufull}
\ee
where
\be
\omega_\pm &=& \sqrt{p_\pm^2+m_2^2}\, ,\nonumber\\
p_\pm &=& \pm\frac{k}{2}\frac{k^2-\omega^2+m_1^2-m_2^2}{\omega^2-k^2} + \frac{\omega}{2}\frac{\sqrt{(k^2-\omega^2+m_1^2-m_2^2)^2-4m_2^2(\omega^2-k^2)}}{\omega^2-k^2}\,\, .\nonumber\\
\ee
In the following we discuss the two special cases of interest. 

\subsubsection{Case I: $\bf m_1=m_2=m$}

This case requires the 4-momentum of the external particle to be space-like, that is $\omega^2-k^2\leq 0$. Expression (\ref{landaufull}) is simplified to
\be
\mbox{Im}~\Pi(\omega,\vec{k}) =  -\frac{g^2}{8\pi}\frac{T}{k}\log{\left(\frac{1-e^{-\frac{\omega_+}{T}}}{1-e^{-\frac{\omega_-}{T}}}\right)}
\label{landau1}
\ee
where
\be
\omega_\pm = \frac{\omega}{2}\pm\frac{k}{2}\sqrt{1+\frac{4m^2}{k^2-\omega^2}}\,\, .
\ee
Expression (\ref{landau1}) vanishes at $T=0$, just as expected. This process is forbidden on mass-shell because it would yield tachyons.

\subsubsection{Case II: $\bf m_1=m$ and $\bf m_2=0$}  

In this case one of the particles in the loop is massless and the other has a kinematic constraint, $\omega^2-k^2\leq m^2$. 
\be
\mbox{Im}~\Pi(\omega,\vec{k}) &=& -\frac{g^2}{16\pi k}\int_{p_l}^\infty dp\left[f(p)-f(\omega+p)\right]\nonumber\\
&=& -\frac{g^2}{16\pi k}\int_{p_l}^{p_l+\omega}dp f(p)\nonumber\\
&=& -\frac{g^2}{16\pi k}T\log\left({\frac{e^{\frac{p_l}{T}}-e^{-\frac{\omega}{T}}}{e^{\frac{p_l}{T}}-1}}\right)\,\, ,
\ee
where
\be
p_l =\frac{k^2-\omega^2+m^2}{2(\omega+k)} \rightarrow \frac{m^2}{4k}~~~\mbox{as}~~~~\omega \rightarrow k\,\, .
\ee
For the self-energy of massless on-shell bosons the result is
\be
\mbox{Im}~\Pi(k) =  -\frac{g^2}{16\pi k}\int_{\frac{m^2}{4k}}^{\frac{m^2}{4k}+k} dp f(p) \,\, .
\ee
When the massless on-shell bosons are soft, too, then
\be
\mbox{Im}~\Pi(k) \simeq -\frac{g^2}{16\pi k}f\left(\frac{m^2}{4k}\right)\,\, .
\label{land}
\ee
%

\section{Sigma Decay}

\begin{figure}[htbp]
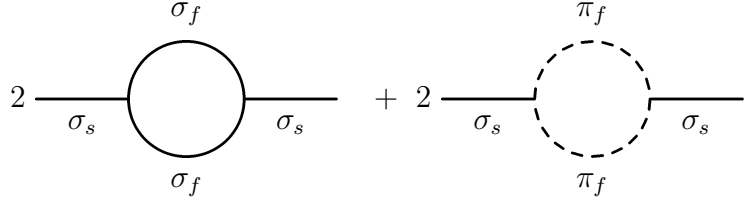

\bc
\be
2~\parbox{4cm}{\begin{fmfchar*}(40,25)
  \fmfleft{i}
  \fmfright{o}
  \fmf{plain,label=$\sigma_s$}{i,v1}
  \fmf{plain,label=$\sigma_s$}{v2,o}
  \fmf{plain,left,tension=.4,label=$\sigma_f$}{v1,v2,v1}  
\end{fmfchar*}}
\hspace*{.4cm}
+ ~2
~\parbox{4cm}{\begin{fmfchar*}(40,25)
  \fmfleft{i}
  \fmfright{o}
  \fmf{plain,label=$\sigma_s$}{i,v1}
  \fmf{plain,label=$\sigma_s$}{v2,o}
  \fmf{dashes,left,tension=.4,label=$\pi_f$}{v1,v2,v1}  
\end{fmfchar*}}\nonumber
\ee

\parbox{12cm}{\caption{\small One-loop self-energy contribution to the long-wavelength sigma $\sigma_s$ with couplings to the hard $\sigma_f$ and $\pi_f$ through $3\lambda v\sigma_s\sigma_f^2$ and $\lambda v\sigma_s\pi_f^2$.
\label{si.fig}}}
\ec
\end{figure}

Contributions to the imaginary part of the sigma meson self-energy come from two different couplings. These were derived in section \ref{sect-coupling}, and are $3\lambda v\sigma_s\sigma_f^2$ and $\lambda v\sigma_s\pi_f^2$. The corresponding coupling constants are $g=3\lambda v$ and $\lambda v$, respectively. The corresponding diagrams are presented in figure \ref{si.fig}. The factor 2 in front of the diagrams is a symmetry factor due to the interchangeability of the lines inside the loop. 
\noindent\\
Let us look at the on-shell processes that are allowed since these determine the dynamics of the decay. Figure \ref{si.fig} shows that $m_1=m_2=m$, where $m=m_\sigma$ or $m=m_\pi$. For an on-shell sigma meson, $\omega^2 = \vec{k}~^2 + m_\sigma^2$, condition (\ref{land-cond}) is never satisfied. Moreover, for $m=m_\sigma$ condition (\ref{thres-cond}) is not fulfilled either. One can say that the kinematics forbids Landau-damping and $\sigma_s\rightarrow\sigma_f\sigma_f$ decay. However, contribution from the decay of soft sigma mesons into hard thermal pions is non-zero above the threshold $m_\sigma^2\geq 4m_\pi^2$. The rate for $\sigma_s\rightarrow\pi_f\pi_f$ is given by equation (\ref{direct})
\be
\Gamma_d(\omega, {\vec k}) = \frac{1}{2\omega}\int\frac{d^3p_1}{(2\pi)^32E_1}\int\frac{d^3p_2}{(2\pi)^32E_2} |{\cal{M}}|^2 (2\pi)^4\delta^4(k-p_1-p_2)(1+f_1)(1+f_2)\nonumber\,~,
\ee
where $E_1 = \sqrt{m_\pi^2+\vec{p}~^2} = p$ and $E_2 = \sqrt{m_\pi^2+(\vec{p}-\vec{k})^2}$. To obtain the total rate one has to take into account the reverse process, the annihilation of two pions from the heat bath into a low momentum sigma, $\pi_f\pi_f\rightarrow\sigma_s$, which has a rate (\ref{inverse})
\be
\Gamma_i(\omega, {\vec k}) = \frac{1}{2\omega}\int\frac{d^3p_1}{(2\pi)^32E_1}\int\frac{d^3p_2}{(2\pi)^32E_2} |{\cal{M}}|^2 (2\pi)^4\delta^4(k-p_1-p_2)f_1 f_2\nonumber\,~.
\ee
The total rate is then 
\be
\Gamma(\omega, {\vec k}) &=& \Gamma_d(\omega, {\vec k})- \Gamma_i(\omega, {\vec k})\nonumber\\
&=& \frac{1}{2\omega}\int\frac{d^3p}{(2\pi)^3}\frac{1}{4E_1 E_2} |{\cal{M}}|^2(2\pi)\delta(\omega-E_1-E_2)(1+f_1+f_2) .
\ee
The scattering amplitude for this process is basically the coupling constant associated with the vertex, $|{\cal{M}}|^2 = g^2 = (\lambda v)^2$. Performing a simple variable change $\vec{p}-\vec{k}=\vec{q}$ in (\ref{imthres}), the identity
\be
\Gamma(\omega, {\vec k}) = -\frac{\mbox{Im}~\Pi(\omega, {\vec k})}{\omega}
\label{Gamma}
\ee
becomes transparent. It is important to be aware of the difference of a factor of 2 between expressions (\ref{Gamma/2}) and (\ref{Gamma}). The first one describes the rate of decay of the amplitude of the wave, $\exp(-\Gamma t/2)$, while the second one represents the loss rate for the number density, $\exp(-\Gamma t)$.

The imaginary part of the self-energy of a soft, $k\ll T$, on-shell sigma meson is given by eq.~(\ref{smallkthres}) as
\be
\mbox{Im}~\Pi_\sigma(\omega) = \mbox{Im}~\Pi_\sigma(\omega,T=0)\left[1+2f\left(\frac{\omega}{2T}\right)\right]\,\, ,
\ee
where the non-zero vacuum contribution is, from (\ref{zerodecay}),
\be
\mbox{Im}~\Pi_\sigma(\omega,T=0) = -\frac{\lambda^2v^2}{8\pi}\sqrt{1-\frac{4m_\pi^2}{m_\sigma^2}}\,\, .
\ee
The rate at which sigmas of energy $\omega$ disappear from the condensate due to their decay into pions is
\be
\Gamma_{\sigma\pi\pi}(\omega) =  \frac{(N-1)}{16\pi}\lambda\frac{m_\sigma^2-m_\pi^2}{\omega}\sqrt{1-\frac{4m_\pi^2}{m_\sigma^2}}\coth{\left(\frac{\omega}{4T}\right)}\,\, ,
\label{sipipi}
\ee
where account is taken for the relation $m_\sigma^2-m_\pi^2=2\lambda v^2$ and that there are $N-1$ pion fields. In the rest-frame of the sigma $\omega = m_\sigma$. When the pions are Goldstone bosons expression (\ref{sipipi}) is reduced to 
\be
\Gamma_{\sigma\pi\pi}(m_\sigma) =  \frac{(N-1)}{16\pi}\lambda m_\sigma\coth{\left(\frac{m_\sigma}{4T}\right)}\,\, .
\label{sipipi0}
\ee
This agrees with equation (23) in \cite{cejk}. At the critical temperature $m_\sigma=0$ and just below this $m_\sigma\ll T$. Thus one can perform a high temperature expansion using 
$$
\coth(x)= \frac{1}{x}+\frac{x}{3}-\frac{x^3}{45}+{\cal{O}}(x^4) ~~~\mbox{as}~~~x\rightarrow 0\,\, .
$$
When keeping the leading term only, eq.~(\ref{sipipi0}) results in the classical limit
\be
\Gamma_{\sigma\pi\pi} \simeq \frac{(N-1)}{16\pi}\lambda T\,\, .
\ee
At low temperatures the sigma is heavy, $T\ll m_\sigma$, and no such expansion is reasonable. In this case 
$$
\coth(x)\rightarrow 1 ~~~\mbox{as}~~~x\rightarrow\infty
$$
and
\be
\Gamma_{\sigma\pi\pi} \simeq  (N-1)\frac{\lambda m_\sigma}{16\pi}\,\, ,
\ee
which is equal to the decay rate calculated at zero temperature. 

\section{Pion Damping}

\begin{figure}[htbp]
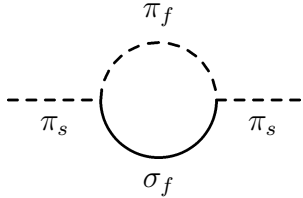

\bc
\begin{fmfchar*}(40,25)
  \fmfleft{i}
  \fmfright{o}
  \fmf{dashes,label=$\pi_s$}{i,v1}
  \fmf{dashes,label=$\pi_s$}{v2,o}
  \fmf{dashes,left,tension=.4,label=$\pi_f$}{v1,v2}  
  \fmf{plain,left,tension=.4,label=$\sigma_f$}{v2,v1}
\end{fmfchar*}

\parbox{9cm}{\caption{\small One-loop self-energy contribution to $\pi_s$ with coupling to $\sigma_s$ and $\pi_s$ through $2\lambda v\pi_s\sigma_f\pi_f$.
\label{pi.fig}}}
\ec
\end{figure}

At one-loop order there is only one diagram contributing to the pion self-energy, the origin of which lies in the $2\lambda v\pi_s\sigma_f\pi_f$ coupling with $g=2\lambda v$. This diagram is shown in figure \ref{pi.fig}.
\noindent\\
At this level, dissipation of the pion condensate occurs provided that the energy and momentum of the soft pion satisfies the condition $\omega^2-k^2\leq (m_\sigma-m_\pi)^2$. Then the transformation of a pion into a sigma when propagating through a thermal medium can happen. Basically, a soft pion from the condensate annihilates with a hard thermal pion producing a hard thermal sigma meson, $\pi_s\pi_f\rightarrow\sigma_f$. The inverse process is the decay of a hard thermal sigma meson into a soft and a hard pion, $\sigma_f\rightarrow\pi_s\pi_f$. The net rate of dissipation is given by eqs.~(\ref{direct}) and (\ref{inverse}) as
\be
\Gamma(\omega, {\vec k}) &=& \Gamma_d(\omega, {\vec k})- \Gamma_i(\omega, {\vec k})\nonumber\\
&=& \frac{1}{2\omega}\int\frac{d^3p_\pi}{(2\pi)^3}\frac{1}{4E_\sigma E_\pi} |{\cal{M}}|^2(2\pi)\delta(\omega+E_\pi-E_\sigma)(f_\pi-f_\sigma) \,~,
\label{gt}
\ee
where $E_\sigma=\sqrt{m_\sigma^2+(\vec{k}+\vec{p})~^2}$, $E_\pi = p$, and the amplitude is $|{\cal{M}}|^2 = g^2 = (2\lambda v)^2$. Note that with (\ref{gt}) and (\ref{imland})
\be
\Gamma(\omega, {\vec k}) = -\frac{\mbox{Im}~\Pi(\omega, {\vec k})}{\omega}\,~.
\ee
The rate of dissipation of massive pions is 
\be
\Gamma_{\pi\pi\sigma}(\omega,\vec{k}) =  -\frac{1}{8\pi}\lambda\frac{T(m_\sigma^2-m_\pi^2)}{k\omega}\left[\log{\left(\frac{e^{\frac{\omega_+}{T}}-1}{e^{\frac{\omega_-}{T}}-1}\right)} + \log{\left(\frac{e^{\frac{\omega_+ +\omega}{T}}-1}{e^{\frac{\omega_- +\omega}{T}}-1}\right)}\right]
\label{pipisi}
\ee
where
\be
\omega_\pm &=& \sqrt{p_\pm^2+m_\pi^2}\, ,\nonumber\\
p_\pm &=& \pm\frac{k}{2}\frac{k^2-\omega^2+m_\sigma^2-m_\pi^2}{\omega^2-k^2} + \frac{\omega}{2}\frac{\sqrt{(k^2-\omega^2+m_\sigma^2-m_\pi^2)^2-4m_\pi^2(\omega^2-k^2)}}{\omega^2-k^2}\,\, .\nonumber\\
&=& \pm\frac{k}{2}\frac{m_\sigma^2-2m_\pi^2}{m_\pi^2} + \frac{\omega}{2}\frac{m_\sigma^2}{m_\pi^2}\sqrt{1-\frac{4m_\pi^2}{m_\sigma^2}} ~~~\mbox{for} ~~~\omega^2=k^2+m_\pi^2\,\, .
\label{ppm}
\ee
The damping rate of massless pions is given by equation (\ref{land}) 
\be
\Gamma_{\pi\pi\sigma}(\omega,\vec{k}) = -\frac{1}{8\pi}\lambda\frac{m_\sigma^2T}{k\omega}\log{\frac{e^{\frac{p_l}{T}}-e^{-\frac{\omega}{T}}}{e^{\frac{p_l}{T}}-1}}\,\, ,
\label{pipisi0}
\ee
where
\be
p_l =\frac{k^2-\omega^2+m_\sigma^2}{2(\omega+k)} \,\, .
\ee
On-shell, $\omega=k$ Goldstone modes are damped as  
\be
\Gamma_{\pi\pi\sigma}(k) = \frac{\lambda m_\sigma^2}{8\pi k}f\left(\frac{m_\sigma^2}{4k}\right)
\ee
In the low temperature limit $T\ll m_\sigma$ the approximation
$$
f(x) = \frac{1}{e^x-1} \rightarrow e^{-x}
$$
can be used, leading to 
\be
\Gamma_{\pi\pi\sigma}(k) \simeq \frac{\lambda m_\sigma^2}{8\pi k} e^{-m_\sigma^2/4kT}\,\, .
\ee
This form clearly shows that the process is exponentially suppressed by the heavy sigma meson. As the momentum vanishes, $k\rightarrow 0$, the damping rate goes to zero as it should in order to satisfy Goldstone's Theorem.  For high temperatures $T\gg m_\sigma$ and $T\rightarrow T_c$ 
$$
f(x)  = \frac{1}{e^x-1} \rightarrow \frac{1}{x}-\frac{1}{2}+\frac{x}{12} ~~~\mbox{as}~~~x\rightarrow 0
$$
resulting in 
\be
\Gamma_{\pi\pi\sigma}(k) \simeq \frac{\lambda m_\sigma^2}{8\pi k}\left(\frac{4kT}{m_\sigma^2}-\frac{1}{2}\right) \simeq  \frac{\lambda T}{2\pi}\,\, .
\label{piclas}
\ee
The last expression corresponds to the classical limit. As pointed out in \cite{magyar}, care has to be taken when taking this limit. For low momentum, $k\ll m_\sigma^2/4T$, the classical approximation would overestimate the damping rate, since this is exponentially suppressed. Moreover, in the classical limit the damping is not zero when the momentum is zero, thus Goldstone's Theorem is not fulfilled. 
\section{Numerical Results}

\begin{figure}[htp]
\begin {center}
\leavevmode
\hbox{%
\epsfysize=11.5cm
\epsffile{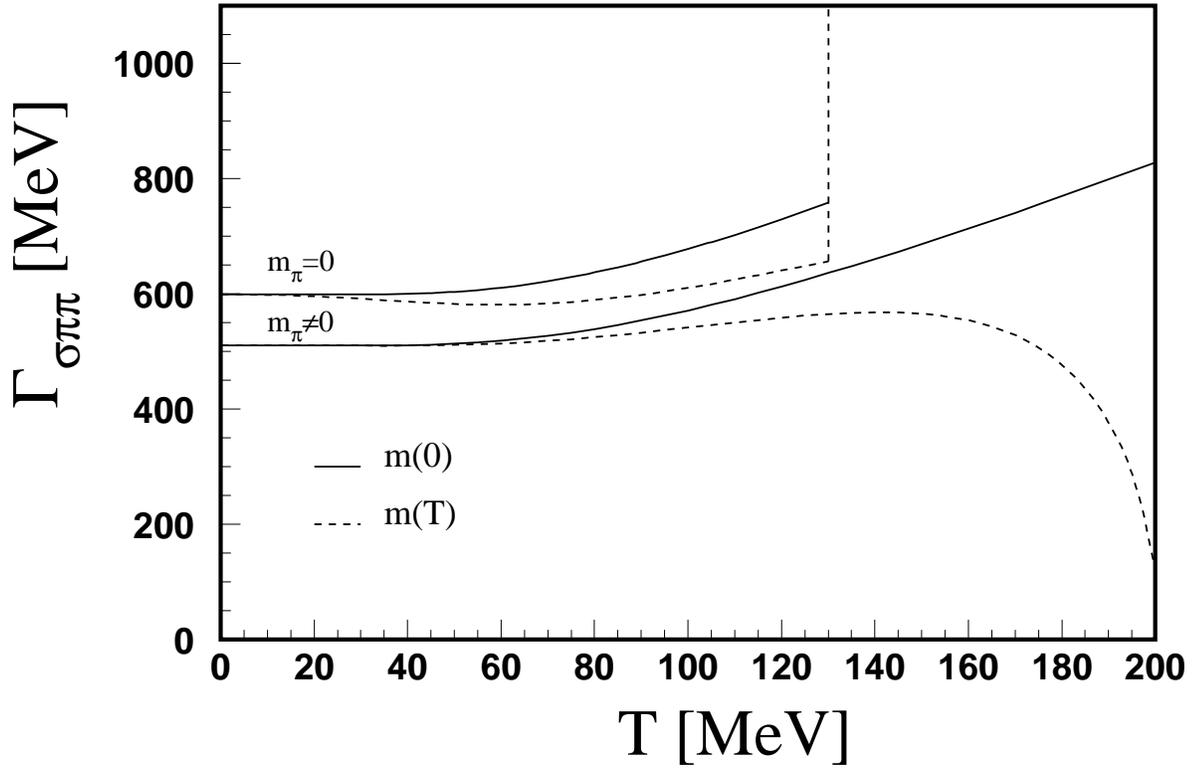} 
}
\vspace*{5cm}
\caption{Temperature dependence of the sigma damping rate at one-loop order calculated in the sigma rest frame in the case of $m_\pi\neq 0$ (lower) and in the chiral limit $m_\pi=0$ (upper) with $T=0$ masses (solid) and T-dependent masses (dashed).\label{sidecay.fig}}
\end{center}
\end{figure}

The temperature dependence of the sigma damping rate, eq.~(\ref{sipipi}), in the rest frame of the sigma is shown in figure \ref{sidecay.fig}. Note that even at zero temperature there is a finite damping, which means that the sigma meson can decay into two pions in the vacuum. Our results show that the dissipation is greater in the theory with exact chiral symmetry (upper curves) compared to the explicitly broken symmetric case (lower curves). This is due to the fact that there is more phase space available when pions are massless than when they are massive.
At $T=0$, for model parameters defined in section \ref{sect-masses-results}, the damping rate is about $\Gamma_{\sigma\pi\pi}=600~$MeV when $m_\pi=0$ and  $\Gamma_{\sigma\pi\pi}=510~$MeV when $m_\pi\neq 0$. These values are of the order of the mass of the sigma, meaning that the width of the sigma resonance is very broad, in other words the sigma meson is overdamped. Figure \ref{sidecay.fig} presents the comparison of results obtained when calculations were done keeping the masses at their zero temperature value (solid lines), and when the temperature-dependence of the masses is taken into account (dashed lines). Both set of curves show a suppression of the sigma decay rate when the resummed masses are used. 
\begin{figure}[htp]
\begin {center}
\leavevmode
\hbox{%
\epsfysize=11.5cm
\epsffile{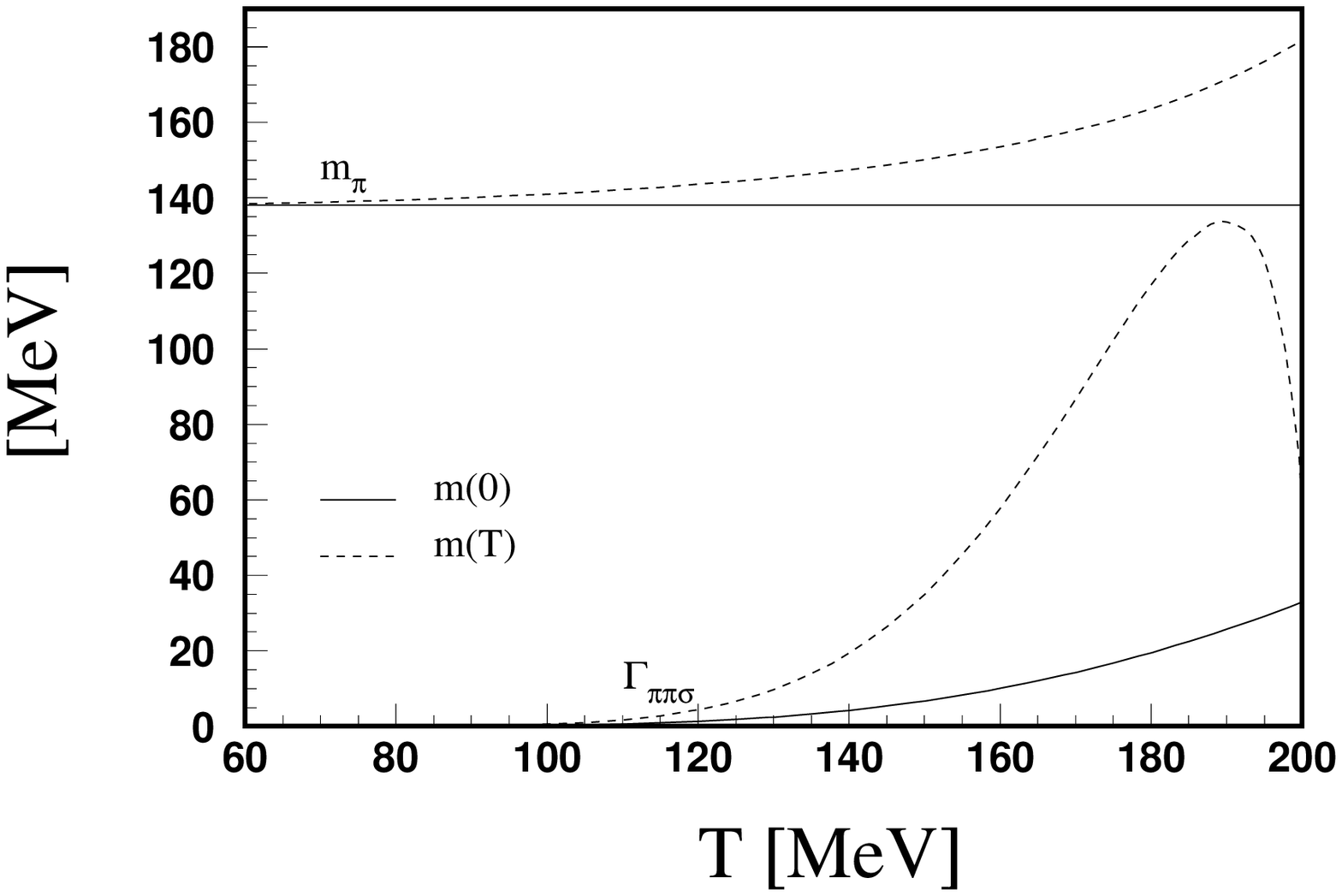} 
}
\vspace*{5cm}
\caption{Temperature dependence of the pion damping rate at one-loop order calculated in the pion rest frame with $T=0$ masses (solid) and T-dependent masses (dashed) and compared to the pion mass.\label{pidecay.fig}}
\end{center}
\end{figure}
This suppression is increasing with $T$. When $m_\pi=0$ the decay rate diverges at the critical temperature on account of large fluctuations. However, this would require further investigation. Without this, one can only say that due to its strong damping the sigma field quickly relaxes to its equilibrium value. In the $m_\pi\neq 0$ case a significant qualitative change is exhibited: the increase of the damping with $T$ is followed by its quite drastic decrease starting from about $T=150~$MeV. This corresponds to the temperature where the sigma mass begins to drop significantly (see figure \ref{step.fig}). It is then natural to expect a decrease of its decay rate into pions with masses approaching that of the sigma. Thus, the strong relaxation of sigma mesons at higher temperatures is not obvious anymore and some oscillations in this field might be expected. As the temperature drops the damping of the sigma mesons will set in.

Figure \ref{pidecay.fig} shows the temperature dependence of the damping rate of massive pions at one-loop order calculated in the pion's rest frame. Unlike the sigma meson, at zero temperature the dissipation is zero. This makes sense because the transformation of pions into sigmas is due to their annihilation with a hard thermal pion in the medium, and so this is an exclusively finite temperature process. Notice that $\Gamma_{\pi\pi\sigma}\ll\Gamma_{\sigma\pi\pi}$. This is the case because the phase space available for the pion decay process is suppressed by the large sigma mass. As the temperature increases, though, the dissipation rate is increasing too. As shown in figure \ref{pidecay.fig} the damping can get quite strong. At $T=170~$MeV, for example, when the pion mass is kept at its vacuum value, $m_\pi=138~$MeV, $\Gamma_{\pi\pi\sigma}=14.4~$MeV, and for $m_\pi=158~$MeV the damping is $\Gamma_{\pi\pi\sigma}=87.0~$MeV. Thus the width of pion excitations is about $10\%$ and $55\%$ of their energy, respectively.  This result should lead us to rethink our initial assumption of $\Gamma\ll\omega$ (see section \ref{sect-genform}).   

\begin{figure}[htp]
\begin {center}
\hspace*{-1.4cm}
\leavevmode
\hbox{%
\epsfysize=11.5cm
\epsffile{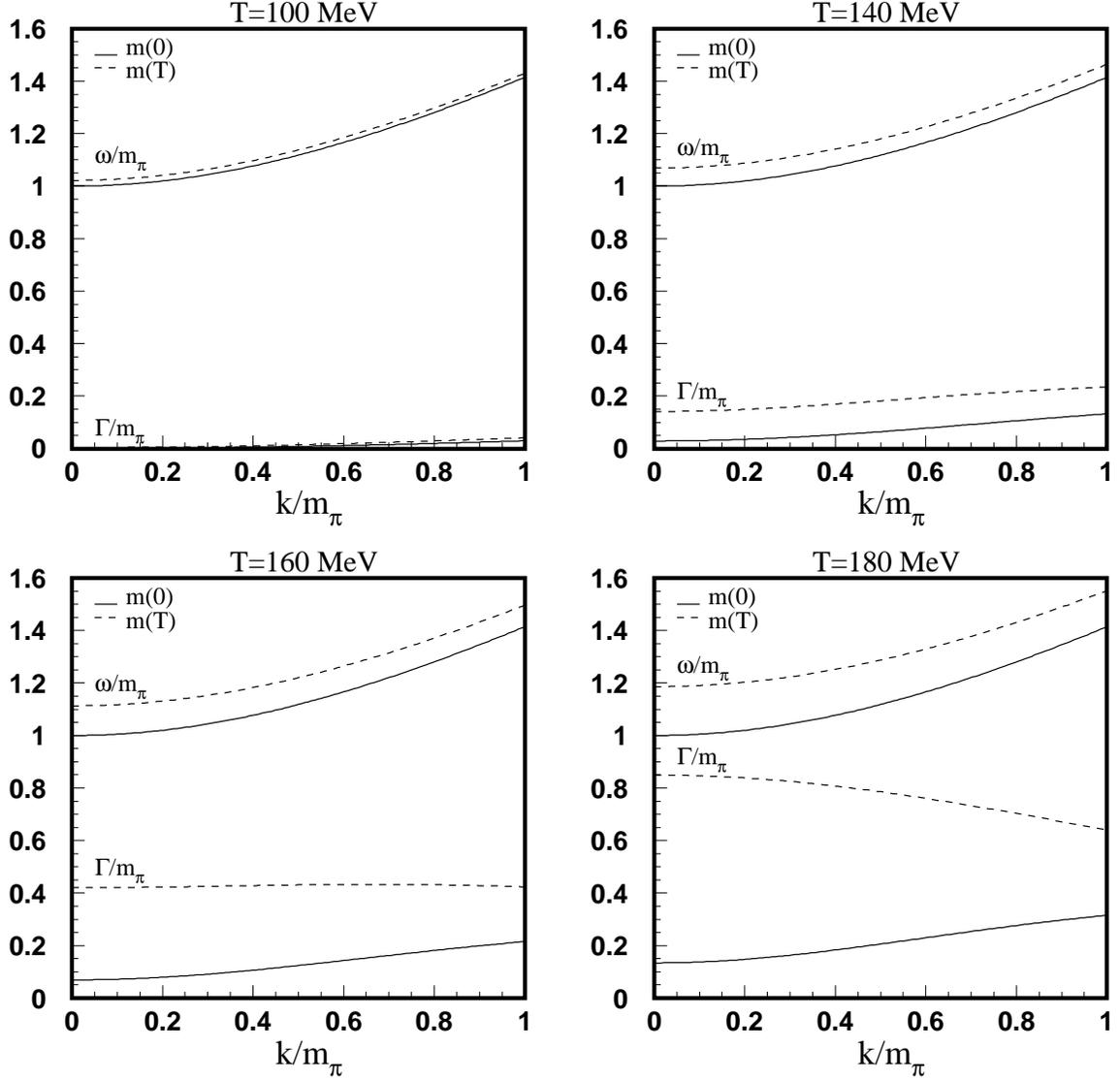} 
}
\vspace*{5cm}
\caption{Momentum dependence of the pion energy (upper) and width (lower) at different temperatures calculated with $T=0$ masses (solid) and T-dependent masses (dashed) and dispersion relation $\omega^2=k^2+m_\pi^2$.\label{hdisipk.fig}}
\end{center}
\end{figure}
\begin{figure}[htp]
\begin {center}
\hspace*{-1.4cm}
\leavevmode
\hbox{%
\epsfysize=11.5cm
\epsffile{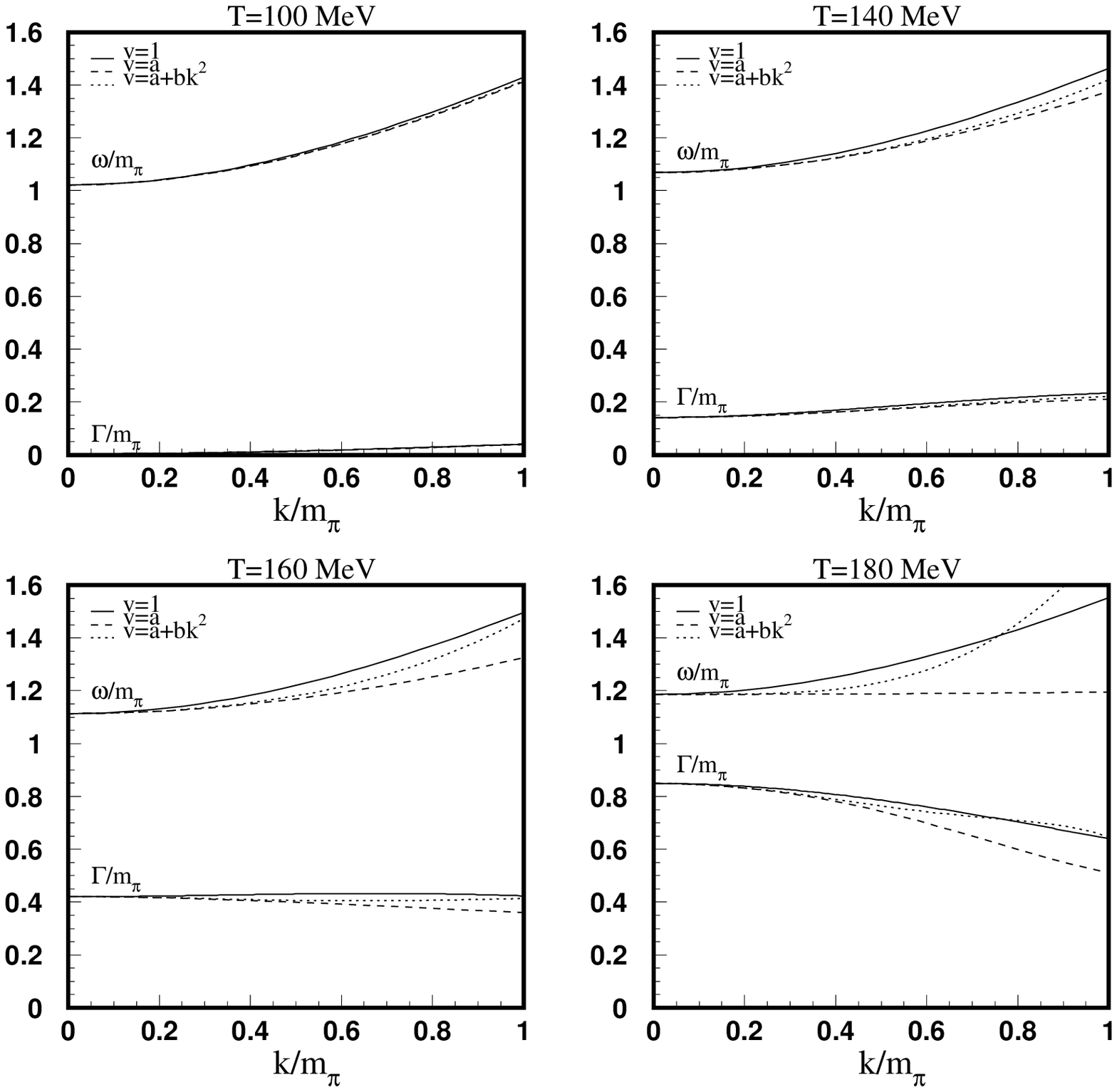} 
}
\vspace*{5cm}
\caption{Momentum dependence of the pion energy and width at different temperatures calculated with temperature-dependent masses and dispersion relation $\omega^2=vk^2+m_\pi^2$ with $v=1$ (solid), $v=a$ (dashed) and $v=a+bk^2$ (dotted).\label{hk.fig}}
\end{center}
\end{figure}

The damping in terms of the momenta of the pion fields is presented in figure \ref{hdisipk.fig} for different temperatures. Results using the dispersion relation $\omega^2 = k^2+m_\pi^2$ with zero temperature masses (solid lines) and with resummed masses (dashed lines) are displayed. Noticeable damping occurs above $T=100~$MeV, and this increases with $T$. One can see that pions that have the vacuum mass and hard momenta are more damped than pions with the same mass and soft momenta. However, when the temperature-dependence of the mass is taken into account such a statement cannot be made. At $T=160~$MeV all modes are equally damped. In other words, the width of pion modes is independent of their momentum. This width is increasing with $T$ and it can be as great as $30\%$ of the energy. We also analyzed the momentum dependence of the energy and width of pions with different dispersion relations. Our results for the vacuum (solid lines), the 2nd-order (dashed lines) and 4th-order (dotted lines) corrected dispersion relations (see section \ref{sect-dispersion}) are presented in figure \ref{hk.fig}. No major difference exists between the vacuum and 4th-order results. Note, though, that the quantities are normalized to the vacuum pion mass, so the true differences are somewhat greater than indicated on the axis. The overall message is that the width of the excitations, and so the damping of the fields, is more pronounced at high temperatures and should not be neglected.   

\begin{figure}[htp]
\begin {center}
\leavevmode
\hbox{%
\epsfysize=11.5cm
\epsffile{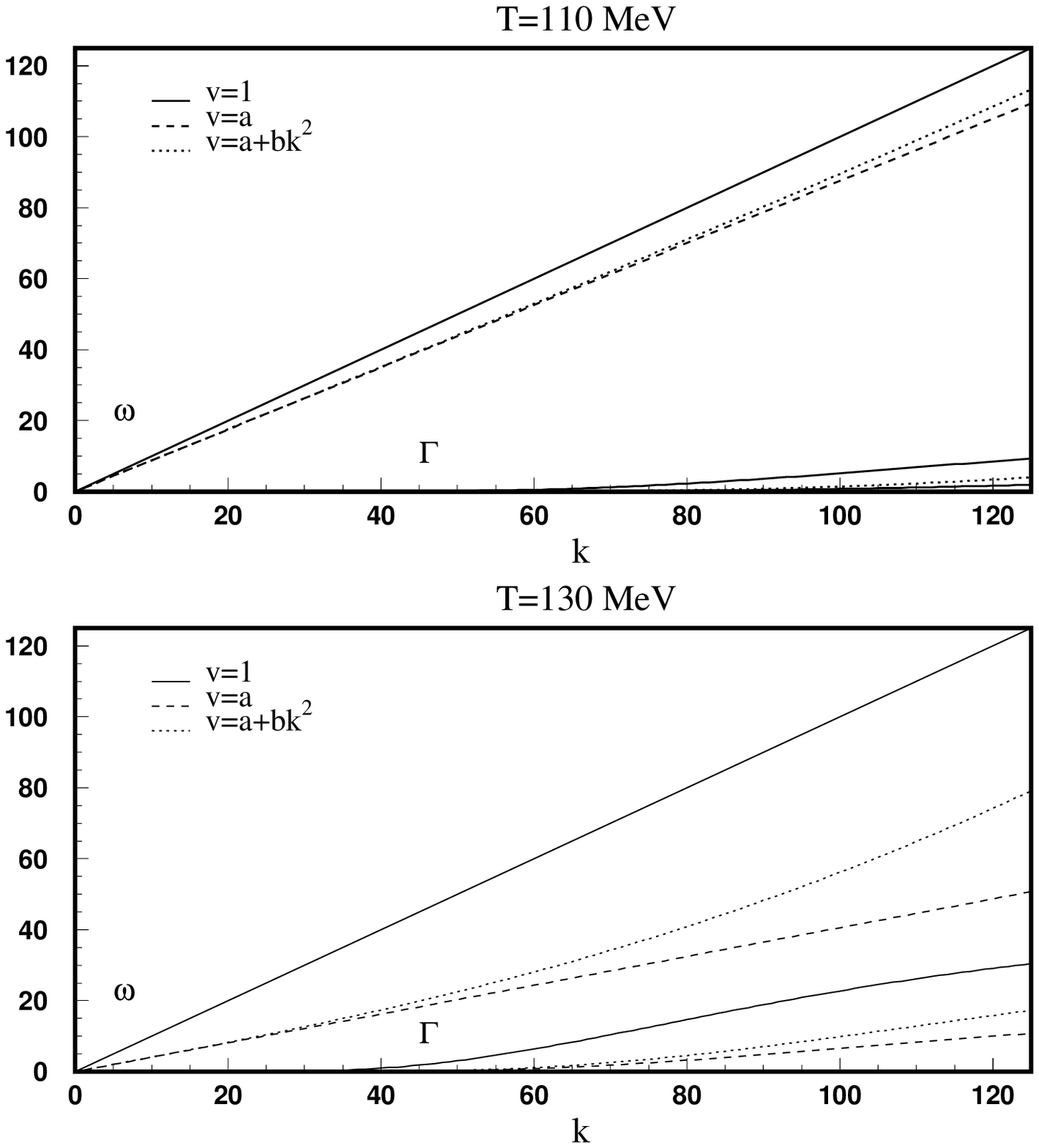} 
}
\vspace*{5cm}
\caption{Momentum dependence of the energy (upper curves) and width (lower curves) of Goldstone modes ($m_\pi=0$) at different temperatures calculated with T-dependent masses and dispersion relation $\omega^2=vk^2$ with $v=1$ (solid), $v=a$ (dashed) and $v=a+bk^2$ (dotted).\label{k.fig}}
\end{center}
\end{figure}

We close this section by presenting the results of our study of the dissipation of soft Goldstone modes which happens through the absorption of a high frequency Golstone boson producing a sigma meson. The momentum dependence of the energy and width of these massless excitations at different temperatures (\ref{pipisi}) is shown in figure \ref{k.fig}. Notice that modes with $k=0$ have $\Gamma=0$, meaning that true Goldstone modes do not decay. Thus homogeneous condensates are undamped even at high temperatures. This result is consistent with \cite{rischke}. Figure \ref{k.fig} shows that the pion modes with higher momenta are significantly damped near $T_c$.  

Our results are contrary to previous beliefs, according to which sigma mesons are quickly damped and, due to their small mass, pion oscillations persist in the small momentum region \cite{rajagopal}. We found that at high temperatures not only is the damping of sigma mesons significant, but also that of the pions. This damping is most accentuated in the phase transition region. Accordingly, we do not expect amplified pion oscillations but rather the decay of non-equilibrium pion fields. Clearly, further investigations are required to determine if there are oscillations of damped waves surviving. Moreover, to determine how fast thermal equilibrium is reestablished the time-scales that are involved should be studied. 
\chapter{Dissipation at Two-Loop Order \label{sect-dissip2}}

This section is dedicated to a detailed discussion of processes in which a meson from the thermal bath (pion or sigma) elastically scatters off a meson from the condensate (pion or sigma). As a consequence of these scatterings mesons get knocked out of the condensate, making this analysis relevant for the study of dissipation.  Contrary to one-loop level damping processes which are present only when certain conditions are satisfied, scattering always occurs. First, I present the general formalism for determining the rate of dissipation. Then, all possible tree-level scattering processes are evaluated. Finally, my numerical results are presented.  

\section{General Formalism}

Two-loop linear response functions can be evaluated following the procedure presented in the previous chapter for one-loop order. However, rather than attempt this sophisticated evaluation, a direct determination from the physical processes is presented. Two-loop contributions correspond to two-particle scatterings with amplitudes evaluated at tree level. The equivalence of these two methods at two-loop order can be shown in a way similar to the proof at one-loop order presented in Appendix \ref{ap-1loop}.  

The contribution from scattering processes to the rate at which the condensate fields decay to their equilibrium values is given in terms of the imaginary parts of their self-energies \cite{weldon},
\be
\Gamma(\omega,\vec{k})=-\frac{\mbox{Im}~\Pi(\omega,\vec{k})}{2\omega}\,~.
\label{weldon}
\ee
Here $\omega$ and $\vec{k}$ are the energy and momentum of the meson. This rate refers to the field amplitude, which is half of what the decrease in number density is. The general form of the self-energy of a particle of mass $m_a$, propagating with four-momentum $k=(\omega,\vec{k})$ through a medium in thermal equilibrium, is given by \cite{shuryak,eletsky} 
\be
\Pi_{ab}(k) = \int\frac{d^3p}{(2\pi)^32E} f(E){\cal M}(s)\,~.
\label{shuryak0}
\ee
Here ${\cal M}$ is the transition amplitude for the process $ab\rightarrow ab$. The thermodynamical weight $f(E)$ is the Bose distribution of thermal mesons of mass $m_b$ and four-momentum $p=(E,\vec{p})$. In terms of the forward scattering amplitude ${\cal M}(s)= -8\pi\sqrt{s}f_{cm}(s)$, where $s=(p+k)^2$. From (\ref{shuryak0}) follows the imaginary part 
\be
\mbox{Im}~\Pi_{ab}(k) &=& -\int\frac{d^3p}{(2\pi)^3}f(E)\sqrt{s}\frac{q_{cm}}{E}\sigma_{total}(s)\, .
\label{shuryak}
\ee
To obtain this equality we applied the standard form of the optical theorem that relates the imaginary part of the forward scattering amplitude and the total cross-section \cite{peskin}:
\be
\mbox{Im}~f_{cm}(s) = \frac{q_{cm}}{4\pi}\sigma_{total}(s)\,~.
\ee
The differential cross section for $a(k)+b(p)\rightarrow a(k')+b(p')$ is given by the general formula \cite{peskin}:
\be
d\sigma_{ab} = \frac{1}{4\sqrt{(p\cdot k)^2-m_a^2m_b^2}}|{\cal M}|^2d\tau_2\,~,
\ee
where $d\tau_2$ is the two-body phase-space
\be
d\tau_2 = (2\pi)^4\delta^4(p+k-p'-k')\frac{d^3p'}{(2\pi)^32E'}\frac{d^3k'}{(2\pi)^32\omega'}\,\,.
\ee
One can simplify this expression by evaluating the phase-space integral in the center-of-mass frame of the two colliding particles. Then the total scattering cross-section is given by
\be
\sigma_{ab} = \frac{1}{S!}\int\left(\frac{d\sigma_{ab}}{d\Omega}\right)_{cm}d\Omega\, ,
\label{cross}
\ee
where 
\be
\left(\frac{d\sigma_{ab}}{d\Omega}\right)_{cm} = \frac{1}{64\pi^2s}|{\cal M}|^2\, .
\label{dcross}
\ee
The symmetry factor $1/S!$ is due to the number $S$ of identical particles in the final state. It is clear that knowing the amplitude ${\cal M}$ of a process readily results in the dissipation rate due to that process. However, we must distinguish between scatterings that involve massive and massless mesons. For the former it is convenient to work in the rest frame of the meson $a$ from the condensate. Then $\vec{k}=0$, and the dispersion is $\omega=m_a$. It makes no sense, though, to talk about the rest frame for massless particles. In this case we evaluate the self-energy in the rest frame of the condensate. Then $\vec{k}$ labels the momentum of the meson with respect to this frame. Expression (\ref{shuryak}) then takes simplified forms:
\be
\mbox{Im}~\Pi_{ab}(\omega=m_a,\vec{k}=0) = -\frac{m_a}{2\pi^2}\int_{m_b}^{\infty}dE(E^2-m_b^2)f(E)\sigma_{ab}(E)\, ,
\label{rest}
\ee
for massive mesons, and 
\be
\mbox{Im}~\Pi_{ab}(\omega,\vec{k}) = -\frac{1}{8\pi^2}\int_{-1}^{1}d\cos{\theta}\int_{0}^{\infty}dp\frac{p^2}{E}(s-m_b^2)f(E)\sigma_{ab}(E)\, ,
\label{restless}
\ee
for Goldstone modes ($m_a=0$). Here $s=m_b^2+2E\omega-2pk\cos{\theta}$ where $\theta$ is the angle between $\vec{k}$ and $\vec{p}$. The dispersion relation of the hard thermal modes is $E=\sqrt{p^2+m_b^2}$. In both of the above formulas there should be a $\Theta(p-\Lambda_c)$ restricting the integration to hard momenta only. The probability of finding long wavelength modes in the heath bath is small leading to negligible contribution from the soft end of the spectrum.

\section{Sigma Scattering \label{sect-si}}

There are two mechanisms that contribute to the removal or addition of a sigma meson to the condensate: elastic scattering of a hard thermal sigma and a hard thermal pion off a sigma from the condensate. 

\subsection{Sigma-Sigma Scattering}

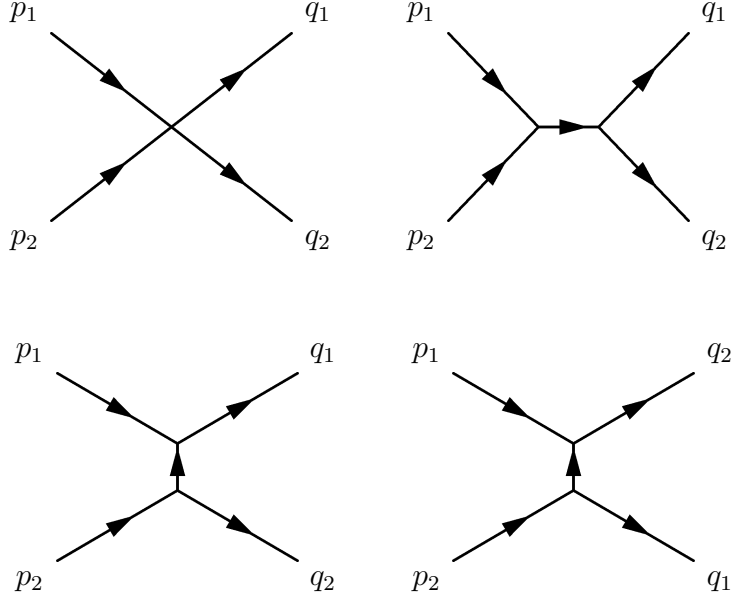
\begin{figure}[t]
\bc
\setlength{\unitlength}{1mm}
\parbox{30mm}{
\begin{fmfgraph*}(40,25)
  \fmfleft{i1,i2}
  \fmfright{o1,o2}
  \fmf{fermion}{i1,v}
  \fmf{fermion}{i2,v}
  \fmf{fermion}{v,o1}
  \fmf{fermion}{v,o2}
\fmflabel{$p_2$}{i1}
\fmflabel{$p_1$}{i2}
\fmflabel{$q_2$}{o1}
\fmflabel{$q_1$}{o2}
\end{fmfgraph*}}
\hspace*{2cm} 
\parbox{30mm}{
\begin{fmfgraph*}(40,25)
  \fmfleft{i1,i2}
  \fmfright{o1,o2}
 \fmf{fermion,tension=1/3}{i1,v1}
 \fmf{fermion,tension=1/3}{i2,v1}
 \fmf{fermion}{v1,v2}
 \fmf{fermion,tension=1/3}{v2,o1}
 \fmf{fermion,tension=1/3}{v2,o2}
\fmflabel{$p_2$}{i1}
\fmflabel{$p_1$}{i2}
\fmflabel{$q_2$}{o1}
\fmflabel{$q_1$}{o2}
\end{fmfgraph*}
}
\vspace*{2cm}

\parbox{30mm}{
\begin{fmfgraph*}(40,25)
  \fmfleft{i1,i2}
  \fmfright{o1,o2}
 \fmf{fermion,tension=1/3}{i1,v1}
 \fmf{fermion,tension=1/3}{i2,v2}
 \fmf{fermion}{v1,v2}
 \fmf{fermion,tension=1/3}{v1,o1}
 \fmf{fermion,tension=1/3}{v2,o2}
\fmflabel{$q_2$}{o1}
\fmflabel{$p_1$}{i2}
\fmflabel{$p_2$}{i1}
\fmflabel{$q_1$}{o2}
\end{fmfgraph*}
}
\hspace*{2cm}
\parbox{30mm}{
\begin{fmfgraph*}(40,25)
  \fmfleft{i1,i2}
  \fmfright{o1,o2}
 \fmf{fermion,tension=1/3}{i1,v1}
 \fmf{fermion,tension=1/3}{i2,v2}
 \fmf{fermion}{v1,v2}
 \fmf{fermion,tension=1/3}{v1,o1}
 \fmf{fermion,tension=1/3}{v2,o2}
\fmflabel{$p_2$}{i1}
\fmflabel{$p_1$}{i2}
\fmflabel{$q_1$}{o1}
\fmflabel{$q_2$}{o2}
\end{fmfgraph*}
}

\vspace*{1cm}
\parbox{9cm}{\caption{\small $\sigma\sigma$ scattering at tree-level. 
Straight lines are $\sigma$s, dashed lines are $\pi$s.\label{sisi.fig}}}
\ec
\end{figure}

To first order in the coupling $\lambda$ there are four diagrams contributing to the process in which a thermal sigma meson knockes out a low momentum sigma from the condensate. These are presented in figure \ref{sisi.fig}. The transition amplitude is the sum of contributions from different diagrams: 
\be 
{\cal M} &=& -6\lambda\left[1+3(m_\sigma^2-m_\pi^2)\left(\frac{1}{s-m_\sigma^2} + \frac{1}{t-m_\sigma^2} +
 \frac{1}{u-m_\sigma^2}\right)\right]\,~,
\label{msisi}
\ee
which reflects the symmetry in the $s=(p_1+p_2)^2$, $t=(p_1-q_1)^2$, $u=(p_1-q_2)^2$ channels. These Mandelstam variables satisfy $s+t+u= 4m_\sigma^2$. The total cross-section is 
\be
\sigma_{\sigma\sigma}(s) &=& \frac{9\lambda^2}{8\pi s}\left[\left(\frac{s+2m_\sigma^2-3m_\pi^2}{s-m_\sigma^2}\right)^2 + \frac{18(m_\sigma^2-m_\pi^2)^2}{m_\sigma^2(s-3m_\sigma^2)} \right. \nonumber\\
&-& \!\!\!\! \left.\frac{12(m_\sigma^2-m_\pi^2)(s^2-3sm_\sigma^2-m_\sigma^4+3m_\sigma^2m_\pi^2)}{(s-4m_\sigma^2)(s-2m_\sigma^2)(s-m_\sigma^2)}\ln\left(\frac{s-3m_\sigma^2}{m_\sigma^2}\right)\right] 
\label{crosssisi}
\ee
%
\noindent
In the low energy limit expand (\ref{crosssisi}) about $s=4m_\sigma^2$. To leading order  
\be
\sigma_{\sigma\sigma} = \frac{9}{32\pi}\lambda^2\frac{(4m_\sigma^2-5m_\pi^2)^2}{m_\sigma^6} \,~,
\label{lowsisi}
\ee
and in the rest-frame of the sigma this gives rise to
\be
\mbox{Im}~\Pi_{\sigma\sigma} = -\frac{9}{32\pi^3}\lambda^2T^2e^{-m_\sigma/T}\frac{(m_\sigma+T)(4m_\sigma^2-5m_\pi^2)^2}{m_\sigma^5}\,~.
\label{imlowsisi}
\ee
In the high energy limit only the four-point vertex contributes to the amplitude, resulting in
\be
\sigma_{\sigma\sigma} = \frac{9\lambda^2}{8\pi s}\,~.
\label{highsisi}
\ee
This gives rise to
\be
\mbox{Im}~\Pi_{\sigma\sigma} = -\frac{3}{64\pi}\lambda^2T^2\,~.
\label{imhighsisi}
\ee
With these two limits one can construct an interpolating formula that describes the whole energy range:
\be
\mbox{Im}~\Pi_{\sigma\sigma}\simeq -\frac{9\lambda^2}{32\pi}\frac{T^2(m_\sigma+T)(4m_\sigma^2-5m_\pi^2)^2}{6(m_\sigma+T)(4m_\sigma^2-5m_\pi^2)^2+\pi^2m_\sigma^5(e^{m_\sigma/T}-1)}\,~.
\label{imhsisi}
\ee
In the theory with exact chiral symmetry $m_\pi=0$ and (\ref{imhsisi}) is simplified to the result obtained in \cite{cejk}
\be
\mbox{Im}~\Pi_{\sigma\sigma}\simeq -\frac{9\lambda^2}{2\pi}\frac{T^2}{96+\pi^2(e^{m_\sigma/T}-1)}\,~.
\label{imsisi}
\ee
The contribution to the rate of decay of the amplitude is then
\be
\Gamma_{\sigma\sigma} = -\frac{\mbox{Im}~\Pi_{\sigma\sigma}}{2m_\sigma} = \frac{9\lambda^2}{4\pi}\frac{T^2}{m_\sigma\left[96+\pi^2(e^{m_\sigma/T}-1)\right]}\,~.
\label{gammasisi}
\ee
%

\subsection{Sigma-Pion Scattering \label{sect-sipi}}

\begin{figure}[t]
\bc
\setlength{\unitlength}{1mm}
\parbox{30mm}{
\begin{fmfgraph*}(40,25)
  \fmfleft{i1,i2}
  \fmfright{o1,o2}
  \fmf{scalar}{i1,v}
  \fmf{fermion}{i2,v}
  \fmf{scalar}{v,o1}
  \fmf{fermion}{v,o2}
\fmflabel{$q_1$}{i1}
\fmflabel{$p_1$}{i2}
\fmflabel{$q_2$}{o1}
\fmflabel{$p_2$}{o2}
\end{fmfgraph*}}
\hspace*{2cm} 
\parbox{30mm}{
\begin{fmfgraph*}(40,25)
  \fmfleft{i1,i2}
  \fmfright{o1,o2}
 \fmf{scalar,tension=1/3}{i1,v1}
 \fmf{fermion,tension=1/3}{i2,v1}
 \fmf{scalar}{v1,v2}
 \fmf{scalar,tension=1/3}{v2,o1}
 \fmf{fermion,tension=1/3}{v2,o2}
\fmflabel{$q_1$}{i1}
\fmflabel{$p_1$}{i2}
\fmflabel{$q_2$}{o1}
\fmflabel{$p_2$}{o2}
\end{fmfgraph*}
}
\vspace*{2cm}

\parbox{30mm}{
\begin{fmfgraph*}(40,25)
  \fmfleft{i1,i2}
  \fmfright{o1,o2}
 \fmf{scalar,tension=1/3}{i1,v1}
 \fmf{fermion,tension=1/3}{i2,v2}
 \fmf{fermion}{v1,v2}
 \fmf{scalar,tension=1/3}{v1,o1}
 \fmf{fermion,tension=1/3}{v2,o2}
\fmflabel{$q_2$}{o1}
\fmflabel{$p_1$}{i2}
\fmflabel{$q_1$}{i1}
\fmflabel{$p_2$}{o2}
\end{fmfgraph*}
}
\hspace*{2cm}
\parbox{30mm}{
\begin{fmfgraph*}(40,25)
  \fmfleft{i1,i2}
  \fmfright{o1,o2}
 \fmf{scalar,tension=1/3}{i1,v1}
 \fmf{fermion,tension=1/3}{i2,v2}
 \fmf{scalar}{v1,v2}
 \fmf{fermion,tension=1/3}{v1,o1}
 \fmf{scalar,tension=1/3}{v2,o2}
\fmflabel{$q_1$}{i1}
\fmflabel{$p_1$}{i2}
\fmflabel{$p_2$}{o1}
\fmflabel{$q_2$}{o2}
\end{fmfgraph*}
}

\vspace*{1cm}
\parbox{11cm}{\caption{\small $\sigma\pi$ scattering at tree-level. Solid 
lines are $\sigma$s, dashed lines are $\pi$s.  
\label{sipi.fig}}}
\ec
\end{figure}
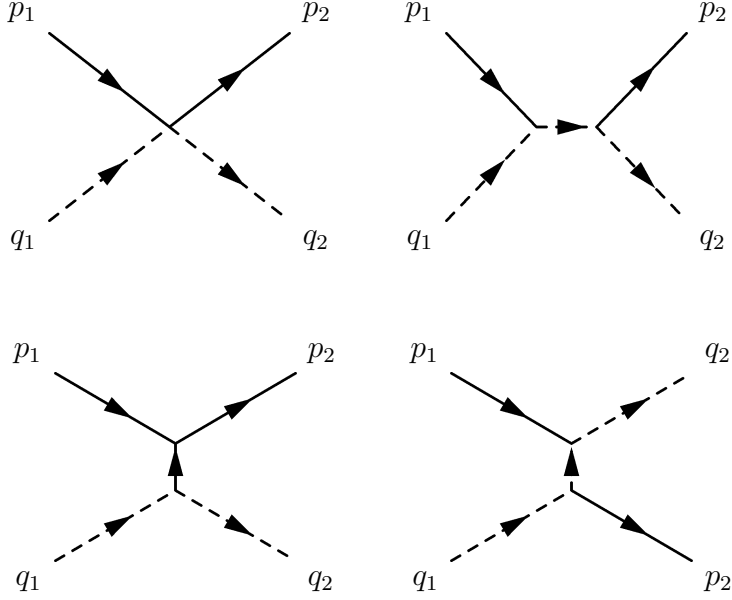

A hard pion in the heat bath can be energetic enough to knock out a sigma-meson from the condensate. The possible tree-level processes may happen according to the four different diagrams shown in figure \ref{sipi.fig}. The transition amplitude obtained from these is
\be 
{\cal M} &=& -2\lambda\left[1+(m_\sigma^2-m_\pi^2)\left(\frac{1}{s-m_\pi^2} + \frac{3}{t-m_\sigma^2} + \frac{1}{u-m_\pi^2}\right)\right]\,~,
\label{msipi}
\ee
where $s=(p_1+q_1)^2$, $t=(p_1-p_2)^2$, $u=(p_1-q_2)^2$ and $s+t+u=2(m_\sigma^2+m_\pi^2)$~.
The total scattering cross-section is
\be
\sigma_{\sigma\pi}(s) &=& \frac{\lambda^2}{4\pi s}\left[\left(\frac{s-2m_\pi^2+m_\sigma^2}{s-m_\pi^2}\right)^2 + \frac{9s(m_\sigma^2-m_\pi^2)^2}{m_\sigma^2(s^2-sm_\sigma^2-2sm_\pi^2+(m_\sigma^2-m_\pi^2)^2)} \right.\nonumber\\
&& \left. +\frac{s(m_\sigma^2-m_\pi^2)^2}{(2m_\sigma^2+m_\pi^2-s)((m_\sigma^2-m_\pi^2)^2-sm_\pi^2) }\right.\nonumber\\
&&\left. +\frac{6s(m_\sigma^2-m_\pi^2)(sm_\sigma^2+2sm_\pi^2-s^2+m_\sigma^4-m_\pi^4-2m_\sigma^2m_\pi^2)}{(s-m_\pi^2)(m_\sigma^2+m_\pi^2-s)(s^2-2s(m_\sigma^2+m_\pi^2)+(m_\sigma^2-m_\pi^2)^2)}\right.\nonumber\\
&& \times\left.\ln{\frac{sm_\sigma^2}{s^2-sm_\sigma^2-2sm_\pi^2+(m_\sigma^2-m_\pi^2)^2}}\right.\nonumber\\
&& \left. -\frac{2s(m_\sigma^2-m_\pi^2)(3sm_\sigma^2-s^2+m_\pi^4+m_\sigma^4-4m_\sigma^2m_\pi^2)}{(s-m_\pi^2)(m_\sigma^2+m_\pi^2-s)(s^2-2s(m_\sigma^2+m_\pi^2)+(m_\sigma^2-m_\pi^2)^2)}\right.\nonumber\\
&& \times\left.\ln{\frac{s(2m_\sigma^2+m_\pi^2-s)}{(m_\sigma^2-m_\pi^2)^2-sm_\pi^2}}\right] \,~.
\label{crosshsipi}
\ee
The low energy limit is obtained by expanding (\ref{crosshsipi}) about $s=(m_\sigma+m_\pi)^2$ and is
\be
\sigma_{\sigma\pi}(s) &=& \frac{9}{4\pi}\lambda^2\frac{m_\pi^4(3m_\sigma^2-4m_\pi^2)^2}{sm_\sigma^2(m_\sigma^2-4m_\pi^2)^2}\,~.
\label{lowhsipi}
\ee
At high temperatures, where only the first diagram of figure \ref{sipi.fig} contributes, (\ref{crosshsipi}) reduces to 
\be
 \sigma_{\sigma\pi}(s) = \frac{\lambda^2}{4\pi s}\,~.
\label{highsipi}
\ee
The contribution to the imaginary part of the self-energy is readily determined in these two limits using (\ref{rest}). At low energy 
\be
\mbox{Im}~\Pi_{\sigma\pi} = -\frac{9}{4\pi^3}\lambda^2T^2e^{-m_\pi/T}F_{\sigma\pi}(m_\sigma,m_\pi)\,~,
\label{imlowhsipi}
\ee
where
\be
F_{\sigma\pi}(m_\sigma,m_\pi) = \frac{m_\pi^4(m_\pi+T)(3m_\sigma^2-4m_\pi^2)^2}{m_\sigma^3(m_\sigma+m_\pi)^2(m_\sigma^2-4m_\pi^2)^2}\,~,
\ee
and at high energy
\be
\mbox{Im}~\Pi_{\sigma\pi} = -\frac{\lambda^2T^2}{96\pi}\,~.
\label{imhighsipi}
\ee
%
\noindent
The two limits can be combined into one approximate expression which is used then to determine the rate of dissipation 
\be
\Gamma_{\sigma\pi} \simeq \frac{9}{8\pi}\lambda^2\frac{T^2}{m_\sigma}\frac{F_{\sigma\pi}}{216F_{\sigma\pi}+\pi^2(e^{m_\pi/T}-1)}
\label{gammahsipi}
\ee
This expression gives the contribution from a single pion. In order to factor in all pions, we must multiply (\ref{gammahsipi}) by $N-1$. 

At this point a distinction should be made between scatterings with massless and massive pions. Notice that the low energy expression (\ref{lowhsipi}) vanishes for zero pion mass. The first nonzero term in the series expansion of the cross section is the 4th order term, which yields 
\be
\sigma_{\sigma\pi}(s) &=& \frac{7}{3\pi}\lambda^2\frac{(s-m_\sigma^2)^4}{sm_\sigma^8}\,~,
\label{lowsipi}
\ee
giving rise to
\be
\mbox{Im}~\Pi_{\sigma\pi} = -\frac{13 440}{\pi^3}\lambda^2\zeta(7)\frac{T^7}{m_\sigma^5}\,~.
\label{imlowsipi}
\ee
The high temperature limit is given by (\ref{imhighsipi}). We combine the two limits again and obtain 
\be
\Gamma_{\sigma\pi} \simeq (N-1)\frac{\lambda^2}{192\pi}\frac{T^2}{m_\sigma}\frac{T^5}{\left[T^5+\left(m_\sigma/10.6\right)^5\right]}\,~,
\label{gammasipi}
\ee
formula identical to the one obtained in \cite{cejk}.

\section{Pion Scattering \label{sect-pi}}

Dissipation of the condensate can arise from scattering of soft pions with hard pions or hard sigma mesons. The damping of massive pions and that of the Goldstone pions is expected to be different, requiring a somewhat different analysis. 

\subsection{Pion-Sigma Scattering}

The possible tree-level diagrams representing the reaction in which a hard thermal sigma knocks out a low momentum pion from the condensate are shown in figure \ref{sipi.fig}. The total cross section is the same as for sigma-pion scattering and is given by expression (\ref{crosshsipi}). The low and high energy limits are also the same and are given by equations (\ref{lowhsipi}) and (\ref{highsipi}). 

When the pions are massive the imaginary part of the self-energy in the low energy limit in the rest frame of the pion is 
\be
\mbox{Im}~\Pi_{\pi\sigma} = -\frac{9}{4\pi^3}\lambda^2T^2e^{-m_\sigma/T}F_{\pi\sigma}\,~,
\label{imlowpisi}
\ee
where
\be
F_{\pi\sigma} = \frac{m_\pi^5(m_\sigma+T)(3m_\sigma^2-4m_\pi^2)^2}{m_\sigma^4(m_\sigma+m_\pi)^2(m_\sigma^2-4m_\pi^2)^2}\,~,
\ee
and for high energies is
\be
\mbox{Im}~\Pi_{\pi\sigma} = -\frac{\lambda^2T^2}{96\pi}\,~.
\label{imhighpisi}
\ee
The total scattering rate due to this process can be parametrized by an interpolating formula between the two known limits:
\be
\Gamma_{\pi\sigma} \simeq \frac{9\lambda^2T^2}{8\pi m_\sigma}\frac{F_{\pi\sigma}}{216F_{\pi\sigma}+\pi^2(e^{m_\sigma/T}-1)}\,~.
\label{gammahpisi}
\ee

When the pions are Goldstone bosons the cross-section (\ref{crosshsipi}) reduces to
\be
\sigma_{\pi\sigma}(s) &=& \frac{\lambda^2}{4\pi s}\left[\frac{(s+m_\sigma^2)^2}{s^2} +
\frac{s}{2m_\sigma^2-s} +\frac{9sm_\sigma^2}{s^2-sm_\sigma^2+m_\sigma^4} \right.\nonumber\\
&& \left. -\frac{2sm_\sigma^2(s^2-3sm_\sigma^2-m_\sigma^4)}{s(s-m_\sigma^2)^3}\ln\left(\frac{s(2m_\sigma^2-s)}{m_\sigma^4}\right)\right.\nonumber\\
&&\left. - \frac{6sm_\sigma^2(s^2-sm_\sigma^2-m_\sigma^2)}{s(s-m_\sigma^2)^3}\ln\left(\frac{s^2-sm_\sigma^2+m_\sigma^4}{sm_\sigma^2}\right)
\right]\, ,
\label{crosspisi}
\ee
with low energy limit obtained from the series expansion of this about $s=m_\sigma^2~$,
\be
\sigma_{\pi\sigma} = \frac{7\lambda^2}{3\pi}\frac{(s-m_\sigma^2)^4}{m_\sigma^{10}}\,~.
\ee
When $T\rightarrow T_c$ the three-point vertex diagrams do not matter and 
\be
\sigma_{\pi\sigma} = \frac{\lambda^2}{4\pi s}\,~.
\ee
The self-energies in these limits are obtained from (\ref{restless}) with $s=m_\sigma^2+2E\omega-2|\vec{p}||\vec{k}|\cos\theta$. It is perhaps worth noting that when evaluating (\ref{restless}) we keep the energy $\omega$ and momentum $\vec{k}$ independent. At high temperatures
\be
\mbox{Im}~\Pi_{\pi\sigma}(\omega, \vec{k}) = -\frac{\lambda^2T^2}{96\pi}\,~.
\ee
This is exactly the same as for the massive pion case. No surprise here because this limit is defined by the temperature being much larger than the masses. The evaluation of the self-energy in the low temperature limit results in 
\be
 \mbox{Im}~\Pi_{\pi\sigma}(\omega, \vec{k}) = -\frac{56\sqrt{\pi}\lambda^2}{3\sqrt{2}\pi^3m_\sigma^3}\sqrt{\frac{T}{m_\sigma}}e^{-m_\sigma/T}u(\omega,k)\omega\,~,
\ee
with the following definition:
\be
u(\omega,k) = 15k^4\frac{T^3}{m_\sigma^3}+10k^2\omega^2\left[\frac{T^2}{m_\sigma^2}-3\frac{T^3}{m_\sigma^3}\right]+\omega^4\left[\frac{T}{m_\sigma}-5\frac{T^2}{m_\sigma^2}\right]\nonumber\\
&&
\label{u}
\ee
An interpolating formula for the contribution to the massless pion damping rate is:
\be
\Gamma_{\pi\sigma} = \frac{\lambda^2T^2}{192\pi}\frac{u(\omega,k)\sqrt{\frac{T}{m_\sigma}}}{\omega u(\omega,k)\sqrt{\frac{T}{m_\sigma}}+T^2\left(\frac{m_\sigma}{6.1}\right)^3(e^{m_\sigma/T}-1)}\,~.
\label{gammapisi}
\ee
%

\subsection{Pion-Pion Scattering}

$\pi\pi$ scattering has been extensively studied during the last couple of decades in a variety of different models and approaches. An incomplete but significant list of references is \cite{weinberg}-\cite{mexico}. 

\begin{figure}[t]
\bc
\setlength{\unitlength}{1mm}
\parbox{25mm}{
\begin{fmfgraph*}(40,25)
  \fmfleft{i1,i2}
  \fmfright{o1,o2}
  \fmf{scalar}{i1,v}
  \fmf{scalar}{i2,v}
  \fmf{scalar}{v,o1}
  \fmf{scalar}{v,o2}
\fmflabel{$p_2$}{i1}
\fmflabel{$p_1$}{i2}
\fmflabel{$q_2$}{o1}
\fmflabel{$q_1$}{o2}
\end{fmfgraph*}}
\hspace*{2.5cm} 
\parbox{30mm}{
\begin{fmfgraph*}(40,25)
  \fmfleft{i1,i2}
  \fmfright{o1,o2}
 \fmf{scalar,tension=1/3}{i1,v1}
 \fmf{scalar,tension=1/3}{i2,v1}
 \fmf{fermion}{v1,v2}
 \fmf{scalar,tension=1/3}{v2,o1}
 \fmf{scalar,tension=1/3}{v2,o2}
\fmflabel{$p_2$}{i1}
\fmflabel{$p_1$}{i2}
\fmflabel{$q_2$}{o1}
\fmflabel{$q_1$}{o2}
\end{fmfgraph*}
}
\vspace*{2cm}

\parbox{30mm}{
\begin{fmfgraph*}(40,25)
  \fmfleft{i1,i2}
  \fmfright{o1,o2}
 \fmf{scalar,tension=1/3}{i1,v1}
 \fmf{scalar,tension=1/3}{i2,v2}
 \fmf{fermion}{v1,v2}
 \fmf{scalar,tension=1/3}{v1,o1}
 \fmf{scalar,tension=1/3}{v2,o2}
\fmflabel{$q_2$}{o1}
\fmflabel{$p_1$}{i2}
\fmflabel{$p_2$}{i1}
\fmflabel{$q_1$}{o2}
\end{fmfgraph*}
}
\hspace*{2.5cm}
\parbox{30mm}{
\begin{fmfgraph*}(40,25)
  \fmfleft{i1,i2}
  \fmfright{o1,o2}
 \fmf{scalar,tension=1/3}{i1,v1}
 \fmf{scalar,tension=1/3}{i2,v2}
 \fmf{fermion}{v1,v2}
 \fmf{scalar,tension=1/3}{v1,o1}
 \fmf{scalar,tension=1/3}{v2,o2}
\fmflabel{$p_2$}{i1}
\fmflabel{$p_1$}{i2}
\fmflabel{$q_1$}{o1}
\fmflabel{$q_2$}{o2}
\end{fmfgraph*}
}

\vspace*{1cm}
\parbox{10.5cm}{\caption{\small $\pi\pi\rightarrow\pi\pi$ scattering at tree-level in $O(2)$ model. Solid lines represent sigmas and dashed lines represent pions.\label{pipi2.fig}}}
\ec
\end{figure}
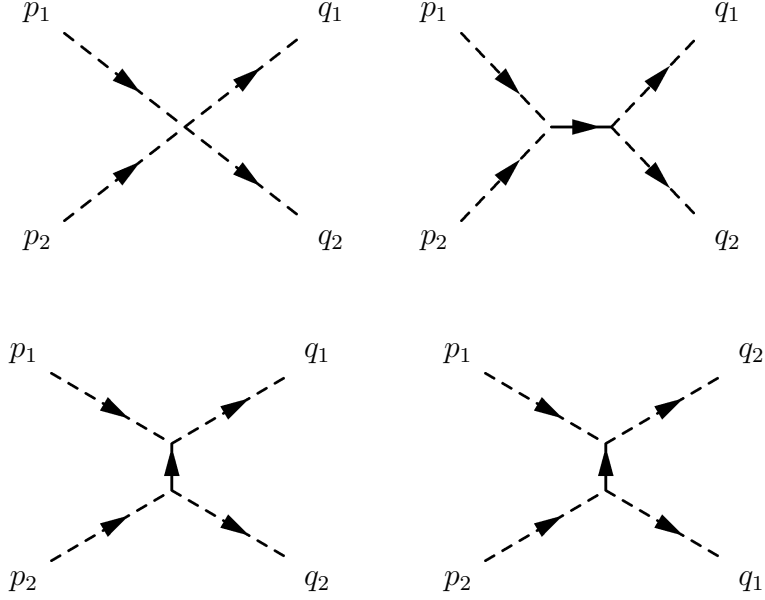

Here we study the scattering of a hard pion off a soft pion at tree-level. In our model with one pion field there are four diagrams contributing to this process: one 4-point vertex diagram and three 3-point vertex contributions involving a sigma exchange in the $s$, $t$ and $u$ channels. These are presented in figure \ref{pipi2.fig}. The transition amplitude is
\be 
{\cal M} = -2\lambda\left[ 3 + (m_\sigma^2-m_\pi^2)\left(\frac{1}{s-m_\sigma^2} + \frac{1}{t-m_\sigma^2} + \frac{1}{u-m_\sigma^2}\right)\right]\, ,
\ee
where $s=(p_1+p_2)^2$, $t=(p_1-q_1)^2$ and $u=(p_1-q_2)^2$, and $s+t+u = 4m_\pi^2$. 
Remember that for pion-sigma scattering all pion species were taken into account simply multiplying the result by an overall factor of $N-1$. In the case of pion-pion scattering the situation is more complicated since additional channels open up. Let us look at the model with $N=4$ for physical pions. With the usual definition of
\be
\pi^\pm = \frac{\pi_1\pm\pi_2}{\sqrt{2}}~~~~\mbox{and}~~~~\pi^0=\pi_3
\ee
figure \ref{pipi2.fig} shows the possible diagrams for $\pi^0\pi^0\rightarrow\pi^0\pi^0$ scattering.
\begin{figure}[htbp]
\bc
\setlength{\unitlength}{1mm}
\parbox{25mm}{
\begin{fmfgraph*}(40,25)
  \fmfleft{i1,i2}
  \fmfright{o1,o2}
  \fmf{scalar}{i1,v}
  \fmf{scalar}{i2,v}
  \fmf{scalar}{v,o1}
  \fmf{scalar}{v,o2}
\fmflabel{$\pi^0(p_1)$}{i1}
\fmflabel{$\pi^0(p_2)$}{i2}
\fmflabel{$\pi^-(q_2)$}{o1}
\fmflabel{$\pi^+(q_1)$}{o2}
\end{fmfgraph*}}
\hspace*{3.7cm} 
\parbox{30mm}{
\begin{fmfgraph*}(40,25)
  \fmfleft{i1,i2}
  \fmfright{o1,o2}
 \fmf{scalar,tension=1/3}{i1,v1}
 \fmf{scalar,tension=1/3}{i2,v1}
 \fmf{fermion}{v1,v2}
 \fmf{scalar,tension=1/3}{v2,o1}
 \fmf{scalar,tension=1/3}{v2,o2}
\fmflabel{$\pi^0(p_2)$}{i1}
\fmflabel{$\pi^0(p_1)$}{i2}
\fmflabel{$\pi^-(q_2)$}{o1}
\fmflabel{$\pi^+(q_1)$}{o2}
\end{fmfgraph*}
}

\vspace*{2cm}
\parbox{25mm}{
\begin{fmfgraph*}(40,25)
  \fmfleft{i2,o2}
  \fmfright{i1,o1}
  \fmf{scalar}{i1,v}
  \fmf{scalar}{i2,v}
  \fmf{scalar}{v,o1}
  \fmf{scalar}{v,o2}
\fmflabel{$\pi^0(-p_2)$}{o2}
\fmflabel{$\pi^+(-q_2)$}{i1}
\fmflabel{$\pi^0(p_1)$}{i2}
\fmflabel{$\pi^+(q_1)$}{o1}
\end{fmfgraph*}}
\hspace*{3.7cm} 
\parbox{30mm}{
\begin{fmfgraph*}(40,25)
  \fmfleft{i1,i2}
  \fmfright{o1,o2}
 \fmf{scalar,tension=1/3}{i1,v1}
 \fmf{scalar,tension=1/3}{i2,v2}
 \fmf{fermion}{v1,v2}
 \fmf{scalar,tension=1/3}{v1,o1}
 \fmf{scalar,tension=1/3}{v2,o2}
\fmflabel{$\pi^0(p_1)$}{i1}
\fmflabel{$\pi^+(-q_2)$}{i2}
\fmflabel{$\pi^0(-p_2)$}{o1}
\fmflabel{$\pi^+(q_1)$}{o2}
\end{fmfgraph*}
}

\vspace*{2cm}
\parbox{25mm}{
\begin{fmfgraph*}(40,25)
  \fmfleft{i2,o2}
  \fmfright{i1,o1}
  \fmf{scalar}{i1,v}
  \fmf{scalar}{i2,v}
  \fmf{scalar}{v,o1}
  \fmf{scalar}{v,o2}
\fmflabel{$\pi^0(-p_2)$}{o2}
\fmflabel{$\pi^-(q_2)$}{o1}
\fmflabel{$\pi^0(p_1)$}{i2}
\fmflabel{$\pi^-(-q_1)$}{i1}
\end{fmfgraph*}}
\hspace*{3.7cm} 
\parbox{30mm}{
\begin{fmfgraph*}(40,25)
  \fmfleft{i1,i2}
  \fmfright{o1,o2}
 \fmf{scalar,tension=1/3}{i1,v1}
 \fmf{scalar,tension=1/3}{i2,v2}
 \fmf{fermion}{v1,v2}
 \fmf{scalar,tension=1/3}{v1,o1}
 \fmf{scalar,tension=1/3}{v2,o2}
\fmflabel{$\pi^0(p_1)$}{i1}
\fmflabel{$\pi^-(q_2)$}{o2}
\fmflabel{$\pi^0(-p_2)$}{o1}
\fmflabel{$\pi^-(-q_1)$}{i2}
\end{fmfgraph*}
}

\vspace*{1cm}
\parbox{13cm}{\caption{\small $\pi^0\pi^0\rightarrow\pi^+\pi^-$, $\pi^0\pi^+\rightarrow\pi^0\pi^+$ and  $\pi^0\pi^-\rightarrow\pi^0\pi^-$ scattering at tree-level. Solid lines represent sigmas and dashed lines represent pions. The 4-momenta are shown in brackets. \label{pipi4.fig}}}
\ec
\end{figure}
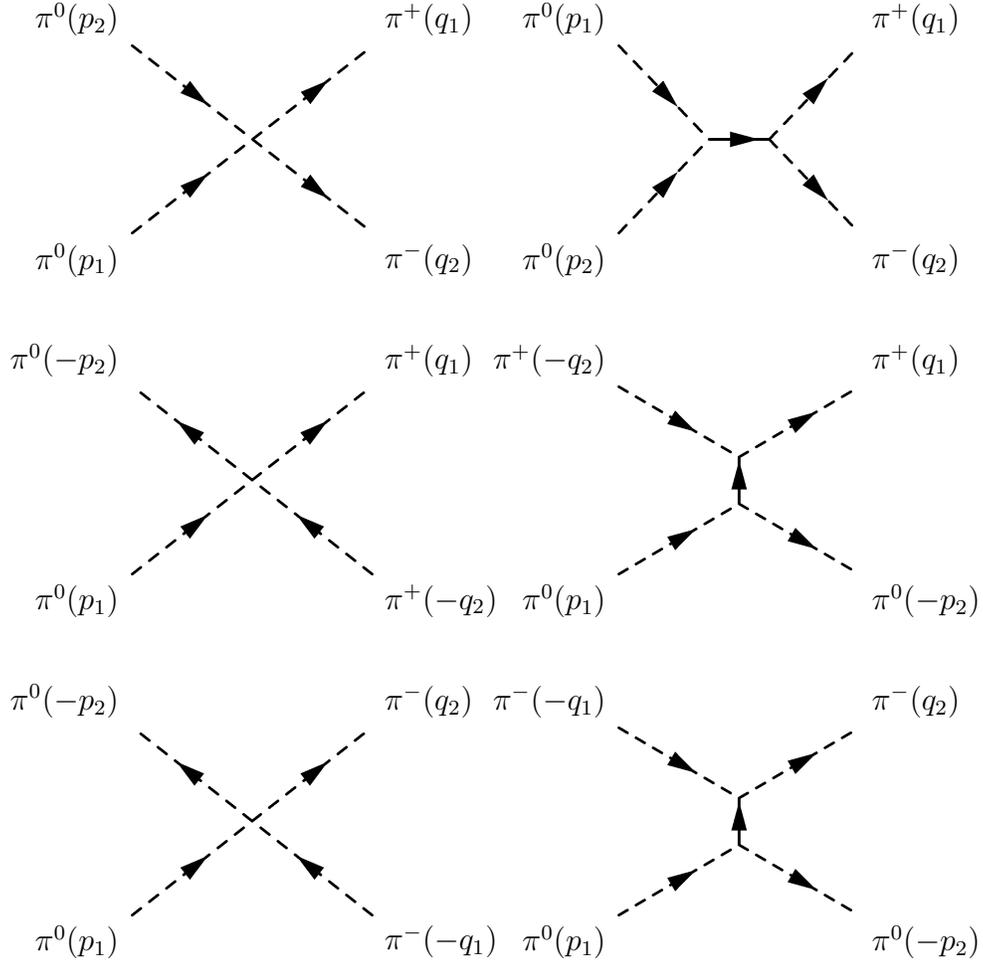 
\noindent
Since we are interested in the dissipation of the scalar condensate the cross-sections for $\pi^0\pi^0\rightarrow\pi^+\pi^-$, $\pi^0\pi^+\rightarrow\pi^0\pi^+$ and  $\pi^0\pi^-\rightarrow\pi^0\pi^-$ processes are evaluated. The possible diagrams are shown in figure \ref{pipi4.fig}. The corresponding transition amplitudes are 
\be 
{\cal M}_{00\rightarrow +-} =  -2\lambda\left[1 + \frac{m_\sigma^2-m_\pi^2}{s-m_\sigma^2}\right]
\ee
and
\be 
{\cal M}_{0+\rightarrow 0+} = {\cal M}_{0-\rightarrow 0-} = -2\lambda\left[1 + \frac{m_\sigma^2-m_\pi^2}{t-m_\sigma^2}\right]\,~.
\ee
Instead of writing long formulas for the cross-sections let us display only the leading term of the expansion of these about $s=4m_\pi^2$. This low energy limit is quite acceptable for $m_\sigma\gg T$. The sum of the contributions from all the processes of figures \ref{pipi2.fig} and \ref{pipi4.fig} is
\be
\sigma_{\pi\pi} = \frac{\lambda^2}{32\pi}\frac{m_\pi^2}{(m_\sigma^2-4m_\pi^2)^2}\left(23-16\frac{m_\pi^2}{m_\sigma^2}+128\frac{m_\pi^4}{m_\sigma^4}\right)\,~.
\label{lowcrosspipi}
\ee
It is important to mention that this is not the isospin averaged cross-section found in other papers \cite{prakash}. We study only scatterings that involve the neutral pion field, because condensates of charged pions would be destroyed by electromagnetic repulsion. Therefore contributions from processes like $\pi^+\pi^-\rightarrow\pi^+\pi^-$ have not been taken into account. In the $T\rightarrow T_c$ limit only 4-point vertices contribute to the amplitude, resulting in 
\be
\sigma_{\pi\pi} = \frac{15\lambda^2}{8\pi s}\,~.
\ee
In the rest frame of the $\pi^0$ the above two limits give
\be
\mbox{Im}~\Pi_{\pi\pi} \simeq -\frac{23}{32\pi^3}\lambda^2T^2e^{-m_\pi/T}\frac{m_\pi^3(m_\pi+T)}{(m_\sigma^2-4m_\pi^2)^2}
\label{lowhpipi}
\ee
and
\be
\mbox{Im}~\Pi_{\pi\pi} = -\frac{5\lambda^2T^2}{64\pi}\,~,
\label{highpipi}
\ee
respectively. The rate of dissipation due to massive pion-pion scattering is then
\be
\Gamma_{\pi\pi} \simeq \frac{23\lambda^2T^2}{64\pi}\frac{m_\pi^2(m_\pi+T)}{\frac{5}{46}m_\pi^3(m_\pi+T)+\pi^2(m_\sigma^2-4m_\pi^2)^2(e^{m_\pi/T}-1)}\,~.
\label{gammahpipi}
\ee

When the pion mass is zero the expression (\ref{lowcrosspipi}) vanishes. The first nonzero terms in the low-energy expansion of the cross-section are
\be
\sigma_{\pi\pi} = \frac{5\lambda^2}{12\pi}\frac{s}{m_\sigma^4} + \frac{\lambda^2}{4\pi}\frac{s^2}{m_\sigma^6} + \frac{23\lambda^2}{10\pi}\frac{s^3}{m_\sigma^8}\,~.
\label{lowpipi}
\ee
The leading term coincides with the current algebra representation for the total cross-section, eqn.~(7.14) with $m_\pi=0$ in \cite{schenk}. In the rest frame of the condensate $s=2E\omega-2pk\cos{\theta}$ and hard modes are considered to be on the light-cone, $E=p$. Then (\ref{restless}) gives to leading order
\be
\mbox{Im}~\Pi_{\pi\pi}(\omega,\vec{k}) \simeq -\frac{\pi}{36}\lambda^2\frac{T^4}{m_\sigma^4}\left(\frac{1}{3}k^2+\omega^2\right)\,~.
\label{lowimpipi}
\ee
Expressions (\ref{lowimpipi}) and (\ref{highpipi}) can be combined into an interpolating formula. However, since the sigma is heavy almost up to the critical temperature it is feasible to write the damping rate of Goldstone modes as
\be
\Gamma_{\pi\pi}(\omega,\vec{k}) = -\frac{\mbox{Im}~\Pi_{\pi\pi}(\omega,\vec{k})}{2\omega} \simeq \frac{\pi}{72}\lambda^2\frac{T^4}{m_\sigma^4}\left(\frac{1}{3}\frac{k^2}{\omega}+\omega\right)\,~.
\label{gammapipi}
\ee
Expressing dispersion relations in the form $\omega=v_\pi(k)k$, this damping rate vanishes linearly as $\vec{k}\rightarrow 0$ in the chiral limit, showing that Goldstone's Theorem is obeyed.

\section{Numerical Results}

All the numerics have been done for the model with $N=4$. 

\subsection{$\bf m_\pi\neq 0$}

\begin{figure}[htbp] 
\begin {center}
\leavevmode
\hbox{
\epsfysize=11.5cm
\epsffile{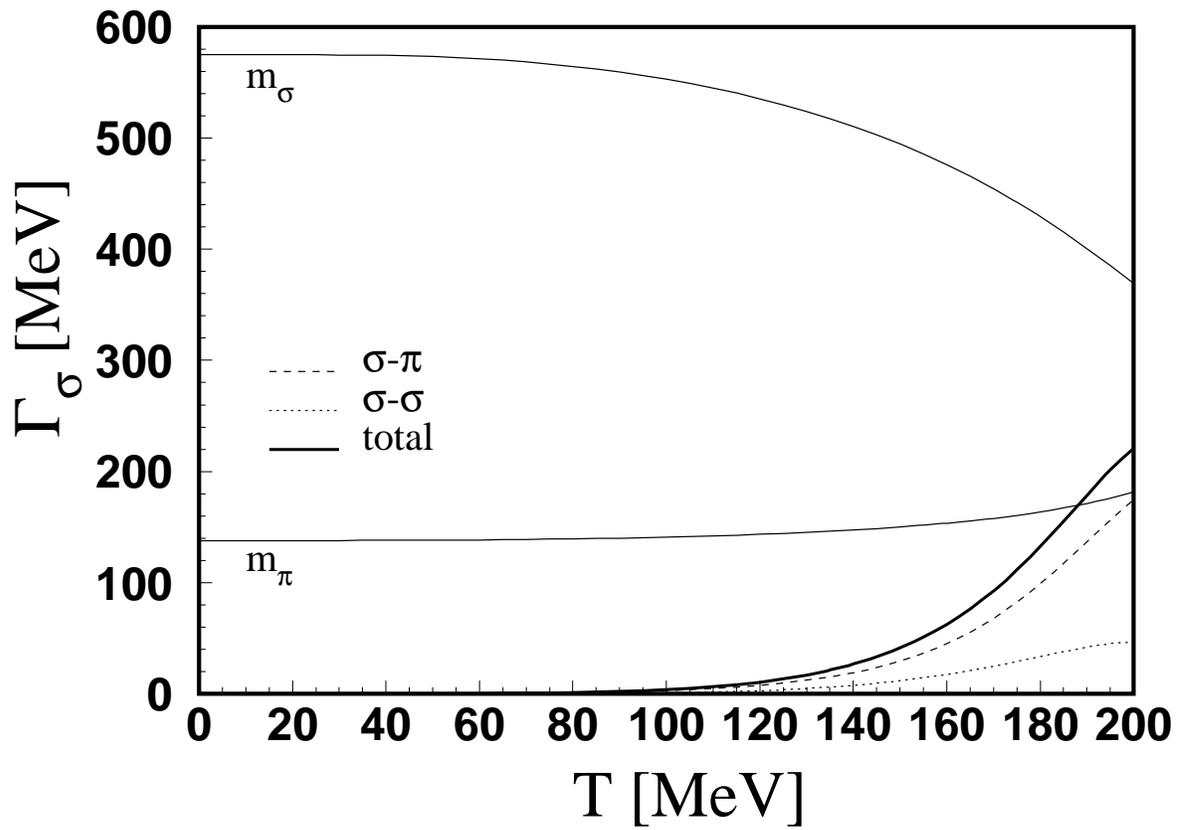}
}
\end{center}
\vspace*{5cm}
\caption{Scattering contribution to the width of the sigma meson as function 
of temperature in the model with explicitly broken chiral symmetry. 
\label{hsiscatt.fig}}
\end{figure}

The temperature dependence of the total scattering rate of the sigma meson, 
$\Gamma_{\sigma\sigma}+\Gamma_{\sigma\pi}$, is shown in figure 
\ref{hsiscatt.fig}. Here expressions (\ref{imhsisi}) and (\ref{gammahsipi}) 
have been used. The masses involved are the self-consistently determined meson
 masses. One can see that scattering is more accentuated at higher 
temperatures. 
For comparison we plotted the energy, which is the sigma mass 
in its rest-frame. The scattering contribution to the sigma damping rates is 
well below the energy. Moreover, dissipation of sigmas from the condensate 
due to their scattering is much smaller than due to their decay. This means 
that the scalar order parameter relaxes to its equilibrium value via the 
production of lighter pion fields. It also means that sigma mesons are so 
unstable that they are more likely to decay before they could ever scatter 
with other particles from the medium. The total damping rate is obtained by 
adding all the components
\be
\Gamma_\sigma = \Gamma_{\sigma\pi\pi} + \Gamma_{\sigma\sigma} + 
\Gamma_{\sigma\pi}\,~.
\ee
This is dominated by $\Gamma_{\sigma\pi\pi}$.
\begin{figure}[htbp] 
\begin {center}
\leavevmode
\hbox{
\epsfysize=11.5cm
\epsffile{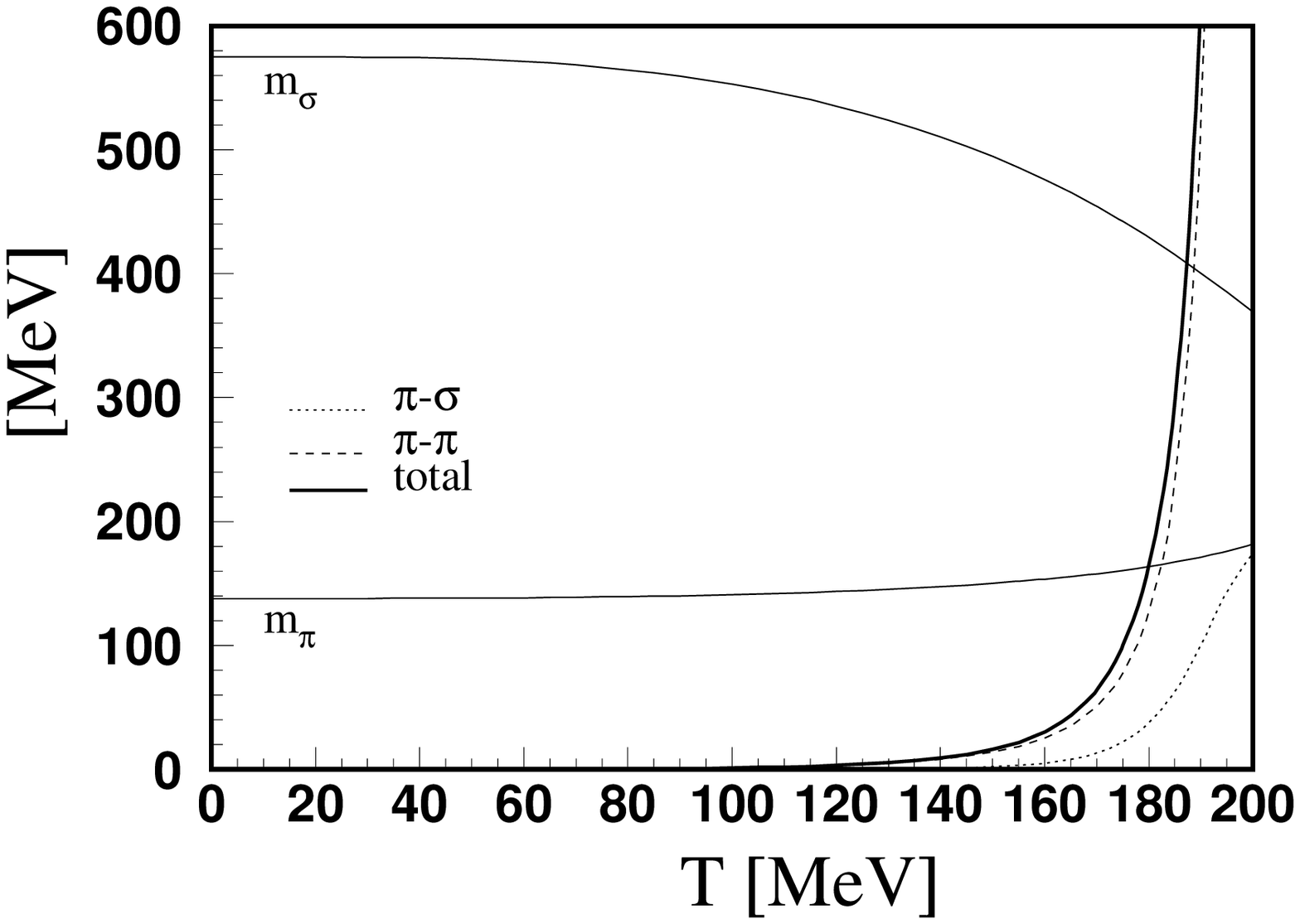}
}
\end{center}
\vspace*{5cm}
\caption{Scattering contribution to the width of the pion as function of 
temperature in the model with explicitly broken chiral symmetry. 
\label{hpi.fig}}
\end{figure} 

The scattering of long-wavelength massive pions according to formulas 
(\ref{gammahpisi}) and (\ref{gammahpipi}) is presented in figure 
\ref{hpi.fig}. Due to the heavy sigma exchange there is a strong suppression 
at low temperatures. When reaching $T\simeq 130~$MeV the scattering rate 
becomes significant and it is comparable in magnitude to the damping at 
1-loop order.

\subsection{$\bf m_\pi=0$}

\begin{figure}[htbp] 
\begin {center}
\leavevmode
\hbox{
\epsfysize=11.5cm
\epsffile{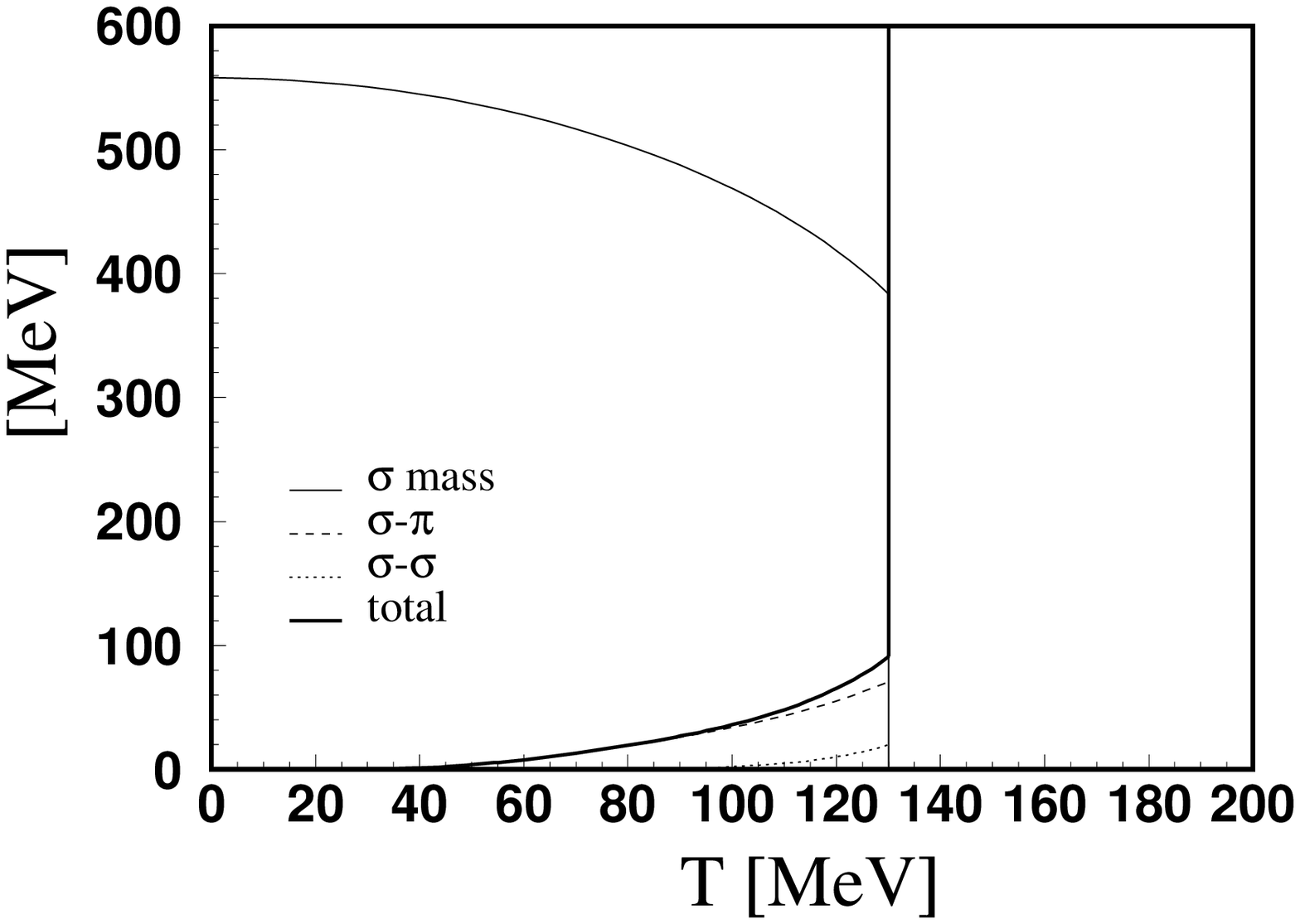}
}
\end{center}
\vspace*{5cm}
\caption{Scattering contribution to the width of the sigma meson as function 
of temperature in the model with exact chiral symmetry.\label{siscatt.fig}}
\end{figure}

Damping of heavy sigma mesons due to their scattering with other heavy sigmas 
and Goldstone bosons is shown as functions of temperature in figure 
\ref{siscatt.fig}. This rate is more important at higher temperatures, 
although it is not dominant over the damping due to sigma decay. 
\begin{figure}[htbp] 
\begin {center}
\leavevmode
\hbox{
\epsfysize=11.5cm
\epsffile{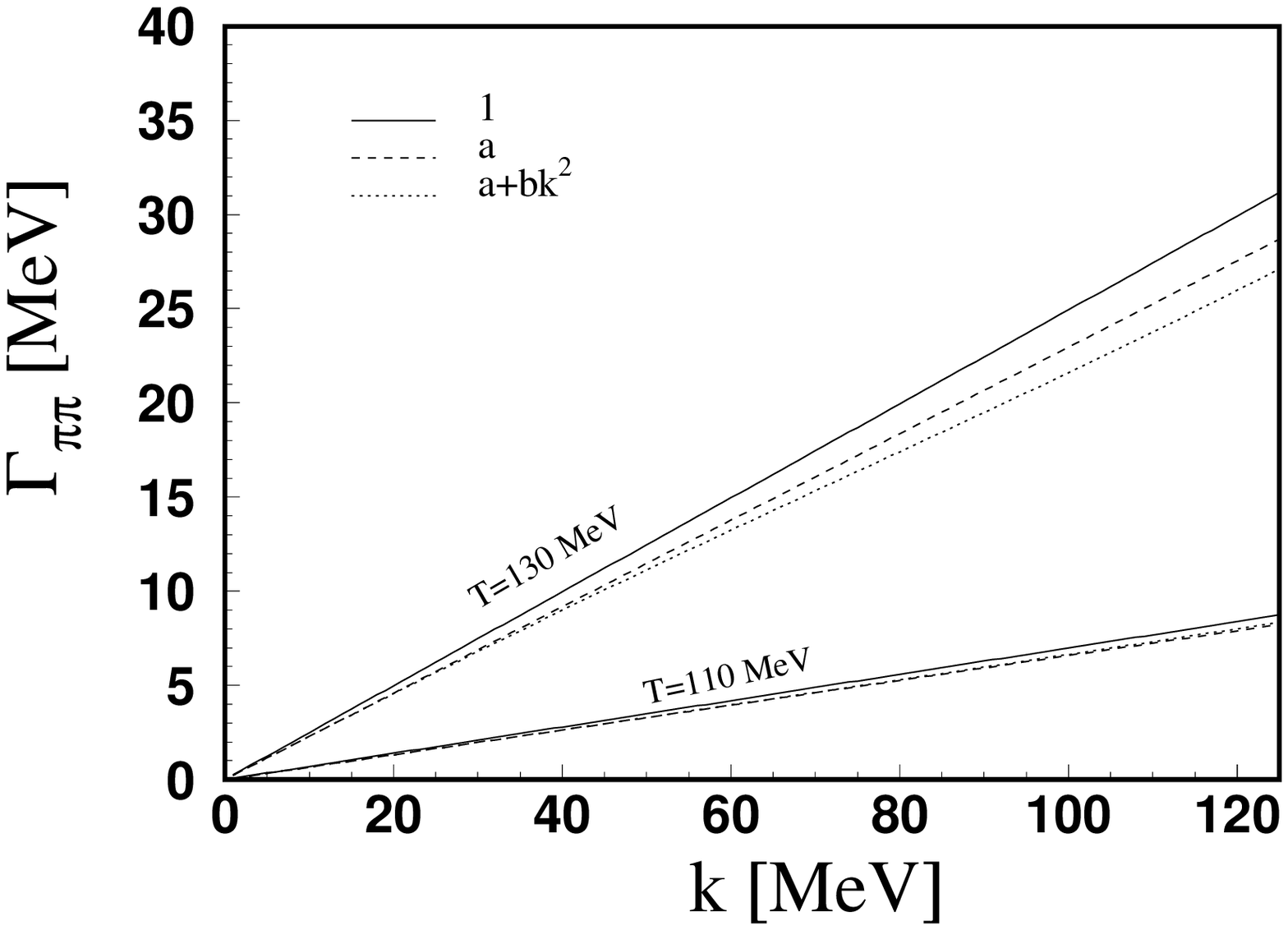}
}
\end{center}
\vspace*{5cm}
\caption{Massless pion-pion scattering rate versus momentum at $T=110~$MeV 
(lower curves) and at $T=130~$MeV (upper curves) with pion dispersion r
elations $\omega^2=v_\pi^2k^2$: $v_\pi=1$ (solid), $v_\pi^2=a$ (dashed) and 
$v_\pi^2=a+bk^2$ (dotted).  \label{pipi.fig}}
\end{figure} 

\begin{figure}[htbp] 
\begin {center}
\leavevmode
\hbox{
\epsfysize=11.5cm
\epsffile{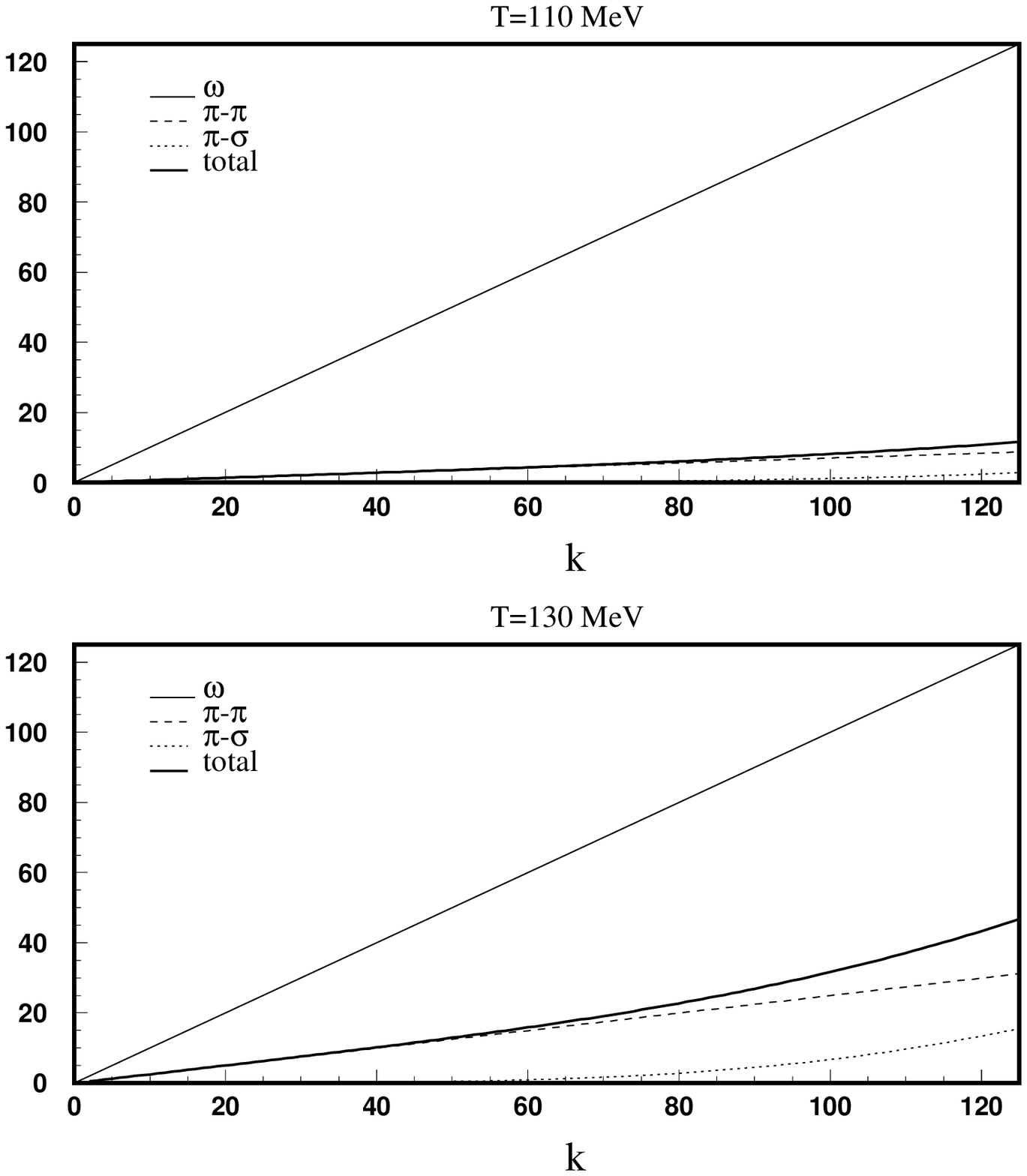}
}
\end{center}
\vspace*{5cm}
\caption{Scattering contribution to the width of the pion as function of temperature in the model with explicitly broken chiral symmetry. \label{pion.fig}}
\end{figure}

In figure \ref{pipi.fig} massless pion scattering processes' contribution to 
the damping of the condensate is presented. Notice that all curves start from 
the origin as they should. This figure is the result of expression 
(\ref{gammapipi}) using different dispersion relations. We have found that 
there is no significant difference in the massless $\pi\pi$ scattering rate 
when different dispersion relations are used. The rate of elastic scatterings 
is not affected by the speed of the pions. Assuming that pions are on the 
light-cone, $\omega=k$, the scattering contribution to the width of the pion 
of momentum $k$ is studied at different temperatures. Figure \ref{pion.fig} 
shows the results at $T=110~$MeV and at $T=130~$MeV. Harder momentum modes 
are damped more strongly. Damping of massless pions due to scattering turns 
out to be almost as significant as Landau damping. Pion modes thus can become 
broad. 

When $T_c$ is approached from above not only the heavy sigma fields but also the much lighter and even the massless pion fields are significantly damped. This fact should certainly be taken into account when considering formation and decay of DCCs.

\chapter{Conclusions}

In this dissertation different aspects of the non-equilibrium chiral/confinement phase transition as well as in-medium properties of the scalar and pseudoscalar mesons were investigated using quantum field theoretical descriptions at nonzero temperatures and densities. The $SU(2)_R\times SU(2)_L$ linear sigma model was used as an effective theory describing the chiral symmetry restoring phase transition of QCD. 

In Chapter \ref{sect-igor} the chiral phase transition at nonzero temperature and baryon chemical potential was studied at mean field level in the sigma model that includes quark degrees of freedom explicitly. Use of such a model was justified since high temperatures are dominated by quarks and low temperatures are dominated by mesons. The model was tuned to reproduce the correct properties in the vacuum. First the order of the phase transition was determined. For small bare quark masses we found a smooth crossover for nonzero temperature and zero chemical potential and a first order transition for zero temperature and nonzero chemical potential. The first order phase transition line in the ($T,\mu$) plane ends in the expected critical point. The behavior of quark and meson masses with changing temperature and chemical potential was also examined. It has been found that the $\sigma$ mass is zero at the critical point. Adiabatic lines were computed. The results of our numerical calculations for the behavior of the adiabats shows a pattern which is in contrast with some earlier predictions on this subject \cite{srs}. In fact, the phase transition found in this model turned out to be of the liquid-gas type. However, we showed that the strength of this transition depends strongly on the choice of coupling constants.

The theory including quarks is well suited to temperatures above the critical temperature. In the rest of the work we kept our focus on the phase below the critical temperature. The model thus was slightly changed: quarks were dropped and fluctuations of meson fields were included beyond the mean field. 

In Chapter \ref{sect-dcc} the concept of disoriented chiral condensates (DCC) was reviewed in the context of their possible formation as a result of a non-equilibrium phase transition from a chiral symmetry restored phase to the symmetry broken phase. Coarse-grained equations of motion for the soft homogeneous and inhomogeneous chiral condensate fields were derived. The sigma and pion field configurations were coupled to a heat bath of hard momentum sigma mesons and pions. Based on linear response theory the fluctuations of the thermal bath, as a response to the presence of the non-thermal condensate, were analyzed. Multiple aspects of the response functions were emphasized and discussed in chapters that followed. Throughout this work the theory with spontaneously broken symmetry and the theory with explicit chiral symmetry bearking were examined separately.

In Chapter \ref{sect-masses} the temperature-dependence of pion and sigma meson masses was studied. These mesons were dressed by the interactions with the thermal medium. Mathematically this was expressed via the response functions. A self-consistent evaluation of the meson masses including all one-loop order contributions was performed. The mass of the sigma meson is completely determined by the equilibrium condensate. Numerical results for the temperature-dependence of this show a first order chiral phase transition in the theory with spontaneously broken symmetry. However, we showed that changing the value of the coupling constant can render the transition second order. Special attention was given to the fulfillment of Goldstone's Theorem when the chiral symmetry is spontaneously broken, according to which the pion is massless at all orders in perturbation theory and at all temperatures. In the model with explicit symmetry breaking a smooth crossover between the symmetry broken and symmetric phases has been found.

In Chapter \ref{sect-dispersion}, the effect of the thermal medium on the velocity of long-wavelength pion modes was computed. Two different approaches were used to derive an expression for the velocity of pions at finite temperatures. Calculating in coordinate space directly from the response functions and in momentum space via the expansion of the pion self-energy about its small frequency and momentum yielded consistent results. Numerical analysis clearly showed a deviation of the velocity of massless pions from the speed of light at finite temperatures. Then the modification of the pion dispersion relation compared to the vacuum relation was discussed. The softening of the dispersion relation due to second-order contributions of the low momentum expansion was found. Going beyond this approximation and taking into account fourth-order contributions resulted in a less drastic softening with respect to the vacuum relation. The effect was found to be increasingly important with temperature for a given momentum mode, and for harder momenta at a given temperature. The limit of validity of our model was determined by requiring causality not to be violated, that is, the group velocity should not become superluminous. 

Finally, in Chapters \ref{sect-dissip1} and \ref{sect-dissip2} the dissipation of the long-wavelength sigma and pion condensate fields was examined at one- and two-loop order, respectively. Dissipation of the condensate occurs because energy can be transferred between the condensate and the heat bath through the interactions of the different degrees of freedom. In Chapter \ref{sect-dissip1} the effect at one-loop order was calculated by evaluating the response functions directly. Dissipation follows from first principles when deriving the equations of motion from the Lagrangian. The direct connection to the physical processes responsible for dissipation was shown. In Chapter \ref{sect-dissip2} the damping of the condensate fields due to elastic scatterings has been examined. These processes were neglected in the literature, based on the argument that they are higher order in coupling. Our results are contrary to previous beliefs according to which sigma mesons are quickly damped and, due to their small mass, pion oscillations persist in the small momentum region \cite{rajagopal}. We found that at high temperatures not only is the damping of sigma mesons significant, but also that of the pions. This damping is most accentuated in the phase transition region. The scattering rate of pions becomes significant at high temperatures and it is even comparable in magnitude to the damping due to the one-loop order Landau damping. Dissipation of sigmas from the condensate due to their scattering was found to be much smaller than their dissipation due to their decay. The results of these two Chapters proved that dissipative processes up to two-loop order are significant when considering possible formation and decay of DCCs.

\appendix
\chapter{Thermodynamics with Small Quark Masses \label{app-igor}}

In the chirally symmetric phase the constituent quark mass $M_q$ is small and it goes to zero in the chiral limit. Therefore, it is instructive to evaluate the thermodynamic potential $\Omega$ for small $M_q$. In this limit $\Omega_{q\bar{q}}$ can be expanded in a power series in $M_q$. The explicit expression is
\be
\Omega_{q\bar{q}}(T,\mu;M_q)=\Omega_0(T,\mu)+\frac{M^2}{2}\left(\frac{\partial^2 \Omega_{q\bar{q}}}{\partial M^2}\right)_{M=0}+...
\label{omqq}
\ee
Here the first term $\Omega_0(T,\mu)\equiv \Omega_{q\bar{q}}(T,\mu,0)$ can be easily calculated for arbitrary $T$ and $\mu$. The well-known result is
\be
\Omega_0(T,\mu)=-\frac{\nu_q}{2\pi^2}\left [\frac{7\pi^4}{180}T^4+\frac{\pi^2}{6}T^2\mu^2+\frac{1}{12}\mu^4 \right ]\,~.
\ee
The quark number and entropy densities for massless fermions are obtained by differentiating $\Omega_0(T,\mu)$ with respect to  $\mu$ and $T$, respectively,
\be
n=\frac{\nu_q}{6\pi^2}(\pi^2T^2\mu+\mu^3)\,~,
\ee
\be
s=\frac{\nu_q}{6\pi^2}\left (\frac{7\pi^4}{15}T^3+\pi^2T\mu^2 \right)\,~.
\ee
The second term in eq. (\ref{omqq}) differs only by a factor of $M_q$ from the scalar density defined in eq. (\ref{scaldens}). A straightforward calculation gives
\begin{equation}
\left (\frac{\partial^2\Omega}{\partial M^2} \right )_{M=0}=\left (\frac{\rho_s}{M} \right )_{M\rightarrow 0}=\nu_q \left (\frac{T^2}{12}+\frac{\mu^2}{4\pi^2} \right ).
\end{equation}
This can be used to estimate the pion and sigma masses at large $T$ and/or $\mu$. Expressing $M_q$ in terms of mean meson fields, equation (\ref{qmass}), and using the definition of effective masses, equation (\ref{mmass}), one arrives at the following asymptotic ($M\rightarrow 0$) expression for the pion and sigma masses
\be
M^2_{\pi}=M^2_{\sigma}=g^2\nu_q \left ( \frac{T^2}{12}+\frac{\mu^2}{4\pi^2}\right)\,~.
\ee
It shows that deep in the chiral-symmetric phase the pion and sigma masses are degenerate and large. At high temperatures $T\gg\mu$ and 
\be
M_\pi=M_\sigma=gT\,~,
\ee
where $g\sim 3$ in our calculations. Therefore, in this limit the contribution of pion and sigma excitations to the thermodynamical potential is suppressed by the Boltzmann factor $e^{-M_\sigma/T}=e^{-g}$.
\chapter{Imaginary Part of the Self-energy \label{ap-1loop}}

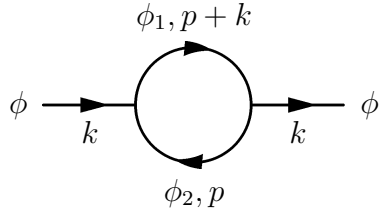
\begin{figure}[htbp]
\bc
\setlength{\unitlength}{1mm}
\parbox{50mm}{
\begin{fmfgraph*}(40,25)
  \fmfleft{i}
  \fmfright{o}
  \fmf{plain_arrow,label=$k$}{i,v1}
  \fmf{plain_arrow,label=$k$}{v2,o}
\fmf{plain_arrow,left,tension=.4,label=$\phi_1,,p+k$}{v1,v2} 
\fmf{plain_arrow,left,tension=.4,label=$\phi_2,,p$}{v2,v1} 
\fmflabel{$\phi$}{i}
\fmflabel{$\phi$}{o}
\end{fmfgraph*}}

\parbox{9cm}{\caption{\small One-loop self-energy contribution to $\phi$ with cubic coupling to $\phi_1$ and $\phi_2$. 
\label{general.fig}}}
\ec
\end{figure}

In the following we compute a very typical one-loop self-energy diagram, shown in figure \ref{general.fig}, in which the bosons have a cubic coupling $g\phi\phi_1\phi_2$. Here $g$ is the coupling constant and $\phi$, $\phi_1$ and $\phi_2$ have masses $M$, $m_1$ and $m_2$, respectively. 
\noindent
Following general finite temperature Feynman rules \cite{kapusta} the contribution to the self-energy of $\phi$, with four-momentum $k$, in imaginary time (Euclidean space) is
\be
\Pi(k) = \Pi(\omega_n,\vec{k}) = g^2T\sum_m\int{\frac{d^3p}{(2\pi)^3}}D_0^1(\omega_m+\omega_n,\vec{p}+\vec{k})D_0^2(\omega_m,\vec{p})\,\, .
\ee
The free imaginary-time propagators have the form
\be
D_0(\omega_n,\vec{p}) = \frac{1}{\omega_n^2+\vec{p}~^2+m^2}\,\, ,
\ee
and so the self-energy is
\be 
\Pi(\omega_n,\vec{k}) = g^2T\sum_m\int{\frac{d^3p}{(2\pi)^3}} \frac{1}{(\omega_m+\omega_n)^2+E_1^2}\frac{1}{\omega_m^2+E_2^2}\,\, ,
\label{matsub}
\ee
where $E_1=\sqrt{m_1^2+(\vec{p}+\vec{k})^2}$ and $E_1=\sqrt{m_2^2+\vec{p}~^2}$ are the on-shell energies of the internal lines. $\vec{k}$ is the momentum of the external boson $\Phi$, and $\omega_n$ are the Matsubara frequencies that take only discrete values, $\omega_n=2\pi nT$, with $n$ integer. It is worth noting that we keep the frequency $\omega_n$ and momentum $\vec{k}$ as independent variables, not requiring boson $\phi$ to be on mass-shell. There are standard ways of evaluating such Matsubara-sums. The interested reader is refered to \cite{kapusta,lebellac}. My preference is  replacing sums with contour integration. For a function $f(p^0=i\omega_n)$ with no singularities along the imaginary $p^0$ axis the sum is given by
\be
T\sum_{n=-\infty}^{\infty}f(p^0) = \frac{1}{2\pi i}\int_{-i\infty+\epsilon}^{i\infty+\epsilon}dp^0\left[f(p^0)+f(-p^0)\right]\left(\frac{1}{2}+\frac{1}{e^{\beta p^0}-1}\right)\,\, .
\ee
This replacement leads to the real time (Minkowski space) form of the self-energy:
\be
\Pi(k^0, \vec{k}) &=& g^2\int\frac{d^3p}{(2\pi)^3} \frac{1}{4E_1E_2}\left[(1+f_1+f_2)\left(\frac{1}{k^0-E_1-E_2}-\frac{1}{k^0+E_1+E_2}\right)\right.\nonumber\\
&& \qquad\qquad\qquad \left. +(f_2-f_1)\left(\frac{1}{k^0-E_1+E_2}-\frac{1}{k^0+E_1-E_2}\right)\right]
\ee
$f_i=f(E_i)=1/(\exp{(E_i/T)}-1)$ are the Bose-Einstein distribution functions of particle species $i$. One can extend this function to the whole complex plane by analytical continuation $k^0=i\omega_n=\omega+i\epsilon$.
\be
&&\!\!\!\!\!\!\!\!\!\!\!\Pi(\omega, \vec{k}) = g^2\int\frac{d^3p}{(2\pi)^3} \frac{1}{4E_1E_2}\left[(1+f_1+f_2)\left(\frac{1}{\omega-E_1-E_2+i\epsilon}-\frac{1}{\omega+E_1+E_2+i\epsilon}\right)\right.\nonumber\\
&& \qquad\qquad\qquad\qquad\left. +(f_2-f_1)\left(\frac{1}{\omega-E_1+E_2+i\epsilon}-\frac{1}{\omega+E_1-E_2+i\epsilon}\right)\right]
\label{general}
\ee
Since the physical interpretation of the various terms becomes more transparent in the imaginary part of the self-energy, in the following we derive the imaginary part of (\ref{general}). Apply the identity known from complex analysis:
\be
\frac{1}{k^0-E\pm i\epsilon} = \mbox{P}\frac{1}{k^0-E}\mp i\pi\delta(k^0-E)\,\, ,
\ee
where P means principle value. Then
\be
&&\!\!\!\!\!\!\!\!\!\!\!\mbox{Im}\Pi(\omega,\vec{k}) = -\pi g^2\int\frac{d^3p}{(2\pi)^3} \frac{1}{4E_1E_2}\left[(1+f_1+f_2)(\delta(\omega-E_1-E_2)-\delta(\omega+E_1+E_2)) \right.\nonumber\\
&& \qquad\qquad\qquad\qquad\qquad\left. +(f_2-f_1)(\delta(\omega-E_1+E_2)-\delta(\omega+E_1-E_2)) \right]
\label{imgen}
\ee
%

\section{Direct Link to Physical Processes}

Based on relativistic kinetic theory one can write the rate for a process to occur as the square of the amplitude of the process weighted with the proper statistical factors, thermal distributions, and integrated over the phase space. Thus, the probability for a particle $\phi$ to propagate through a $T\neq 0$ medium with energy $\omega\leq 0$ and momentum $\vec{k}$ will decrease with a rate $\Gamma_d(\omega)$ and will increase with a rate $\Gamma_i(\omega)$ given by \cite{weldon}:
\be
\Gamma_d(\omega, {\vec k}) = \frac{1}{2\omega}\int d\Omega_{ab}|{{\cal{M}}(\Phi 1...a\rightarrow 1'...b)}|^2f_1...f_a(1+f_{1'})...(1+f_b)
\label{direct}
\ee
\be
\Gamma_i(\omega, {\vec k}) = \frac{1}{2\omega}\int d\Omega_{ab}|{{\cal{M}}( 1'...b\rightarrow\Phi 1...a)}|^2f_{1'}...f_b(1+f_1)...(1+f_a)
\label{inverse}
\ee
the phase-space 
\be
 d\Omega_{ab}=dp_1...dp_adp_{1'}...dp_b (2\pi)^4\delta^4\left(k+\sum_{i=1}^ap_i-\sum_{j=1'}^bp_j\right)
\ee
\be
dp_i = \frac{d^3p_i}{(2\pi)^32E_i}
\ee

\end{fmffile}

\begin{thebibliography}{99}


\bibitem{gross} D.J. Gross and F. Wilczek, Phys. Rev. Lett. {\bf 30}, 1343 (1973); Phys. Rev. D {\bf 8}, 3633 (1973); {\it ibid.} {\bf 9}, 980 (1974).

\bibitem{politzer} H.D. Politzer, Phys. Rev. Lett. {\bf 30}, 1346 (1973).

\bibitem{linde} A.D. Linde, Rep. Prog. Phys. {\bf 42}, 389 (1979).

\bibitem{dolan} L. Dolan and R. Jackiw, Phys. Rev. D {\bf 9}, 3320 (1972). 

\bibitem{lattice} F. Karsch, hep-ph/0103314 (2001).

\bibitem{latticelong}  F. Karsch, hep-lat/0106019 (2001).

\bibitem{su2} M. Gell-Mann and M. Levy, Nuovo Cimento {\bf 16}, 705 (1960).

\bibitem{su3} M. Levy, Nuovo Cimento {\bf 52}, 23 (1967).

\bibitem{njl} Y. Nambu and G. Jona-Lasinio, Phys. Rev. {\bf 122}, 345 (1961).

\bibitem{bose} {\it Bose-Einstein Condensation}, ed. by A. Griffin, D.W. Snoke and S. Stringari, Cambridge University Press, Cambridge (1995). 

\bibitem{landau} L.D. Landau and E.M. Lifshitz, {\it Statistical Physics}, 3rd edition, Butterworth-Heinemann, Oxford (1980).

\bibitem{sm} J.F. Donoghue, E. Golowich and B.R. Holstein, {\it Dynamics of the Standard Model}, Cambridge University Press, Cambridge (1992).

\bibitem{oldlattice} See for instance F. Karsch, Nucl. Phys. Proc. Suppl. {\bf 83}, 14 (2000).

\bibitem{goldi} J. Goldstone, A. Salam and S. Weinberg, Phys. Rev. {\bf 127}, 965 (1962).

\bibitem{lenris} J.~T.~Lenaghan, D.~H.~Rischke and J.~Schaffner-Bielich, Phys. Rev D {\bf 62}, 085008 (2000).

\bibitem{boyavega} D.~Boyanovsky and H.~J.~de~Vega, Proceedings of the VI\`{e}me Colloque Cosmologie, Paris, 16-18 June 1999, and hep-ph/9909372 (1999).

\bibitem{workshop} Proceedings of RIKEN BNL Research Center Workshop on `{\it `Equilibrium $\&$ Non-Equilibrium Aspects of Hot, Dense QCD''}, July 17-30, 2000, RIKEN BNL Research Center (2000).


\bibitem{greg} G.~W.~Carter, {\it A Chiral Effective Approach to Nuclear Phenomena}, Ph. D. Dissertation, University of Minnesota (1997).

\bibitem{bhaduri} R.~K.~Bhaduri, {\it Models of the Nucleon: From Quarks to Soliton}, Addison-Wesley Publishing Company, Inc. (1988). 

\bibitem{lee} B.~W.~Lee, {\it Chiral Dynamics}, Gordon and Breach Science Publishers, New York (1972).

\bibitem{hands} S. Hands, Contemp. Phys. {\bf 42}, 209 (2001); physics/0105022 (2001).



\bibitem{qgpreviews} J.~Harris and B.~M\"uller, Ann.~Rev.~Nucl.~Part.~Sci.~{\bf 46}, 71 (1996).

\bibitem{raju} J. Bjoraker and R. Venugopalan, Phys. Rev. C {\bf 63}, 024609 (2001). 

\bibitem{jd} J.D. Bjorken, Phys. Rev. D {\bf 27}, 140 (1982).

\bibitem{rhic} For related information see the RHIC website: http://www.bnl.gov/RHIC/

\bibitem{jonathan} J.~T.~Lenaghan, {\it Effective Theories for the Chiral Symmetry Restoring Phase Transition in Quantum Chromodynamics}, Ph. D. Dissertation, Yale University (2000).

\bibitem{hu} Z. Fodor and S.D. Katz, hep-lat/0106002 (2001).

\bibitem{igor} I.~N.~Mishustin, Phys. Rev. Lett. {\bf 82}, 4779 (1999).

\bibitem{ms} I.~N.~Mishustin and O.~Scavenius, Phys. Rev. Lett. {\bf 83}, 3134 (1999).

\bibitem{oa} O.~Scavenius and A.~Dumitru, Phys.\ Rev.\ Lett.\  {\bf 83}, 4697 (1999).

\bibitem{srs} M.~Stephanov, K.~Rajagopal and E.~Shuryak, Phys.\ Rev.\ Lett.\  {\bf 81}, 4816 (1998).

\bibitem{critpoint} K. Rajagopal, {\it How to Find the QCD Critical Point}, in Proceedings of Conference on Strong and Electroweak Matter (SEWM 98), Copenhagen, Denmark, 2-5 Dec 1998, p. 112-129; hep-ph/9903547 (1999). 

\bibitem{halasz} M.~A.~Halasz, A.~D.~Jackson, R.~E.~Shrock, M.~A.~Stephanov and J.~J.~Verbaarschot, Phys.\ Rev.\ D {\bf 58}, 096007 (1998).

\bibitem{klevansky} T.~M.~Schwarz, S.~P.~Klevansky and G.~Papp, Phys.\ Rev.\ C {\bf 60}, 055205 (1999).

\bibitem{arw} M.~Alford, K.~Rajagopal and F.~Wilczek, Phys.\ Lett.\ B {\bf 422}, 247 (1998).

\bibitem{piswil} R.~D.~Pisarski and F.~Wilczek, Phys.\ Rev.\ D {\bf 29}, 338 (1984).

\bibitem{hsu} S.D.H. Hsu and M. Schwetz, Phys.\ Lett.\ B {\bf 432}, 203 (1998).

\bibitem{paper} O. Scavenius, \'A. M\'ocsy, I.N. Mishustin and D.H. Rischke, Phys.\ Rev.\ C {\bf 64}, 045202 (2001).

\bibitem{randrup} J.~Randrup,  Phys.\ Rev.\ D {\bf 55}, 1188 (1997).

\bibitem{csernai} L.~P.~Csernai and I.~N.~Mishustin, Phys.\ Rev.\ Lett.\ {\bf 74}, 5005 (1995).

\bibitem{mishusat} I.~N.~Mishustin, L.~M.~Satarov, H.~St\"ocker and W.~Greiner, Phys. Rev. C {\bf 62}, 034901 (2000).

\bibitem{leewick} T.~D.~Lee and G.~C.~Wick, Phys.\ Rev.\ D {\bf 9}, 2291 (1974).

\bibitem{ovead} O. Scavenius, A. Dumitru, E.S. Fraga, J.T. Lenaghan, A.D. Jackson, Phys.\ Rev.\ D {\bf 63}, 116003 (2001).

\bibitem{boya00} D.~Boyanovsky, H.J. de Vega and M. Simionato, Phys. Rev. D {\bf 63}, 045007 (2001). 

\bibitem{rw} K. Rajagopal and F. Wilczek, Nucl. Phys. B {\bf 399}, 395 (1993); {\it ibid.} {\bf 404}, 577 (1993). 

\bibitem{cent1} S.G. Bayburina {\it et al.}, Nucl. Phys. B {\bf 191}, 1 (1981).

\bibitem{cent2} C.M.G. Lattes, Y. Fujimoto and S. Hasegawa, Phys. Rep. {\bf 65}, 151 (1980).

\bibitem{rajagopal} See for example the review by K. Rajagopal, in {\it Quark-Gluon Plasma 2}, edited by R.C. Hwa, World Scientific, Singapore (1995) and references therein.

\bibitem{anselm} A. Anselm, Phys. Lett. B {\bf 217}, 169 (1989).

\bibitem{jdb} J.D. Bjorken, Int. J. Mod. Phys. A {\bf 7}, 4189 (1992); Acta Phys. Pol B {\bf 23}, 561 (1992). 

\bibitem{blaizot} J. P. Blaizot and A. Krzywicki, Phys. Rev. D {\bf 46}, 246 (1992). 

\bibitem{alaska} J.D. Bjorken, K.L. Kowalski and C.C. Taylor, in {\it Results and Perspectives in Particle Physics, 1993} Proceedings of Les Rencontres de Physique de la Vall\'ee d'Aoste, Le Thuile, Italy, edited by M. Greco, Editions Frontieres, Gif-sur-Yvette, France (1993).

\bibitem{gavingosch} S. Gavin, A. Gocksch and R. Pisarski, Phys. Rev. Lett. {\bf 72}, 2143 (1994).

\bibitem{gavinmuller} S. Gavin and B. Muller, Phys. Lett. B {\bf 329}, 486 (1994).

\bibitem{huang} Z. Huang and X. N. Wang, Phys. Rev. D {\bf 49}, 4335 (1994).

\bibitem{cooper} F. Cooper, Y. Kluger, E. Mottola and J.P. Paz, Phys. Rev. D {\bf 51}, 2377 (1995).

\bibitem{bialas} A. Bialas, W. Czyz and M. Gymrek, Phys. Rev. D {\bf 51}, 3739 (1995).

\bibitem{boya95} D.~Boyanovsky, H.~J. de Vega and R.~Holman, Phys. Rev. {\bf D 51}, 734 (1995); D. Boyanovsky, H.J. de Vega, R. Holman, D.S. Lee and A. Singh, {\it ibid.}, 4419 (1995). 

\bibitem{csmm} L.P. Csernai, I.N. Mishustin and \'A. M\'ocsy, Heavy Ion Phys. {\bf 3}, 151 (1996).

\bibitem{denes} T. Bir\'o, D. Moln\'ar, Z. Feng and L.P. Csernai, Phys. Rev. D {\bf 55}, 6900 (1997).

\bibitem{randrup-dcc} J. Randrup, Nucl. Phys. A {\bf 616}, 531 (1997); {\it ibid.} {\bf 630}, 468c (1998).

\bibitem{asakawa} Asakawa, Z. Huang and X. N. Wang, Phys. Rev. Lett. {\bf 74}, 3126 (1995).

\bibitem{french} A. Krzywicki and J. Serreau, Phys. Lett. B {\bf 448}, 257 (1999). 

\bibitem{axel} J. Kapusta and A. Vischer, Z. Phys. C. {\bf 75}, 507 (1997). 

\bibitem{boya96} D.~Boyanovsky, M.~D'Attanasio, H.~J. de Vega and R.~Holman, Phys. Rev. {\bf D 54}, 1748 (1996). 

\bibitem{biro} T. Bir\'o and C. Greiner, Phys. Rev. Lett. {\bf 79}, 3138 (1997).

\bibitem{greinermuller} C. Greiner and B. M\"uller, Phys. Rev. D {\bf 55}, 1026 (1997).

\bibitem{rischke} D. H. Rischke, Phys. Rev. C {\bf 58}, 2331 (1998).

\bibitem{magyar} Sz. Bors\'anyi, A.~Patk\'os and Zs.~Sz\'ep, Phys. Rev. B {\bf 469}, 188 (1999); A.~Jakov\'ac, A.~Patk\'os, P.~Petreczky and Zs.~Sz\'ep, Phys. Rev. D {\bf 61}, 025006 (2000).

\bibitem{cejk} L. P. Csernai, P. J. Ellis, S. Jeon and J. I. Kapusta,  Phys. Rev. C {\bf 61}, 054901 (2000).

\bibitem{bochkarev} A. Bochkarev and J. Kapusta, Phys. Rev. D {\bf 54}, 4066 (1996).

\bibitem{bodecker} D. B\"{o}decker, L. McLerran and A. Smilga, Phys. Rev. D {\bf 52}, 4675 (1995).

\bibitem{peskin} M. E. Peskin and D. V. Schroeder, {\it An Introduction to Quantum Field Theory}, Addison-Wesley (1995).

\bibitem{walecka} A. L. Fetter and J.D. Walecka, {\it Quantum Theory of Many-Particle Systems}, McGraw-Hill (1965).

\bibitem{kapusta} J. I. Kapusta, {\it Finite-temperature Field Theory}, Cambridge University Press, Cambridge (1989). 

\bibitem{lebellac} M. LeBellac, {\it Thermal Field Theory}, Cambridge University Press, Cambridge (1996).

\bibitem{renorm} J.T. Lenaghan and D.H. Rischke, J. Phys. G {\bf 26}, 431 (2000). 

\bibitem{rpa} Z. Aouissat and M. Belkacem, Phys. Rev. C {\bf 60}, 065211 (1999).

\bibitem{largeN} For a comprehensive introduction to the 1/N expansion, see E. Witten, Physics Today, July (1980).

\bibitem{baym} G. Baym and G. Grinstein, Phys. Rev. D {\bf 15}, 2897 (1977).

\bibitem{rob} R. D. Pisarski and M. Tytgat, Phys. Rev. D {\bf 54}, R2989 (1996); hep-ph/9606459.

\bibitem{ayala} A. Ayala, S. Sahu and M. Napsuciale, Phys. Lett. B {\bf 479}, 156 (2000); A. Ayala and S. Sahu, Phys. Rev. D {\bf 62}, 056007 (2000).

\bibitem{gale} C. Gale and J. Kapusta, Phys. Rev. C {\bf 35}, 2107 (1987).

\bibitem{leutwyler} J. Gasser and H. Leutwyler, Phys. Lett. B {\bf 188}, 477 (1987).

\bibitem{goity} J.L. Goity and H. Leutwyler, Phys. Lett. B {\bf 184}; 83 (1989); {\bf 228}, 425 (1989).

\bibitem{song} C. Song, Phys. Rev. D {\bf 49}, 1556 (1994). 

\bibitem{kapwong} J. I. Kapusta and S. M. H. Wong, hep-th/0103065 (2001).

\bibitem{weldon} H. A. Weldon,  Phys. Rev. D {\bf 28}, 2007 (1983). 

\bibitem{shuryak} E. V. Shuryak, Nucl. Phys. A {\bf 533}, 761 (1991).  

\bibitem{eletsky} V.L. Eletsky and J.I. Kapusta, Phys. Rev. C {\bf 59}, 2757 (1999).

\bibitem{adler} S. Adler, Phys. Rev. {\bf 139}, B 1638 (1965)

\bibitem{weinberg} S. Weinberg, Phys. Rev. Lett. {\bf 17}, 616 (1966).

\bibitem{basdevant} J.L. Basdevant and B.W. Lee, Phys. Rev. D {\bf 2}, 1680 (1970).

\bibitem{donoghue} J.F. Donoghue, C. Ramirez and G. Valencia, Phys. Rev. D {\bf 38}, 2195 (1988).

\bibitem{schenk} A. Schenk, Phys. Rev. D {\bf 47}, 5138 (1993).

\bibitem{mexico} J.L. Lucio M., M Napsuciale and M. Ruiz-Altaba, hep-ph/9903420 (1999).

\bibitem{prakash} M. Prakash, M. Prakash, R. Venugopalan and G. Welke, Phys. Report {\bf 227}, 321 (1993).

\bibitem{brownian} L. P. Csernai, S. Jeon and J. I. Kapusta,  Phys. Rev. E {\bf 56}, 6668 (1997). 

\bibitem{fetter} A.L.~Fetter and J.D.~Walecka, {\it Theoretical Mechanics of Particles and Continua}, McGraw-Hill, Inc. (1980). 

\end{thebibliography}
\end{document}